\documentclass[prd, aps, floatfix,nofootinbib, 10 pt]{revtex4}

\usepackage{amsmath,graphicx,color,epsfig}

\begin{document}

\title{Analysis of $B_{c}\rightarrow D_{s}^{\ast }\ell^{+}\ell^{-}$ in the
Standard Model Beyond Third Generation}
\author{Ishtiaq Ahmed$^{1,2}$}
\email{ishtiaq@ncp.edu.pk}
\author{M. Ali Paracha$^{1,2}$}
\email{paracha@phys.qau.edu.pk}
\author{M. Junaid$^{1,2}$}
\email{junaidmuhamad@gmail.com}
\author{Aqeel Ahmed$^{1}$}
\email{aqeel@ncp.edu.pk}
\author{Abdur Rehman$^{1}$}
\email{rehman@ncp.edu.pk}
\author{M. Jamil Aslam$^{2}$}
\email{jamil@phys.qau.edu.pk; jamil@ncp.edu.pk}
\affiliation{$^{1}$National Centre for Physics, Quaid-i-Azam University Campus, Islamabad 45320, Pakistan\\
$^{2}$Physics Department, Quaid-i-Azam University, Islamabad 45320, Pakistan}

\begin{abstract}
We study the FCNC $B_{c} \to D_{s}^{\ast} \ell^{+} \ell^{-}$ ($\ell=\mu, \,\,
\tau$) transition in the Standard Model with fourth generation (SM4).
Taking fourth generation quark mass $m_{t^{\prime }}$ of about $300$ to $600$ GeV with the
CKM matrix elements $\left\vert V_{t^{\prime }b}^{\ast }V_{t^{\prime }s}\right\vert
$ in the range $(0.03-1.2)\times 10^{-2}$ and using the new $CP$ odd phase $(\phi_{sb})$ to be $90^{o}$,
 the analysis of decay rates,
forward-backward asymmetries (FBA), lepton polarization asymmetries $(P_{L,N,T})$ and
the helicity fractions of $D_{s}^{\ast}$ meson $(f_{L,T})$ in
$B_{c} \to D_{s}^{\ast} \ell^{+} \ell^{-}$ ($\ell=\mu, \,\,
\tau$) is made. It is found that in the fourth generation
parameter space the above mentioned physical observables deviates sizably from their
SM values in $B_{c} \to D_{s}^{\ast} \mu^{+} \mu^{-}$. Furthermore, an optimum shift
in the zero position of the FBA in this decay has also been pointed out.
Compared to the dimuon case, the SM4 effects are somewhat mild in the
decay rate and in the FBA for $B_{c} \to D_{s}^{\ast} \tau^{+} \tau^{-}$ decay.
However, they came up in a distinguishing way in
longitudinal and transverse lepton polarizations and also in the helicity fractions of the $D_{s}^{\ast}$
meson which
differ distinctively from their SM values. Thus the study of these physical observables
will provide us useful
information to probe new physics and helps us to search the fourth generation of quarks $%
(t^{\prime },b^{\prime })$ in an indirect way.
\end{abstract}

\maketitle

\date{\today}


\section{ Introduction}

It is well known that the Standard Model (SM) includes three generations of
fermions, but it does not prohibit the fourth generation. The restrictions
on the number of fermion generations come from the QCD asymptotic freedom
which constraint them to nine. Therefore, shortly after the measurement of the third
generation, a fourth generation was an obvious extension.

Interest in the fourth generation Standard Model (SM4) was fairly high in
the 1980s until the electroweak precision data seemed to rule it out. The
other reason which shook the interest in the fourth generation was
the measurement of the number of light neutrinos at
the $Z$ pole that showed only three light neutrinos could exist. However,
the discovery of neutrino oscillations suggested the possibility of a mass
scale beyond the SM, and the models with the sufficiently massive neutrino
became acceptable \cite{Pree}. Though the early study of the EW precision
measurements ruled out a fourth generation \cite{PDG}, however it was
subsequently pointed out \cite{Maltoni} that if the fourth generation masses
are not degenerate, then the EW precision data do not prohibit the fourth
generation \cite{Kible}. Therefore, the SM can be simply extended with a
sequential repetition as four quark and four lepton left handed doublets and
corresponding right handed singlets.

The possible sequential fourth generation may play an important role in
understanding the well known problem of CP violation and flavor structure of
standard theory \cite{3, 4, 5, 6, 7, 8, 9}, electroweak symmetry breaking
\cite{10, 11, 12, 13}, hierarchies of fermion mass and mixing angle in
quark/lepton sectors \cite{14, 15, 16, 17, 18}. A thorough discussion on the
theoretical and experimental aspects of the fourth generation can be found in
ref. \cite{19}.

On the experimental side, recent searches by the CDF collaboration for direct production of fourth
generation up-type quark $(t^{\prime})$ and down-type quark $(b^{\prime})$
found $m_{t^\prime}>335$ GeV \cite{20} and $m_{b^\prime}>385$ GeV \cite{21},
assuming $Br(t^{\prime}\to Wq, ({q = d, s, b}))=100\%$ and $Br(b^{\prime} \to Wt)
= 100\%$ respectively. This indeed suggest that the fourth generation
fermion must be heavy which supports the scenario of compositeness. The
underlying assumption to perform these searches is that $m_{t^\prime}-m_{b^%
\prime}<M_{W}$ and negligible mixing of the $(t^{\prime}, b^{\prime})$
states with the two lightest quark generations. To account for EW precision data
such conditions are
generally required for the SM4 with the one Higgs doublet \cite{22}.
Moreover, when a fourth generation of fermions is
embedded in theories beyond the SM, the large splitting case $%
(m_{t^\prime}-m_{b^\prime}>M_{W})$ and the inverted scenario $%
(m_{t^\prime}<m_{b^\prime})$ have not been excluded. Recently, it has also been shown
\cite{23} that the precision EW data can accommodate $%
(m_{t^\prime}-m_{b^\prime}>M_{W})$ if there are two Higgs doublets. Thus
there is no uniquely interesting set of assumptions under which experimental
data must be interpreted \cite{24} and the determination of allowed
parameter space of fourth generation fermions will be an important goal of
the LHC era. The large values of the masses of fourth generation would
provide special advantage to new interactions originating at a higher scale
and the precise determination of the fourth generation quark properties may
present the existence of physics beyond the SM.

It is necessary to mention here that these new particles are heavy in nature, consequently
they are hard to produce in the accelerators. Therefore, we have to go for some alternate
scenarios where we can find their influence. In this regard,
the Flavor Changing Neutral Current (FCNC) transitions provide an ideal plateform to establish new physics (NP).
This is because of the fact that FCNC transitions are not allowed at
tree level in the SM and are allowed at loop level through GIM mechanism
which can get contributions of NP from newly proposed particles via loop diagrams. Among
different FCNC transitions the one $b\rightarrow s$ transition plays a
pivotal role to perform efficient tests of NP scenarios \cite{25, 26,
27, 28, 29, 30, 31, 32, 33}. It is also the fact that $CP$ violation in $%
b\rightarrow s $ transitions is predicted to be very small in the SM, thus,
any experimental evidence for sizable $CP$ violating effects in the $B$
system would clearly point towards a NP scenario. However, among the other
NP scenarios such as the Littlest Higgs model with T-parity (LHT) and
Randall-Sundrum (RS) models, the study of FCNC transitions in SM4 contain much fewer
parameters and has the possibility of having simultaneously sizable NP
effects in the $K$ and $B$ systems compared to the above mentioned NP
models. In this context the constraint on the mixing between the fourth and
third generation by using FCNC processes
have been studied \cite{34} along with the effects of the sequential fourth generation
on different physical observables in $B_{(d,s)}$, $%
K$ and $D$ decays, see ref. \cite{reviews} for an incomplete list. The study
of the $B$ system will be even more complete if one consider the similar
decays of the charmed $B$ mesons $(B_{c})$. As $B_{c}$
is a bound state of two heavy quarks $b$ and $c$, therefore it is rich in
phenomenology compared to other $B$ mesons. In literature, some of the
possible radiative and semileptonic exclusive decays of $B_{c}$ mesons like $%
B_{c}\rightarrow \left( \rho ,K^{\ast },D_{s}^{\ast },B_{u}^{\ast }\right)
\gamma ,B_{c}\rightarrow \ell\nu \gamma $ $,B_{c}\rightarrow B_{u}^{\ast
}\ell^{+}\ell^{-},B_{c}\rightarrow D_{1}^{0}\ell\nu ,B_{c}\rightarrow D_{s0}^{\ast
}\ell^{+}\ell^{-}$ and $B_{c}\rightarrow D_{s,d}^{\ast }\ell^{+}\ell^{-}$ have been
studied using the frame work of relativistic constituent quark model \cite%
{49}, QCD Sum Rules and Light Cone Sum Rules \cite{48}. In this work we
will focus on the semileptonic $B_{c}\rightarrow D_{s}^{\ast }\ell^{+}\ell^{-}$
decays in SM4.

The special feature of the semileptonic $B_{c}\rightarrow D_{s}^{\ast }\ell ^{+}%
\ell ^{-}$ decays compared to the other $B$ meson decays such as $%
B^{0}\rightarrow (K^{\ast },K_{1},\rho ,\pi )\ell^{+}\ell^{-}$ is that this decay
can occur in two distinct ways i.e. through FCNC transitions and through Weak
Annihilations (WA). In the ordinary charged $B$ meson decays the WA contributions are
very small and can be safely ignored. However, for $B_c$ meson these WA
contributions are proportional to the CKM matrix elements $V_{cb}V^{*}_{cs}$
and hence can not be ignored. \textit{The decay under discussion here has already been studied in the
Universal Extra Dimension (UED) model \cite{51} where it has been seen that
the WA contribution suppress the NP effects coming through UED model in different physical observables.
Therefore, it is interesting to
see how SM4 effects in different physical observables will behave in the presence of these WA contributions}.

While working on the exclusive $B$-meson
decays the main job is to calculate the form factors which are the non
perturbative quantities and are the scalar functions of the square of
momentum transfer $(q^2)$. In literature the form factors for $B_{c}\rightarrow
D_{s}^{\ast }\ell^{+}\ell^{-}$ ($\ell=\mu,\tau)$ decays were calculated using different approaches,
such as light front constituent quark models, a relativistic quark model,
QCD sum rules and through Ward identities \cite{49, 50, 51, 52, 53a, 53b}. In this
work we will use QCD Sum rules form factors \cite{53a, 53b} to
study the fourth generation effects on different physical observables such as
branching ratio, forward-backward asymmetry, polarization
asymmetries of leptons and the helicity fractions of final state meson $%
(D_{s}^{\ast })$ for these decays. As the expected number of events for the
production of $B_{c}$ meson at the Large Hadron Collider (LHC)
are about $10^{8}-10^{10}$ per year \cite{52} therefore, we hope that
the phenomenological study of the $B_{c}$ meson could provide us a valuable
tool for distinguishing the SM4 effects from the SM and other NP scenarios.

The paper is organized as follows. In Sec. II, we discuss the two different
contributions to the amplitude of $B_{c}\rightarrow D_{s}^{\ast } \ell^{+} \ell^{-}$
decays which were named as WA and penguin contributions. These can be
parameterized in terms of the form factors, where the values of the
form factors appearing in the calculation of WA amplitude will be used from \cite%
{53a} and for the penguin contributions will be taken from \cite{53b}%
. Section III presents the basic formulas for physical observables like
decay rate, lepton forward-backward asymmetry, helicity fractions of $D_{s}^{\ast }$ meson
and the polarization asymmetries of the final state lepton. The
numerical study of these observables will be given in Section V and the Section
VI gives the summary of the main outcomes of this study.

\section{Theoretical framework for $B_{c}\rightarrow D_{s}^{\ast } \ell^{+} \ell^{-}$
decays}

\subsection{Weak Annihilation Amplitude}

The charmed $B$-meson $(B_{c})$ is made up of two different heavy flavors, $b$%
-quark and $c$-quark, which brings WA contributions into the play. Following
the procedure given in refs. \cite{54, 55} for $B_{c}\rightarrow D_{s}^{\ast
}\gamma $ decay the WA amplitude for the decay $B_{c}\rightarrow D_{s}^{\ast
}\ell ^{+}\ell ^{-}$ can be written as
\begin{equation}
\mathcal{M}^{\text{WA}}\mathcal{=}\frac{G_{F}\alpha }{2\sqrt{2}\pi }%
\frac{f_{D_s^{\ast}}f_{B_c} }{q^2}V_{cb}V_{cs}^{\ast }\left[ -i\epsilon _{\mu \nu \alpha \beta }\varepsilon
^{\ast \nu }p^{\alpha }q^{\beta }F_{V}^{D_{s}^{\ast }}(q^{2})+\left(
\varepsilon \cdot qp_{\mu }+p\cdot q\varepsilon _{\mu }\right)
F_{A}^{D_{s}^{\ast }}(q^{2})\right] \bar{l}\gamma ^{\mu }l  \label{1a}
\end{equation}%
where $f_{B_c}$ and $f_{D_s^{\ast}}$ are the decay
constants of $B_{c}$ and $D_s^{\ast}$ mesons, respectively. The functions $F_{V}^{D_{s}^{\ast }}(q^{2})$
and $F_{A}^{D_{s}^{\ast }}(q^{2})$ are
the weak annihilation form factors which are calculated in QCD
Sum Rules \cite{53a} and can be parameterized as:
\begin{equation}
F_{V,A}^{D_{s}^{\ast }}(q^{2})=\frac{F_{V,A}^{D_{s}^{\ast }}(0)}{1+\alpha
\hat{q}+\beta \hat{q}^{2}}  \label{1b}
\end{equation}
where $\hat{q}=q^{2}/M_{B_{c}}^{2}.$ The values of the form factors $%
F_{V}^{D_{s}^{\ast }}(0)$ and $F_{A}^{D_{s}^{\ast }}(0)$ along with the
given values of parameters $\alpha $ and $\beta $ were calculated by Azizi
et al. \cite{53a} which are summarized in Table I.
\begin{table}[tbh]
\caption{$B_{c}\rightarrow D_{s}^{\ast }$ form factors corresponding to WA
in the QCD Sum Rules.
$F(0)$ denotes the value of form factors at $q^{2}=0$ while $\alpha$ and $\beta$
are the parameters in the parameterizations shown in Eq. (\protect\ref%
{ff-param})\protect\cite{53a}.}
\label{di-fit B to Dstar}%
\begin{tabular}{cccc}
\hline\hline
 $F(q^{2})$ & $\hspace{2cm}F(0)$ & $\hspace{2cm}\alpha$ & $\hspace{%
2cm}\beta$ \\ \hline
 $F_{V}^{D_{s}^{\ast }}\left( q^{2}\right) $ & $\hspace{2cm}0.23$ & $%
\hspace{2cm}-1.25$ & $\hspace{2cm}-0.097$ \\ \hline
 $F_{A}^{D_{s}^{\ast }}(q^{2})$ & $\hspace{2cm}0.25$ & $\hspace{2cm}%
-0.10$ & $\hspace{2cm}-0.097$ \\ \hline
\end{tabular}%
\end{table}
\subsection{Penguin Amplitude}

At quark level the decay $B_{c}\rightarrow D_{s}^{\ast} \ell^{+}\ell^{-}$ $(\ell=\mu,\tau)$
is governed by the transition $b\rightarrow \ s \ell^{+} \ell^{-}$ for which the
effective Hamiltonian can be written as
\begin{equation}
H_{eff}=-\frac{4G_{F}}{\sqrt{2}}V_{tb}^{\ast }V_{ts}{\sum\limits_{i=1}^{10}}%
C_{i}({\mu })O_{i}({\mu }),  \label{effective hamiltonian 1}
\end{equation}%
where $O_{i}({\mu })$ $(i=1,\ldots ,10)$ are the four-quark operators and $%
C_{i}({\mu })$ are the corresponding Wilson\ coefficients at the energy
scale ${\mu }$ and the explicit expressions of these in the SM at NLO and
NNLL are given in \cite{Buchalla, Buras, Ball, Ali, Kim, Kruger,Grinstein, Cella,
Bobeth, Asatrian, Misiak, Huber}. The operators responsible for $%
B_{c}\rightarrow D_{s}^{\ast }\ell^{+}\ell^{-}$ are $O_{7}$, $O_{9}$ and $O_{10}$ and
their form is given by
\begin{eqnarray}
O_{7} &=&\frac{e^{2}}{16\pi ^{2}}m_{b}\left( \bar{s}\sigma _{\mu \nu
}P_{R}b\right) F^{\mu \nu },\,  \notag \\
O_{9} &=&\frac{e^{2}}{16\pi ^{2}}(\bar{s}\gamma _{\mu }P_{L}b)(\bar{l}\gamma
^{\mu }l),\,  \label{op-form} \\
O_{10} &=&\frac{e^{2}}{16\pi ^{2}}(\bar{s}\gamma _{\mu }P_{L}b)(\bar{l}%
\gamma ^{\mu }\gamma _{5}l),  \notag
\end{eqnarray}%
with $P_{L,R}=\left( 1\pm \gamma _{5}\right) /2$.

The sequential fourth generation model with an additional up-type quark $t^{\prime }$ and down-type
quark $b^{\prime }$ , a heavy charged lepton $\tau^{\prime}$ and an
associated neutrino $\nu^{\prime}$ is a simple and
non-supersymmetric extension of the SM, and as such does not add any new dynamics
to the SM, . Being a simple extension of the SM it
retains all the properties of the SM where the new top quark $t^{\prime }$
like the other up-type quarks, contributes to $b\to s$ transition at the
loop level. Therefore, the effect of fourth generation displays itself by
changing the values of Wilson coefficients $C_{7}\left( \mu \right) $, $%
C_{9}\left( \mu \right) $ and $C_{10}$ via the virtual exchange of fourth
generation up-type quark $t^{\prime }$ which then take the form;%
\begin{equation}
\lambda _{t}C_{i}\rightarrow \lambda _{t}C_{i}^{SM}+\lambda _{t^{\prime
}}C_{i}^{new},  \label{wilson-modified}
\end{equation}%
where $\lambda _{f}=V_{fb}^{\ast }V_{fs}$ and the explicit forms of the $%
C_{i}$'s can be obtained from the corresponding expressions of the Wilson
coefficients in the SM by substituting $m_{t}\rightarrow m_{t^{\prime }}$. By
adding an extra family of quarks, the CKM matrix of the SM is extended by
another row and column which now becomes $4\times 4$. The unitarity of which
leads to%
\begin{equation*}
\lambda _{u}+\lambda _{c}+\lambda _{t}+\lambda _{t^{\prime }}=0.
\end{equation*}%
Since $\lambda _{u}=V_{ub}^{\ast }V_{us}$ has a very small value compared to
the others, therefore, we will ignore it. Then $\lambda _{t}\approx -\lambda
_{c}-\lambda _{t^{\prime }}$ and from Eq. (\ref{wilson-modified}) we have%
\begin{equation}
\lambda _{t}C_{i}^{SM}+\lambda _{t^{\prime }}C_{i}^{new}=-\lambda
_{c}C_{i}^{SM}+\lambda _{t^{\prime }}\left( C_{i}^{new}-C_{i}^{SM}\right) .
\label{wilson-modified1}
\end{equation}%
One can clearly see that under $\lambda _{t^{\prime }}\rightarrow 0$ or $%
m_{t^{\prime }}\rightarrow m_{t}$ the term $\lambda _{t^{\prime }}\left(
C_{i}^{new}-C_{i}^{SM}\right) $ vanishes which is the requirement of GIM
mechanism. Taking the contribution of the $t^{\prime }$ quark in the loop
the Wilson coefficients $C_{i}$'s can be written in the following form%
\begin{eqnarray}
C_{7}^{tot}\left( \mu \right) &=&C_{7}^{SM}\left( \mu \right) +\frac{\lambda
_{t^{\prime }}}{\lambda _{t}}C_{7}^{new}\left( \mu \right) ,  \notag \\
C_{9}^{tot}\left( \mu \right) &=&C_{9}^{SM}\left( \mu \right) +\frac{\lambda
_{t^{\prime }}}{\lambda _{t}}C_{9}^{new}\left( \mu \right) ,
\label{wilson-tot} \\
C_{10}^{tot} &=&C_{10}^{SM}+\frac{\lambda _{t^{\prime }}}{\lambda _{t}}%
C_{10}^{new},  \notag
\end{eqnarray}%
where we factored out $\lambda _{t}=V_{tb}^{\ast }V_{ts}$ term in the
effective Hamiltonian given in Eq. (\ref{effective hamiltonian 1}) and the
last term in these expressions corresponds to the contribution of the $%
t^{\prime }$ quark to the Wilson Coefficients. $\lambda _{t^{\prime }}$ can
be parameterized as:
\begin{equation}
\lambda _{t^{\prime }}=\left\vert V_{t^{\prime }b}^{\ast }V_{t^{\prime
}s}\right\vert e^{i\phi _{sb}}
\end{equation}%
where $\phi_{sb}$ is the new $CP$ odd phase.

In terms of the above Hamiltonian, the free quark decay amplitude for $%
b\rightarrow s$ $\ell^{+}\ell^{-}$ in SM4 can be derived as:
\begin{eqnarray}
\mathcal{M}(b &\rightarrow &s\ell^{+}\ell^{-})=-\frac{G_{F}\alpha }{\sqrt{2}\pi }%
V_{tb}V_{ts}^{\ast }\bigg\{C_{9}^{tot}(\bar{s}\gamma _{\mu }P_{L}b)(\bar{l}%
\gamma ^{\mu }l)+C_{10}^{tot}(\bar{s}\gamma _{\mu }P_{L}b)(\bar{l}\gamma
^{\mu }\gamma _{5}l)  \notag \\
&&-2m_{b}C_{7}^{tot}(\bar{s}i\sigma _{\mu \nu }\frac{q^{\nu }}{q^{2}}P_{R}b)(%
\bar{l}\gamma ^{\mu }l)\bigg\},  \label{quark-amplitude}
\end{eqnarray}%
where $q^{2}$ is the square of momentum transfer. The operator $O_{10}$ can
not be induced by the insertion of four-quark operators because of the
absence of the $Z$ -boson in the effective theory. Therefore, the Wilson
coefficient $C_{10}$ does not renormalize under QCD corrections and hence it
is independent on the energy scale. In addition to this, the above quark
level decay amplitude can receive contributions from the matrix element of
four-quark operators, $\sum_{i=1}^{6}\langle \ell^{+}\ell^{-}s|O_{i}|b\rangle $,
which are usually absorbed into the effective Wilson coefficient $%
C_{9}^{SM}(\mu )$ and can usually be called $C_{9}^{eff}$, that one can
decompose into the following three parts
\begin{equation*}
C_{9}^{SM}=C_{9}^{eff}(\mu )=C_{9}(\mu )+Y_{SD}(z,s^{\prime
})+Y_{LD}(z,s^{\prime }),
\end{equation*}%
where the parameters $z$ and $s^{\prime }$ are defined as $%
z=m_{c}/m_{b},\,\,\,s^{\prime }=q^{2}/m_{b}^{2}$. $Y_{SD}(z,s^{\prime })$
describes the short-distance contributions from four-quark operators far
away from the $c\bar{c}$ resonance regions, which can be calculated reliably
in the perturbative theory. The long-distance contributions $%
Y_{LD}(z,s^{\prime })$ from four-quark operators near the $c\bar{c}$
resonance cannot be calculated from first principles of QCD and are usually
parameterized in the form of a phenomenological Breit-Wigner formula making
use of the vacuum saturation approximation and quark-hadron duality. We will
not incorporate the long-distance contributions in this work. The
expressions for $Y_{SD}(z,s^{\prime })$ can be manifestly written as \cite%
{Buras}
\begin{eqnarray}
Y_{SD}(z,s^{\prime }) &=&h(z,s^{\prime })(3C_{1}(\mu )+C_{2}(\mu
)+3C_{3}(\mu )+C_{4}(\mu )+3C_{5}(\mu )+C_{6}(\mu ))  \notag \\
&&-\frac{1}{2}h(1,s^{\prime })(4C_{3}(\mu )+4C_{4}(\mu )+3C_{5}(\mu
)+C_{6}(\mu ))  \notag \\
&&-\frac{1}{2}h(0,s^{\prime })(C_{3}(\mu )+3C_{4}(\mu ))+{\frac{2}{9}}%
(3C_{3}(\mu )+C_{4}(\mu )+3C_{5}(\mu )+C_{6}(\mu )),  \label{short-distance}
\end{eqnarray}%
with
\begin{eqnarray}
h(z,s^{\prime }) &=&-{\frac{8}{9}}\mathrm{ln}z+{\frac{8}{27}}+{\frac{4}{9}}x-%
{\frac{2}{9}}(2+x)|1-x|^{1/2}\left\{
\begin{array}{l}
\ln \left\vert \frac{\sqrt{1-x}+1}{\sqrt{1-x}-1}\right\vert -i\pi \quad
\mathrm{for}{{\ }x\equiv 4z^{2}/s^{\prime }<1} \\
2\arctan \frac{1}{\sqrt{x-1}}\qquad \mathrm{for}{{\ }x\equiv
4z^{2}/s^{\prime }>1}%
\end{array}%
\right. ,  \notag \\
h(0,s^{\prime }) &=&{\frac{8}{27}}-{\frac{8}{9}}\mathrm{ln}{\frac{m_{b}}{\mu
}}-{\frac{4}{9}}\mathrm{ln}s^{\prime }+{\frac{4}{9}}i\pi \,\,.  \label{hzs}
\end{eqnarray}

\subsection{Parameterizations of matrix elements and form factors in QCD Sum
Rules}

The exclusive $B_{c}\rightarrow D_{s}^{\ast }\ell^{+}\ell^{-}$ decay involves the
hadronic matrix elements which can be obtained by sandwiching the quark
level operators given in Eq. (\ref{quark-amplitude}) between initial state $%
B_{c}$ meson and final state $D_{s}^{\ast }$ meson. These can be
parameterized in terms of the form factors which are the scalar functions of the
square of the four momentum transfer($q^{2}=(p-k)^{2}).$ The non vanishing
matrix elements for the process $B_{c}\rightarrow D_{s}^{\ast }$ can be
parameterized in terms of the seven form factors as follows%
\begin{eqnarray}
\left\langle D_{s}^{\ast }(k,\varepsilon )\left\vert \bar{s}\gamma _{\mu
}b\right\vert B_{c}(p)\right\rangle &=&\frac{2A_{V}(q^{2})}{%
M_{B_{c}}+M_{D_{s}^{\ast}}}\epsilon _{\mu \nu \alpha \beta }\varepsilon
^{\ast \nu }p^{\alpha }k^{\beta }  \label{10} \\
\left\langle D_{s}^{\ast }(k,\varepsilon )\left\vert \bar{s}\gamma _{\mu
}\gamma _{5}b\right\vert B_{c}(p)\right\rangle &=&i\left(
M_{B_{c}^{-}}+M_{D_{s}^{\ast }}\right) \varepsilon ^{\ast \mu }A_{0}(q^{2})
\notag \\
&&-i\frac{A_{+}\left( q^{2}\right) }{M_{B_{c}}+M_{D_{s}^{\ast }}}%
(\varepsilon ^{\ast }\cdot q)\left( p+k\right) ^{\mu }  \notag \\
&&-i\frac{A_{-}\left( q^{2}\right) }{M_{B_{c}}+M_{D_{s}^{\ast }}}%
(\varepsilon ^{\ast }\cdot q)q^{\mu }  \notag \\
&&  \label{11}
\end{eqnarray}%
where $p$ is the momentum of $B_{c}$ meson and , $\varepsilon $ $(k)$ are the
polarization vector (momentum) of the final state $D_{s}^{\ast }$ meson.

In addition to the above form factors there are some penguin form factors,
which we can write as
\begin{eqnarray}
\left\langle D_{s}^{\ast }(k,\varepsilon )\left\vert \bar{s}\sigma _{\mu \nu
}q^{\nu }b\right\vert B_{c}(p)\right\rangle &=&2iT_{1}(q^{2})\epsilon _{\mu
\nu \alpha \beta }\varepsilon ^{\ast \nu }p^{\alpha }k^{\beta }  \label{13a}
\\
\left\langle D_{s}^{\ast }(k,\varepsilon )\left\vert \bar{s}\sigma _{\mu \nu
}q^{\nu }\gamma ^{5}b\right\vert B_{c}(p)\right\rangle &=&\left[ \left(
M_{Bc}^{2}-M_{D_{s}^{\ast }}^{2}\right) \varepsilon _{\mu }^{\ast
}-(\varepsilon ^{\ast }\cdot q)(p+k)_{\mu }\right] T_{2}(q^{2})  \notag \\
&&  \label{13b} \\
&&+(\varepsilon ^{\ast }\cdot q)\left[ q_{\mu }-\frac{q^{2}}{%
M_{Bc}^{2}-M_{D_{s}^{\ast }}^{2}}(p+k)_{\mu }\right] T_{3}(q^{2}).  \notag
\end{eqnarray}%
The form factors $A_{V}\left( q^{2}\right) ,~A_{0}\left( q^{2}\right) ,$ $%
~A_{+}\left( q^{2}\right) ,~A_{-}\left( q^{2}\right) ,~T_{1}\left(
q^{2}\right) ,~T_{2}\left( q^{2}\right) ,~T_{3}\left( q^{2}\right) $ are the
non-perturbative quantities and to calculate them one has to rely on some
non-perturbative approaches and in our numerical analysis we use the form
factors calculated by using QCD Sum Rules \cite{53b}. The dependence of these
form factors on square of the momentum transfer $(q^{2})$ can be written as%
\begin{equation}
F\left( q^{2}\right) =\frac{F\left( 0\right) }{1+a \frac{q^{2}}{M_{B_{c}}^{2}%
}+b \frac{q^{4}}{M_{B_{c}}^{4}}}.  \label{ff-param}
\end{equation}%
where the values of the parameters $F\left( 0\right) $, $\alpha $ and $%
\beta $ are given in Table II.

\begin{table}[tbh]
\caption{$B_{c}\rightarrow D_{s}^{\ast }$ form factors corresponding to penguin
contributions in the QCD Sum Rules.
$F(0)$ denotes the value of form factors at $q^{2}=0$ while $a$ and $b$
are the parameters in the parameterizations shown in Eq. (\protect\ref%
{ff-param})\protect\cite{53b}.}
\label{di-fit B to Dsstarpenguin}%
\begin{tabular}{cccc}
\hline\hline
 $F(q^{2})$ & $\hspace{2cm}F(0)$ & $\hspace{2cm}a$ & $\hspace{%
2cm}b$ \\ \hline
 $A_{V}\left( q^{2}\right) $ & $\hspace{2cm}0.54\pm 0.018$ & $%
\hspace{2cm}-1.28$ & $\hspace{2cm}-0.23$ \\ \hline
 $A_{0}(q^{2})$ & $\hspace{2cm}0.30\pm 0.017$ & $\hspace{2cm}%
-0.13$ & $\hspace{2cm}-0.18$ \\ \hline
 $A_{+}(q^{2})$ & $\hspace{2cm}0.36\pm 0.013$ & $\hspace{2cm}%
-0.67$ & $\hspace{2cm}-0.066$ \\ \hline
 $A_{-}(q^{2})$ & $\hspace{2cm}-0.57\pm 0.04$ & $\hspace{2cm}%
-1.11$ & $\hspace{2cm}-0.14$ \\ \hline
 $T_{1}(q^{2})$ & $\hspace{2cm}0.31\pm 0.017$ & $\hspace{2cm}%
-1.28$ & $\hspace{2cm}-0.23$ \\ \hline
 $T_{2}(q^{2})$ & $\hspace{2cm}0.33\pm 0.016$ & $\hspace{2cm}%
-0.10$ & $\hspace{2cm}-0.097$ \\ \hline
 $T_{3}(q^{2})$ & $\hspace{2cm}0.29\pm 0.034$ & $\hspace{2cm}%
-0.91$ & $\hspace{2cm}0.007$ \\ \hline\hline
\end{tabular}%
\end{table}

Now in terms of these form factors and from Eq. (\ref{quark-amplitude}) it is
straightforward to write the penguin amplitude as
\begin{equation*}
\mathcal{M}^{\text{PENG}}=-\frac{G_{F}\alpha }{2\sqrt{2}\pi }%
V_{tb}V_{ts}^{\ast }\left[ T_{\mu }^{1}(\bar{l}\gamma ^{\mu }l)+T_{\mu
}^{2}\left( \bar{l}\gamma ^{\mu }\gamma ^{5}l\right) \right]
\end{equation*}%
where%
\begin{eqnarray}
T_{\mu }^{1} &=&f_{1}(q^{2})\epsilon _{\mu \nu \alpha \beta }\varepsilon
^{\ast \nu }p^{\alpha }k^{\beta }-if_{2}(q^{2})\varepsilon _{\mu }^{\ast
}+if_{3}(q^{2})(\varepsilon ^{\ast }\cdot q)P_{\mu }  \label{60} \\
T_{\mu }^{2} &=&f_{4}(q^{2})\epsilon _{\mu \nu \alpha \beta }\varepsilon
^{\ast \nu }p^{\alpha }k^{\beta }-if_{5}(q^{2})\varepsilon _{\mu }^{\ast
}+if_{6}(q^{2})(\varepsilon ^{\ast }\cdot q)P_{\mu } +if_{0}(q^{2})(\varepsilon ^{\ast }\cdot q)q_{\mu } \label{61}
\end{eqnarray}

The functions $f_{0}$ to $f_{6}$ in Eq. (\ref{60}) and Eq. (\ref{61}) are
known as auxiliary functions, which contain both long distance (form
factors) and short distance (Wilson coefficients) effects and these can be
written as

\begin{eqnarray}
f_{1}(q^{2}) &=&4(m_{b}+m_{s})\frac{C_{7}^{tot}}{q^{2}}%
T_{1}(q^{2})+2C_{9}^{tot}\frac{A_{V}(q^{2})}{M_{B_{c}}+M_{D_{s}^{\ast }}}
\nonumber \\
f_{2}(q^{2}) &=&\frac{2C_{7}^{tot}}{q^{2}}(m_{b}-m_{s})T_{2}(q^{2})\left(
M_{B_{c}}^{2}-M_{D_{s}^{\ast }}^{2}\right) +C_{9}^{tot}A_{0}(q^{2})\left(
M_{B_{c}}+M_{D^{\ast }}\right)   \nonumber \\
f_{3}(q^{2}) &=&\left[ 4\frac{C_{7}^{tot}}{q^{2}}(m_{b}-m_{s})\left(
T_{2}(q^{2})+q^{2}\frac{T_{3}(q^{2})}{\left( M_{B_{c}}^{2}-M_{D_{s}^{\ast
}}^{2}\right) }\right) +C_{9}^{tot}\frac{A_{+}(q^{2})}{M_{B_{c}}+M_{D_{s}^{%
\ast }}}\right]   \nonumber \\
f_{4}(q^{2}) &=&C_{10}^{tot}\frac{2A_{V}(q^{2})}{M_{B_{c}}+M_{D_{s}^{\ast }}}
\nonumber \\
f_{5}(q^{2}) &=&C_{10}^{tot}A_{0}(q^{2})\left( M_{B_{c}}+M_{D_{s}^{\ast
}}\right)   \nonumber \\
f_{6}(q^{2}) &=&C_{10}^{tot}\frac{A_{+}(q^{2})}{M_{B_{c}}+M_{D_{s}^{\ast }}}
\nonumber \\
f_{0}(q^{2}) &=&C_{10}^{tot}\frac{A_{-}(q^{2})}{M_{B_{c}}+M_{D_{s}^{\ast }}}
\label{62}
\end{eqnarray}%

\section{Physical Observables for $B_{c}\rightarrow D_{s}^{\ast} \ell^{+}\ell^{-}$}

In this section we will present the calculations of the physical observables
like the decay rates, leptons forward-backward asymmetry, the helicity
fractions of $D_{s}^{\ast }$ meson and the final state lepton polarizations.
We use both the weak annihilation (WA) amplitude and the penguin amplitude
to study these observables.

\subsection{The Differential Decay Rate of $B_{c}\rightarrow D_{s}^{\ast } \ell
^{+} \ell ^{-}$}

In the rest frame of $B_{c}$ meson the differential decay width of $%
B_{c}\rightarrow D_{s}^{\ast }\ell^{+}\ell^{-}$ can be written as
\begin{equation}
\frac{d\Gamma (B_{c}\rightarrow D_{s}^{\ast }\ell^{+}\ell^{-})}{dq^{2}}=\frac{1}{%
\left( 2\pi \right) ^{3}}\frac{1}{32M_{B_{c}}^{3}}%
\int_{-u(q^{2})}^{+u(q^{2})}du\left\vert \mathcal{M}\right\vert ^{2}
\label{62a}
\end{equation}%
where
\begin{eqnarray}
\mathcal{M} &=&\mathcal{M}^{\text{WA}}+\mathcal{M}^{\text{PENG}}  \notag \\
q^{2} &=&(p_{l^{+}}+p_{l^{-}})^{2}  \label{62b} \\
u &=&\left( p-p_{l^{-}}\right) ^{2}-\left( p-p_{l^{+}}\right) ^{2}  \notag
\end{eqnarray}%
Now the limits on $q^{2}$ and $u$ are
\begin{eqnarray}
4m_{l}^{2} &\leq &q^{2}\leq (M_{B_{c}}-M_{D_{s}^{\ast }})^{2}  \label{62d} \\
-u(q^{2}) &\leq &u\leq u(q^{2})  \label{62e}
\end{eqnarray}%
with%
\begin{equation}
u(q^{2})=\sqrt{\lambda \left( 1-\frac{4m_{l}^{2}}{q^{2}}\right) }
\label{62f}
\end{equation}%
and%
\begin{equation*}
\lambda \equiv \lambda (M_{B_{c}}^{2},M_{D_{s}^{\ast
}}^{2},q^{2})=M_{B_{c}}^{4}+M_{D_{s}^{\ast
}}^{4}+q^{4}-2M_{B_{c}}^{2}M_{D_{s}^{\ast }}^{2}-2M_{D_{s}^{\ast
}}^{2}q^{2}-2q^{2}M_{B_{c}}^{2}
\end{equation*}%
Here $m_{l}$ corresponds to the mass of the lepton which for our case are
the $\mu $ and $\tau $. The total decay rate of $B_{c}\rightarrow
D_{s}^{\ast }\ell^{+}\ell^{-}$ can be expressed in terms of WA, penguin amplitude
and their interference, which takes the form \cite{51}\
\begin{equation}
\frac{d\Gamma }{dq^{2}}=\frac{d\Gamma ^{\text{WA}}}{dq^{2}}+\frac{d\Gamma
^{\,\text{PENG}}}{dq^{2}}+\frac{d\Gamma ^{\text{WA-PENG}}}{dq^{2}}
\label{62g}
\end{equation}%
with
\begin{eqnarray}
\frac{d\Gamma ^{\text{WA}}}{dq^{2}} &=&\frac{G_{F}^{2}\left\vert
V_{cb}V_{cs}^{\ast }\right\vert ^{2}\alpha ^{2}}{2^{11}\pi
^{5}3M_{B_{c}}^{3}M_{D_{s}^{\ast }}^{2}q^{2}}u(q^{2})\times g\left(
q^{2}\right)  \label{62h} \\
\frac{d\Gamma ^{\,\text{PENG}}}{dq^{2}} &=&\frac{G_{F}^{2}\left\vert
V_{tb}V_{ts}^{\ast }\right\vert ^{2}\alpha ^{2}}{2^{11}\pi
^{5}3M_{B_{c}}^{3}M_{D_{s}^{\ast }}^{2}q^{2}}u(q^{2})\times h\left(
q^{2}\right)  \label{62i} \\
\frac{d\Gamma ^{\text{WA-PENG}}}{dq^{2}} &=&\frac{G_{F}^{2}\left\vert
V_{cb}V_{cs}^{\ast }\right\vert \left\vert V_{tb}V_{ts}^{\ast }\right\vert
\alpha ^{2}}{2^{11}\pi ^{5}3M_{B_{c}}^{3}M_{D_{s}^{\ast }}^{2}q^{2}}%
u(q^{2})\times I\left( q^{2}\right) .  \label{62j}
\end{eqnarray}%
The function $u(q^{2})$ is defined in Eq. (\ref{62f}) and $g(q^{2})$, $%
h(q^{2})$ and $I(q^{2})$ are
\begin{eqnarray}
g\left( q^{2}\right) &=&\frac{1}{2}\left(
2m_{l}^{2}+q^{2}\right)\kappa^{2} \left[ 8\lambda M_{D_{s}^{\ast
}}^{2}q^{2}\left(F_{V}^{D_{s}^{\ast
}}(q^{2})\right)^{2}+\left(F_{A}^{D_{s}^{\ast
}}(q^{2})\right)^{2}[12M_{D_{s}^{\ast }}^{2}q^{2}(\lambda
+4M_{B_{c}}^{2}q^{2})+\lambda ^{2}+\lambda (\lambda
+4q^{2}M_{D_{s}^{\ast
}}^{2}+4q^{4})]\right]  \notag \\
h(q^{2}) &=&24\left\vert f_{0}(q^{2})\right\vert ^{2}m_{l}^{2}M_{D_{s}^{\ast
}}^{2}\lambda +8M_{D_{s}^{\ast }}^{2}q^{2}\lambda
(2m_{l}^{2}+q^{2})\left\vert f_{1}(q^{2})\right\vert
^{2}-(4m_{l}^{2}-q^{2})\left\vert f_{4}(q^{2})\right\vert ^{2}]  \notag \\
&&+\lambda (2m_{l}^{2}+q^{2})\left\vert
f_{2}(q^{2})+(M_{B_{c}}^{2}-M_{D_{s}^{\ast
}}^{2}-q^{2})f_{3}(q^{2})\right\vert ^{2}-(4m_{l}^{2}-q^{2})\left\vert
f_{5}(q^{2})+(M_{B_{c}}^{2}-M_{D_{s}^{\ast
}}^{2}-q^{2})f_{6}(q^{2})\right\vert ^{2}]  \notag \\
&&+4M_{D_{s}^{\ast }}^{2}q^{2}[(2m_{l}^{2}+q^{2})(3\left\vert
f_{2}(q^{2})\right\vert ^{2}-\lambda \left\vert f_{3}(q^{2})\right\vert
^{2})-(4m_{l}^{2}-q^{2})(3\left\vert f_{5}(q^{2})\right\vert ^{2}-\lambda
\left\vert f_{6}(q^{2})\right\vert ^{2})]  \label{hfunction} \\
I(q^{2}) &=&2\kappa[f_{2}(q^{2})F_{A}^{D_{s}^{\ast
}}(q^{2})q^{2}(2m_{l}^{2}+q^{2})(\lambda +6M_{D_{s}^{\ast
}}^{2}(M_{B_{c}}^{2}-M_{D_{s}^{\ast }}^{2}+q^{2}))  \notag \\
&&-(\lambda (2f_{1}(q^{2})F_{V}^{D_{s}^{\ast }}(q^{2})M_{D_{s}^{\ast
}}^{2}q^{4}+f_{3}(q^{2})F_{A}^{D_{s}^{\ast
}}(q^{2})(2m_{l}^{2}+q^{2})(\lambda +q^{4}+4M_{B_{c}}M_{D_{s}^{\ast }}))].
\notag
\end{eqnarray}
where
\begin{equation}
\kappa = \frac{8\pi^{2}M_{D_{s}^{\ast }}f_{B_{c}}f_{D_{s}^{\ast
}}}{(m_{c}^{2}-m_{s}^{2})q^{2}}\frac{\left\vert V_{tb}V_{ts}^{\ast
}\right\vert }{\left\vert V_{cb}V_{cs}^{\ast }\right\vert }.
\label{kappa}
\end{equation}

\subsection{Forward-Backward Asymmetry of $%
B\rightarrow D^{\ast}_s \ell^{+} \ell^{-}$ decay}
In this section we investigate the forward-backward asymmetry (FBA) of
leptons. The measurement of the FBA at LHC is significant due to the minimal
dependence on the form factors \cite{Ball}, and as such this observable is of greater importance to check
the more clear signals of any NP than the other observables such as
branching ratio etc. In the context of fourth generation, the FBA can also
play a crucial role because it is driven by the loop top quark and so it is
sensitive to the fourth generation up type quark $t^{\prime}$ \cite{Ali}.
Now to calculate the forward-backward asymmetry, we consider
the following double differential decay rate formula for the process $%
B_{c}\rightarrow D_{s}^{\ast }\ell^{+} \ell^{-}$
\begin{equation}
{\frac{d^{2}\Gamma (q^{2},\cos \theta )}{dq^{2}d\cos \theta }}={\frac{1}{%
(2\pi )^{3}}}{\frac{1}{32m_{B}^{3}}}u\left( q^{2}\right) |\mathcal{M}%
_{B_{c}\rightarrow D_{s}^{\ast } \ell^{+}\ell^{-}}|^{2},
\end{equation}%
where $\theta $ is the angle between the momentum of $B_{c}$ meson and $\ell^{-}$
in the dilepton rest frame.
Following Refs.
\cite{Ali}, the differential and
normalized FBAs for the semi-leptonic decay $B_{c}\rightarrow D_{s}^{\ast }\ell^{+} \ell^{-}$
are defined as
\begin{equation}
{\frac{dA_{FB}(q^{2})}{dq^{2}}}=\int_{0}^{1}d\cos \theta {\frac{d^{2}\Gamma
(q^{2},\cos \theta )}{dq^{2}d\cos \theta }}-\int_{-1}^{0}d\cos \theta {\frac{%
d^{2}\Gamma (q^{2},\cos \theta )}{dq^{2}d\cos \theta }}
\end{equation}%
and
\begin{equation}
A_{FB}(q^{2})={\frac{\int_{0}^{1}d\cos \theta {\frac{d^{2}\Gamma (q^{2},\cos
\theta )}{dq^{2}d\cos \theta }}-\int_{-1}^{0}d\cos \theta {\frac{d^{2}\Gamma
(q^{2},\cos \theta )}{dq^{2}d\cos \theta }}}{\int_{0}^{1}d\cos \theta {\frac{%
d^{2}\Gamma (q^{2},\cos \theta )}{dq^{2}d\cos \theta }}+\int_{-1}^{0}d\cos \theta {%
\frac{d^{2}\Gamma (q^{2},\cos \theta )}{dq^{2}d\cos \theta }}}}.
\end{equation}
Following the procedure used for the differential decay rate, one
can easily get the expression for the forward-backward asymmetry which can be written as
\begin{equation}
\mathcal{A}_{FB} = \frac{1}{d\Gamma /dq^{2}}\frac{G_{F}^{2}\alpha ^{2}}{%
2^{11}\pi ^{5}m_{B}^{3}}\left\vert V_{tb}V_{ts}^{\ast }\right\vert
^{2}q^{2}u(q^{2})[2{Re}[f_{4}]\kappa F_{A}^{D_{s}^{\ast }}\left(
q^{2}\right) (M_{B_{c}}^{2}-M_{D_{s}^{\ast
}}^{2}+q^{2})+4{Re}[f_{2}^{\ast }f_{4}+f_{1}^{\ast
}f_{5}]+2{Re}[f_{5}]\kappa F_{V}^{D_{s}^{\ast }}\left( q^{2}\right)
)] \label{fb-asymmetry}
\end{equation}
where $\kappa$ is defined in Eq. (\ref{kappa}) and $d\Gamma /dq^{2}$
is given in Eq. (\ref{62g}).

\subsection{Helicity Fractions of $D_{s}^{\ast }$ in $B_{c}\rightarrow
D_{s}^{\ast }\ell^{+} \ell^{-}$ decay}

We now discuss helicity fractions of $D_{s}^{\ast }$ meson in $B_{c}\rightarrow
D_{s}^{\ast } \ell^{+} \ell^{-}$ decay which are almost
independent of the uncertainties arising due to form factors and other input
parameters. Therefore, the study of these observables will provide us a good testing ground for the SM4.
The explicit expression of the decay rate for $%
B_{c}^{-}\rightarrow D_{s}^{\ast} \ell^{+} \ell^{-}$ decay in terms
of longitudinal $(\Gamma _{L})$ and transverse $(\Gamma _{T})$ components of decay rate can be written as
\cite{51}
\begin{eqnarray}
\frac{d\Gamma _{L}(q^{2})}{dq^{2}} &=&\frac{d\Gamma _{L}^{\text{WA}}(q^{2})}{%
dq^{2}}+\frac{d\Gamma _{L}^{\text{PENG}}(q^{2})}{dq^{2}}+\frac{d\Gamma _{L}^{%
\text{WA-PENG}}(q^{2})}{{dq^{2}}} \\
\frac{d\Gamma _{\pm }(q^{2})}{dq^{2}} &=&\frac{d\Gamma _{\pm }^{\text{WA}%
}(q^{2})}{dq^{2}}+\frac{d\Gamma _{\pm }^{\text{PENG}}(q^{2})}{dq^{2}}+\frac{%
d\Gamma _{\pm }^{\text{WA-PENG}}(q^{2})}{{dq^{2}}}  \label{65b} \\
\frac{d\Gamma _{T}(q^{2})}{dq^{2}} &=&\frac{d\Gamma _{+}(q^{2})}{dq^{2}}+%
\frac{d\Gamma _{-}(q^{2})}{dq^{2}}.
\end{eqnarray}%
where
\begin{eqnarray}
\frac{d\Gamma _{L}^{\text{WA}}(q^{2})}{dq^{2}} &=&\frac{G_{F}^{2}\left\vert
V_{cb}V_{cs}^{\ast }\right\vert ^{2}\alpha ^{2}}{2^{11}\pi ^{5}}\frac{%
u(q^{2})}{M_{B_{c}}^{3}}\times \frac{1}{3}A_{L}^{\text{WA}}  \label{65c} \\
\frac{d\Gamma _{L}^{\text{PENG}}(q^{2})}{dq^{2}} &=&\frac{%
G_{F}^{2}\left\vert V_{tb}V_{ts}^{\ast }\right\vert ^{2}\alpha ^{2}}{%
2^{11}\pi ^{5}}\frac{u(q^{2})}{M_{B_{c}}^{3}}\times \frac{1}{3}A_{L}^{\text{%
PENG}}  \label{65d} \\
\frac{d\Gamma _{L}^{\text{WA-PENG}}(q^{2})}{dq^{2}} &=&\frac{%
G_{F}^{2}\left\vert V_{cb}V_{cs}^{\ast }\right\vert \left\vert
V_{tb}V_{ts}^{\ast }\right\vert \alpha ^{2}}{2^{11}\pi ^{5}}\frac{u(q^{2})}{%
M_{B_{c}}^{3}}\times \frac{1}{3}A_{L}^{\text{WA-PENG}}  \label{65e} \\
\frac{d\Gamma _{\pm }^{\text{WA}}(q^{2})}{dq^{2}} &=&\frac{%
G_{F}^{2}\left\vert V_{cb}V_{cs}^{\ast }\right\vert ^{2}\alpha ^{2}}{%
2^{11}\pi ^{5}}\frac{u(q^{2})}{M_{B_{c}}^{3}}\times \frac{2}{3}A_{\pm }^{%
\text{WA}}  \label{65f} \\
\end{eqnarray}
\begin{eqnarray}
\frac{d\Gamma _{\pm }^{\text{PENG}}(q^{2})}{dq^{2}} &=&\frac{%
G_{F}^{2}\left\vert V_{tb}V_{ts}^{\ast }\right\vert ^{2}\alpha ^{2}}{%
2^{11}\pi ^{5}}\frac{u(q^{2})}{M_{B_{c}}^{3}}\times \frac{4}{3}A_{\pm }^{%
\text{PENG}}  \label{65g} \\
\frac{d\Gamma _{\pm }^{\text{WA-PENG}}(q^{2})}{dq^{2}} &=&\frac{%
G_{F}^{2}\left\vert V_{cb}V_{cs}^{\ast }\right\vert \left\vert
V_{tb}V_{ts}^{\ast }\right\vert \alpha ^{2}}{2^{11}\pi ^{5}}\frac{u(q^{2})}{%
M_{B_{c}}^{3}}\times \frac{2}{3}A_{\pm }^{\text{WA-EP}}.  \label{65h}
\end{eqnarray}%
The different functions appearing in above equation can be expressed
in terms of auxiliary functions (c.f. Eq. (\ref{62})) as
\begin{align}
A_{L}^{\text{WA}} &=\frac{\kappa^{2}}{4 q^{2} M_{D_{s}^{\ast }}^{2}}%
\bigg[\left( { F_{V}^{D_{s}^{\ast }}}(q^{2})\right)^{2}
\left\{q^{2}\lambda(\lambda +4q^{2}M_{D_{s}^{\ast
}}^{2})-4M^{2}\lambda(2\lambda+8q^{2}M_{D_{s}^{\ast }}^{2})-q^{2}
\left(M_{B_{c}}^{2}-M_{D_{s}^{\ast
}}^{2}-q^{2}\right)^{2}\left(\lambda-2u^{2}(q^{2})\right)\right\} \notag\\
&+\left( { F_{A}^{D_{s}^{\ast}}}(q^{2})\right)^{2}\bigg\{12\lambda
q^{2}((M_{B_{c}}^{2}-M_{D_{s}^{\ast}}^{2})^{2}-M_{D_{s}^{\ast}}^{2})
-\lambda^{2}(q^{2}-4m^{2})+q^{2}(8q^{2}M_{D_{s}^{\ast }}^{2}-\lambda
)(M_{B_{c}}^{2}-M_{D_{s}^{\ast }}^{2}+q^{2})^{2} \notag\\
&-2u^{2}(q^{2})q^{2}((M_{B_{c}}^{2}
-M_{D_{s}^{\ast}}^{2})^{2}+q^{4})
+4m^{2}((M_{B_{c}}^{2}-M_{D_{s}^{\ast
}}^{2})^{2}-q^{4})^{2}\bigg\}\bigg] \notag \\
A_{L}^{\text{PENG}} &=\frac{1}{2M_{D_{s}^{\ast
}}^{2}q^{2}}[24\left\vert f_{0}(q^{2})\right\vert
^{2}m_{l}^{2}M_{D_{s}^{\ast }}^{2}\lambda
+(2m_{l}^{2}+q^{2})\left\vert (M_{B_{c}}^{2}-M_{D_{s}^{\ast
}}^{2}-q^{2})f_{2}(q^{2})+\lambda f_{3}(q^{2})\right\vert ^{2} \notag \\
&+(q^{2}-4m_{l}^{2})\left\vert (M_{B_{c}}^{2}-M_{D_{s}^{\ast
}}^{2}-q^{2})f_{5}(q^{2})+\lambda f_{6}(q^{2})\right\vert ^{2}] \notag \\
A_{L}^{\text{WA-PENG \ }} &=\frac{\kappa}{ q^{2} M_{D_{s}^{\ast
}}^{2}}\bigg[Re(f_{1}\left(q^{2}\right)F_{V}^{D_{s}^{\ast}}(q^{2}))\left\{(\lambda
+4M_{D_{s}^{\ast }}^{2}q^{2})\left(8m^{2}\sqrt{\lambda}+q^{2}(2u(q^{2})-\sqrt{\lambda})\right)-4M_{D_{s}^{\ast }}^{2}q^{2}\lambda \right\}  \notag \\
&+Re(f_{2}\left(q^{2}\right)F_{A}^{D_{s}^{\ast}}(q^{2}))\bigg\{q^{2}u^{2}(q^{2})(M_{B_{c}}^{2}-M_{D_{s}^{\ast
}}^{2}-q^{2})+6q^{2}\lambda(M_{D_{s}^{\ast
}}^{2}-M_{B_{c}}^{2})\notag \\
&+q^{2}(\lambda-8q^{2}M_{D_{s}^{\ast}}^{2})(M_{B_{c}}^{2}-M_{D_{s}^{\ast
}}^{2}+q^{2})-4m^{2}q^{2}(4q^{2}M_{D_{s}^{\ast }}^{2}+\lambda) \bigg\}\notag \\
&
+Re(f_{3}\left(q^{2}\right)F_{A}^{D_{s}^{\ast}}(q^{2}))\bigg\{\lambda^{2}(4m^{2}-q^{2})+q^{4}(q^{2}u(q^{2})\sqrt{\lambda}-6\lambda(M_{B_{c}}^{2}+M_{D_{s}^{\ast
}}^{2}))+q^{2}(M_{B_{c}}^{2}-M_{D_{s}^{\ast
}}^{2})(6\lambda-u^{2}(q^{2})) \bigg\}\bigg] \notag \\
A_{\pm }^{\text{WA }} &=\kappa^{2}\bigg[\left( 2m^{2}+q^{2}\right)
\bigg[\lambda \left( { F_{V}^{D_{s}^{\ast
}}}(q^{2})\right)^{2}+\left( { F_{A}^{D_{s}^{\ast }}}(q^{2})\right)
^{2}\left( \lambda+4M_{D^{\ast}_{s}}^{2}q^{2}\right)\bigg]\bigg] \notag \\
A_{\pm}^{\text{PENG}} &=(q^{2}-4m_{l}^{2})\left\vert
f_{5}(q^{2})\mp\sqrt{\lambda }f_{4}(q^{2})\right\vert ^{2}+\left(
q^{2}+2m_{l}^{2}\right)
\left\vert f_{2}(q^{2})\pm\sqrt{\lambda }f_{1}(q^{2})\right\vert ^{2} \notag \\
A_{\pm }^{\text{WA -PENG}}
&=-\kappa\bigg\{2\sqrt{\lambda}(q^{2}-4m^{2})
Re(f_{2}\left(q^{2}\right)F_{V}^{D_{s}^{\ast}}(q^{2}))+4\lambda(q^{2}+2m^{2})
Re(f_{1}\left(q^{2}\right)F_{V}^{D_{s}^{\ast}}(q^{2})) \nonumber \\
& \pm2(q^{2}+2m^2)(M_{B_{c}}^{2}-M_{D_{s}^{\ast
}}^{2}+q^{2})[2Re[(f_{1}\left(q^{2}\right)F_{A}^{D_{s}^{\ast}}(q^{2}))]\sqrt{\lambda}\mp2Re{(f_{2}\left(q^{2}\right)F_{V}^{D_{s}^{\ast}}(q^{2}))}]\bigg\}
\label{helcity-expression}
\end{align}%
Finally the longitudinal and transverse helicity amplitude becomes
\begin{eqnarray}
f_{L}(q^{2}) &=&\frac{d\Gamma _{L}(q^{2})/dq^{2}}{d\Gamma (q^{2})/dq^{2}}
\notag \\
f_{\pm }(q^{2}) &=&\frac{d\Gamma _{\pm }(q^{2})/dq^{2}}{d\Gamma
(q^{2})/dq^{2}}  \notag \\
f_{T}(q^{2}) &=&f_{+}(q^{2})+f_{-}(q^{2})
\end{eqnarray}%
so that \ the sum of the longitudinal and transverse helicity amplitudes is
equal to one i.e. $f_{L}(q^{2})+f_{T}(q^{2})=1$ for each value of $q^{2}$.

\subsection{Lepton Polarization asymmetries of $B_c\rightarrow D^{\ast}_s \ell^{+} \ell^{-}$}

In the rest frame of the lepton $\ell^{-}$, the unit vectors along
longitudinal, normal and transversal component of the $\ell^{-}$ can be defined
as:
\begin{eqnarray}
s_{L}^{-\mu } &=&(0,\vec{e}_{L})=\left( 0,\frac{\vec{p}_{-}}{\left| \vec{p}%
_{-}\right| }\right) ,  \notag \\
s_{N}^{-\mu } &=&(0,\vec{e}_{N})=\left( 0,\frac{\vec{k}\times
\vec{p}_{-}}{\left| \vec{k}\times \vec{p}_{-}\right| }\right) ,
\label{p-vectors} \\
s_{T}^{-\mu } &=&(0,\vec{e}_{T})=\left( 0,\vec{e}_{N}\times \vec{e}%
_{L}\right) ,  \notag
\end{eqnarray}
where $\vec{p}_{-}$ and $\vec{k}$ are the respective three-momenta of the
lepton $\ell^{-}$ and $D_{s}^{*}$ meson in the center mass
(CM) frame of $l^{+}l^{-}$ system. Lorentz transformation is used to boost
the longitudinal component of the lepton polarization to the CM frame of the
lepton pair as
\begin{equation}
\left( s_{L}^{-\mu }\right) _{CM}=\left( \frac{|\vec{p}_{-}|}{m_{l}},\frac{%
E_{l}\vec{p}_{-}}{m_{l}\left| \vec{p}_{-}\right| }\right)
\label{bossted component}
\end{equation}
where $E_{l}$ and $m_{l}$ are the energy and mass of the lepton. The normal
and transverse components remain unchanged under the Lorentz boost.

The longitudinal ($P_{L}$), normal ($P_{N}$) and transverse ($P_{T}$)
polarizations of lepton can be defined as:
\begin{equation}
P_{i}^{(\mp )}(q^{2})=\frac{\frac{d\Gamma }{dq^{2}}(\vec{\xi}^{\mp }=\vec{e}%
^{\mp })-\frac{d\Gamma }{dq^{2}}(\vec{\xi}^{\mp }=-\vec{e}^{\mp })}{\frac{%
d\Gamma }{dq^{2}}(\vec{\xi}^{\mp }=\vec{e}^{\mp })+\frac{d\Gamma }{dq^{2}}(%
\vec{\xi}^{\mp }=-\vec{e}^{\mp })}  \label{polarization-defination}
\end{equation}%
where $i=L,\;N,\;T$ and $\vec{\xi}^{\mp }$ is the spin direction along the
leptons $l^{\mp }$. The differential decay rate for polarized lepton $l^{\mp
}$ in $B_{c}\rightarrow D_{s}^{\ast } \ell^{+} \ell^{-}$ decay along any
spin direction $\vec{\xi}^{\mp }$ is related to the unpolarized decay rate (%
\ref{62g}) with the following relation
\begin{equation}
\frac{d\Gamma (\vec{\xi}^{\mp })}{dq^{2}}=\frac{1}{2}\left( \frac{d\Gamma }{%
dq^{2}}\right) [1+(P_{L}^{\mp }\vec{e}_{L}^{\mp }+P_{N}^{\mp }\vec{e}%
_{N}^{\mp }+P_{T}^{\mp }\vec{e}_{T}^{\mp })\cdot \vec{\xi}^{\mp }].
\label{polarized-decay}
\end{equation}%
Using these inputs we  can achieve the expressions of longitudinal, normal and transverse
lepton polarizations for $B_{c}\rightarrow D_{s}^{\ast } \ell^{+} \ell^{-}$
decays.
The expression of the numerator of longitudinal lepton
polarization is
\begin{eqnarray}
P_{L}(q^2) &\propto &\frac{4\lambda}{3M_{D_{s}^{\ast }}^{2}}\sqrt{\frac{q^2-4m_{l}^{2}}{q^{2}}}\times  \bigg\{2Re(f_{2}f_{5}^{\ast})+\lambda Re(f_{3}f_{6}^{\ast})+4\sqrt{q^{2}}Re(f_{1}f_{4}^{\ast})\left(1+\frac{12q^2M_{D_{s}^{\ast}}^{2}}{\lambda}\right) \notag \\
&& +\left(-M_{B_{c}}^{2}+M_{D_{s}^{\ast }}^{2}+q^{2}\right)\left[Re(f_{3}f_{5}^{\ast})+Re(f_{2}f_{6}^{\ast})\right]\bigg\}. \label{long-polarization}
\end{eqnarray}
Similarly the numerator of normal lepton polarization can be written as
\begin{eqnarray}
P_{N}(q^2) &\propto & \frac{m_{l}\pi}{M_{D_{s}^{\ast }}^{2}}\sqrt{\frac{\lambda}{q^{2}}}\times \bigg\{-\lambda q^{2}Re(f_{3}f_{0}^{\ast})+\lambda(M_{B_{c}}^{2}-M_{D_{s}^{\ast }}^{2})Re(f_{3}f_{6}^{\ast})-\lambda Re(f_{3}f_{5}^{\ast})\notag\\
&&+\left(-M_{B_{c}}^{2}+M_{D_{s}^{\ast }}^{2}+q^{2}\right)\left[q^{2}Re(f_{2}f_{0}^{\ast})\right]\notag\\
&& - 8q^{2}M_{D_{s}^{\ast }}^{2}Re(f_{1}f_{2}^{\ast}) +\kappa^{2}F_{V}^{D_{s}^{\ast}}F_{A}^{D_{s}^{\ast}}(M^{2}_{B_{c}}-M^{2}_{D_{s}^{\ast}}+q^{2})\bigg\}  \label{normal-polarization}
\end{eqnarray}
and that of the transverse leptons polarization is given by
\begin{eqnarray}
P_{T}\left( q^{2}\right)  &\propto & i\frac{m_{l}\pi \sqrt{ \left( q^{2}-\frac{4m_{l}^{2}}{q^{2}}\right) \lambda }}{M_{D_{s}^{\ast }}^{2}}\bigg\{ M_{B_{c}}^{2}Im(f_{5}f_{6}^{\ast})+M_{D_{s}^{\ast}}\left[4Im(f_{2}f_{4}^{\ast})+4Im(f_{1}f_{5}^{\ast})+3Im(f_{5}f_{6}^{\ast})\right]\notag\\
&&+\left(-M_{B_{c}}^{2}+M_{D_{s}^{\ast }}^{2}+q^{2}\right)\left[Im(f_{0}f_{5}^{\ast})\right]+Im (f_{5}f_{6}^{\ast})+Im(f_{4}f_{6}^{\ast})+2\kappa Im[f_5]F_{V}^{D_{s}^{\ast}}+2\kappa M_{D_{s}^{\ast}}Im[f_4]F_{A}^{D_{s}^{\ast}}\bigg\}\notag\\
&&
\label{Transverse-polarization}
\end{eqnarray}%
where $\kappa$ is defined in Eq. (\ref{kappa}) along with auxiliary functions $f_{0},...,f_{6}$ and the form factors $F_{V,A}^{D_{s}^{\ast }}$
are the ones defined in Eq. (\ref{62}) and Eq. (\ref{1b}), respectively. Here we have dropped out the constant
factors which are understood.

\section{Numerical Analysis:}

In this section, we would like to present the numerical analysis of
decay rates, FBAs of leptons, helicity fractions of final state $D_{s}^{\ast}$ meson and
different lepton polarization asymmetries both in the SM and SM4. The numerical values of Wilson coefficients and
other input parameters used in our analysis are collected in Tables II and III.
\begin{table}[tbh]
\caption{{}Values of input parameters used in our numerical analysis}
\label{Input parameters}%
\begin{tabular}{cc}
\hline\hline
$G_{F}=1.166\times 10^{-2}$ GeV$^{-2}$ & $\left| V_{ts}\right|
=41.61_{-0.80}^{+0.10}\times 10^{-3}$ \\
$\left| {V_{tb}}\right| =0.9991$ & $m_{b}=\left( 4.68\pm 0.03\right) $ GeV
\\
{$m_{c}\left( m_{c}\right) =1.275_{-0.015}^{+0.015}$ GeV} & $m_{s}\left( 1%
\text{ GeV}\right) =\left( 142\pm 28\right) $ MeV \\ \hline
$M_{B_{c}}=6.26$ GeV & $M_{D_{s}^{\ast} }=2.12$ GeV \\ \hline
$f_{B_{c}}=0.35$ GeV$ $ & $f_{D_{s}^{\ast}
}=0.30$ GeV$$ \\ \hline\hline
\end{tabular}%
\end{table}
\begin{table*}[ht]
\caption{The Wilson coefficients $C_{i}(\mu)$ at the scale $\mu\sim m_{b}$ in the SM.}
\begin{tabular}{cccccccccc}
\hline\hline
$C_{1}$&$C_{2}$&$C_{3}$&$C_{4}$&$C_{5}$&$C_{6}$&$C_{7}$&$C_{9}$&$C_{10}$
\\ \hline
 \ \ \ 1.107 \ \ \ & \ \ \ -0.248 \ \ \ & \ \ \ -0.011 \ \ \ & \ \ \ -0.026 \ \ \ & \ \ \ -0.007 \ \ \ & \ \ \ -0.031 \ \ \ & \ \ \ -0.313 \ \ \ & \ \ \ 4.344 \ \ \ & \ \ \ -4.669 \ \ \ \\
\hline\hline
\end{tabular}
\label{wc table}
\end{table*}
It has already been mentioned that in $B_c$ to $D_{s}^{\ast}$ transition the WA contributions are
proportional to $V_{cb}V^{*}_{cs}$ and hence can not be ignored like in the ordionary $B_{u,d,s}$ to light meson decays.
Using the values of the form factors given in Table I and II along with the value of input parameters
from Tables III and IV, the numerical result of the branching ratios for the decays $B_{c}\rightarrow
D_{s}^{\ast }\ell^{+}\ell^{-}$ containing contributions from penguin, WA and both amplitudes are given in Table
V \cite{51}.
\begin{table}[tbh]
\caption{Branching ratio for $B_{c}\rightarrow D_{s}^{\ast }\protect\mu ^{+}%
\protect\mu ^{-}(\protect\tau ^{+}\protect\tau ^{-})$ decay using
form factors calculated in QCD sum rules \cite{53a,53b}.}%
\begin{tabular}{|r|r|r|}
\hline\hline
$BR^{\text{(PENG)}}(B_{c}\rightarrow D_{s}^{\ast }\mu ^{+}\mu ^{-}(\tau^{+}\tau^{-}))$ & $%
BR^{\text{(WA)}}(B_{c}\rightarrow D_{s}^{\ast }\mu ^{+}\mu ^{-}(\tau^{+}\tau^{-}))$ & $BR^{%
\text{(Total)}}(B_{c}\rightarrow D_{s}^{\ast }\mu ^{+}\mu ^{-}(\tau^{+}\tau^{-}))$ \\
\hline
$2.57\times 10^{-7}\left( 1.13\times 10^{-8}\right) $ & $%
2.20\times 10^{-6}\left( 0.35\times 10^{-9}\right) $ & $2.46\times
10^{-6}\left( 1.49\times 10^{-8}\right) $ \\ \hline\hline
\end{tabular}%
\label{Branching ratios}
\end{table}
Here, one can see that the WA contribution to the branching
ratio for $B_{c}\rightarrow D_{s}^{\ast }\mu^{+}\mu ^{-}$ decay is almost an order of magnitude larger than the penguin ones
therefore, we have to include it in the analysis of $B_{c}\rightarrow D_{s}^{\ast } \ell^{+}\ell^{-}$ in SM4.

Regarding the parameters of the SM4, recently CDF collaboration has
given the lower bound on the mass of the $t^{\prime }$ quark to be
$m_{t^{\prime }}\geq 335$ GeV at $95\%$ CL \cite{CDFNEW}. These
bounds are little higher than the ones quoted in Ref. \cite{CDF} of
$m_{t^{\prime}}\gtrsim 256$ GeV.
On the other hand, the perturbativity of the Yukawa coupling implies that $%
m_{t^{\prime}}\lesssim \sqrt{2\pi }\left\langle v\right\rangle
\approx 600$ GeV, where $\left\langle v\right\rangle $ is the vacuum
expectation value of the Higgs boson \cite{Londonnewsm4}. Thus, the
mass $m_{t^{\prime}}$ is constrained in a band, $m_{t^{\prime
}}=335-600$ GeV, which increases the predictability of SM4. Keeping
in the view that these bounds will be considerably improved at LHC, we
set $m_{t^{\prime }}=300-600$ GeV in our numerical
calculation. In addition to the masses of the sequential fourth
generation of quarks the other important parameters are the CKM4
matrix elements, where $|V_{t^{\prime }s}|$ and $|V_{t^{\prime }b}|$
are of the main interest for present study. The experimental upper
bounds on these CKM matrix elements are $|V_{t^{\prime }s}|<0.11$
and $|V_{t^{\prime
}b}|<0.12$ \cite{boundsCKM,Samitra}. By taking the CKM unitarity condition, $%
\sum\limits_{i}V_{is}^{\ast }V_{ib},~(i=u,c,t,t^{\prime })$ together
with the present measurements of the $3\times 3$ CKM matrix
\cite{boundsSM3}, the bounds for CKM4 matrix elements are obtained
to be \cite{boundsSM4,Samitra}
\begin{equation}
|V_{t^{\prime }s}^{\ast }V_{t^{\prime }b}|\leq 1.5\times 10^{-2}.
\label{CKMbounds}
\end{equation}%
Incorporating these constraints on the fourth generation parameter space, the NP effects origination from SM4 on different physical observables
are shown in Figs. 1-14.

\begin{figure}[tbp]
\begin{center}
\begin{tabular}{ccc}
\vspace{-0.3cm} \includegraphics[scale=0.6]{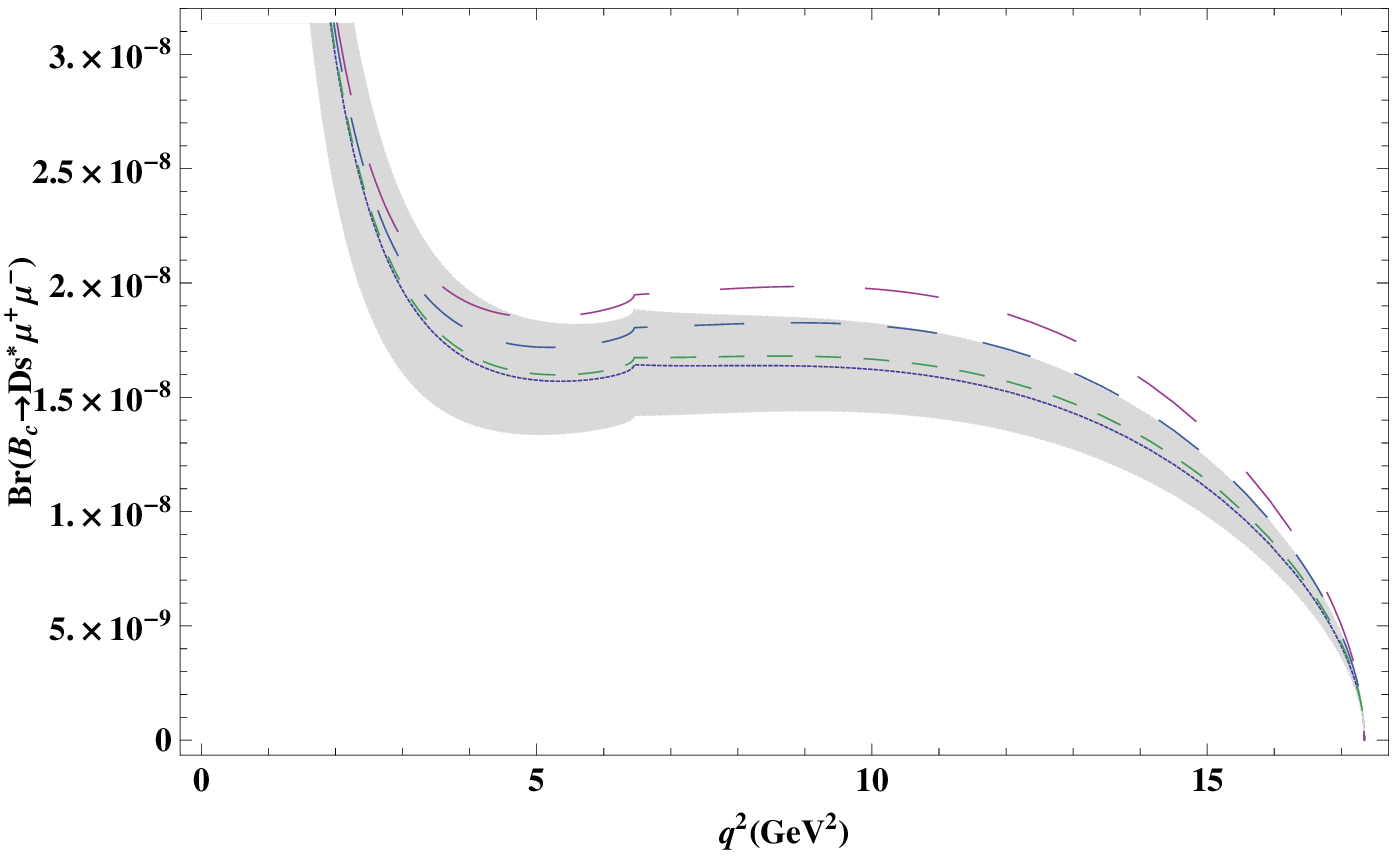} %
\includegraphics[scale=0.6]{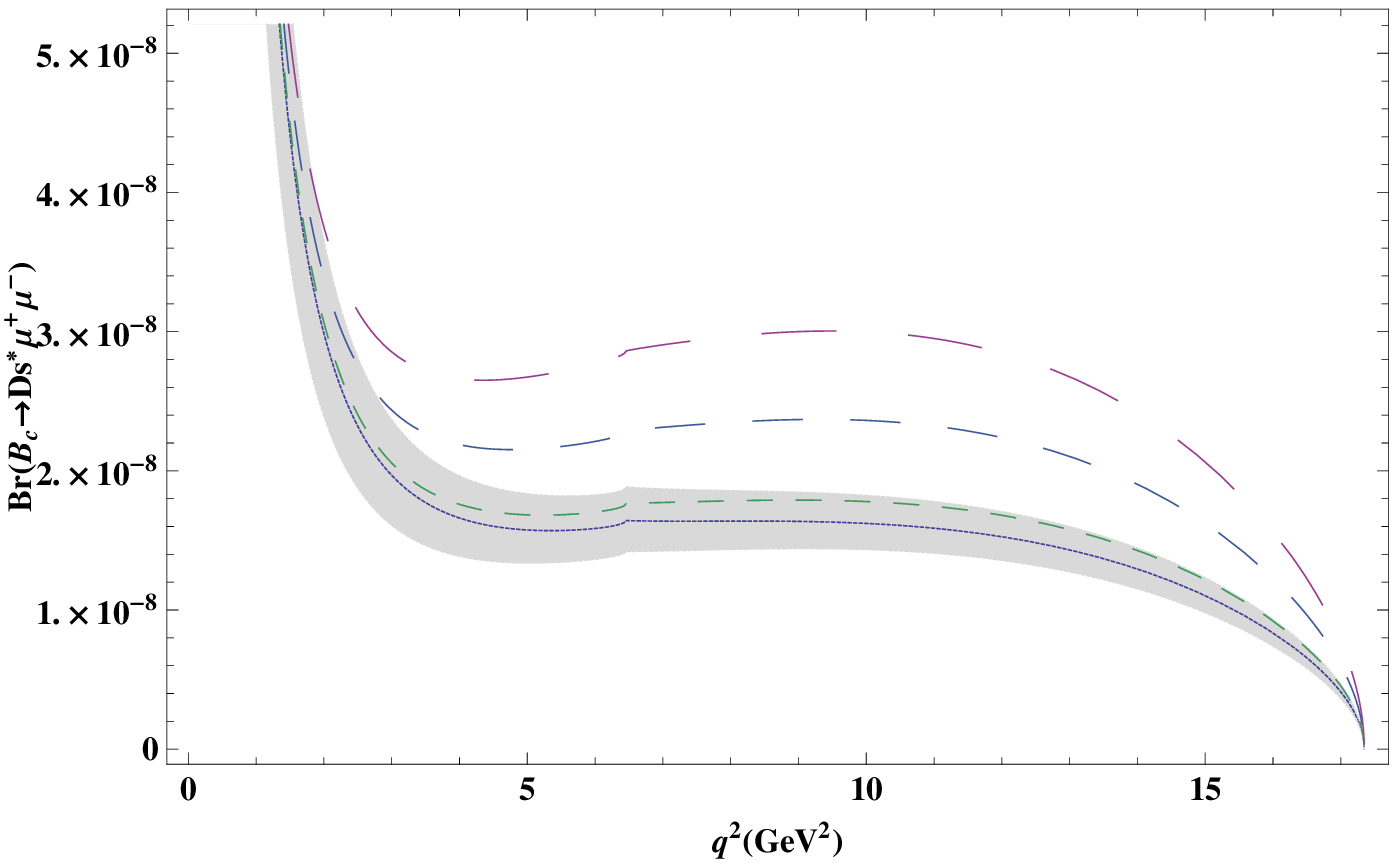} &  &  \\
\includegraphics[scale=0.6]{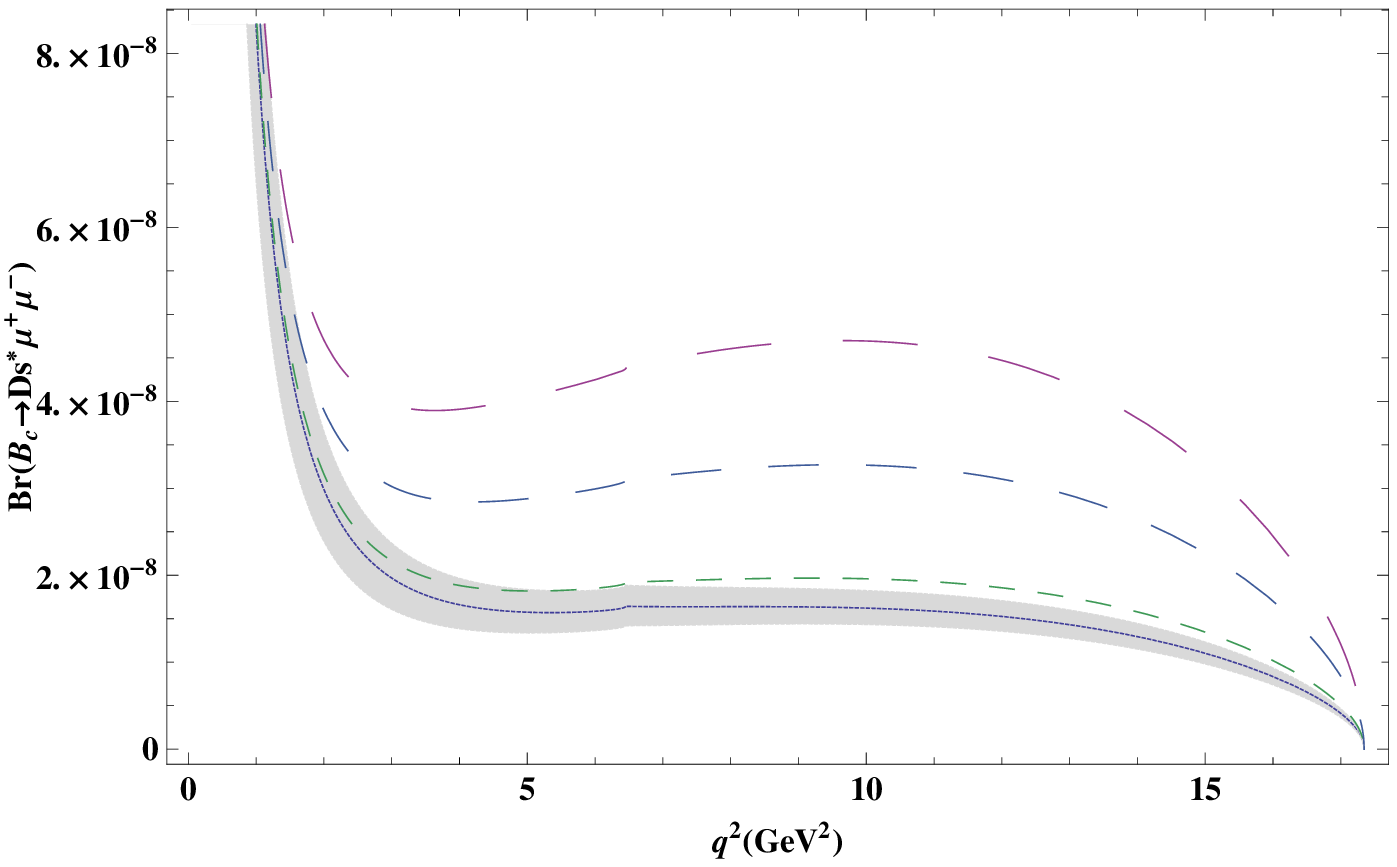} \includegraphics[scale=0.6]{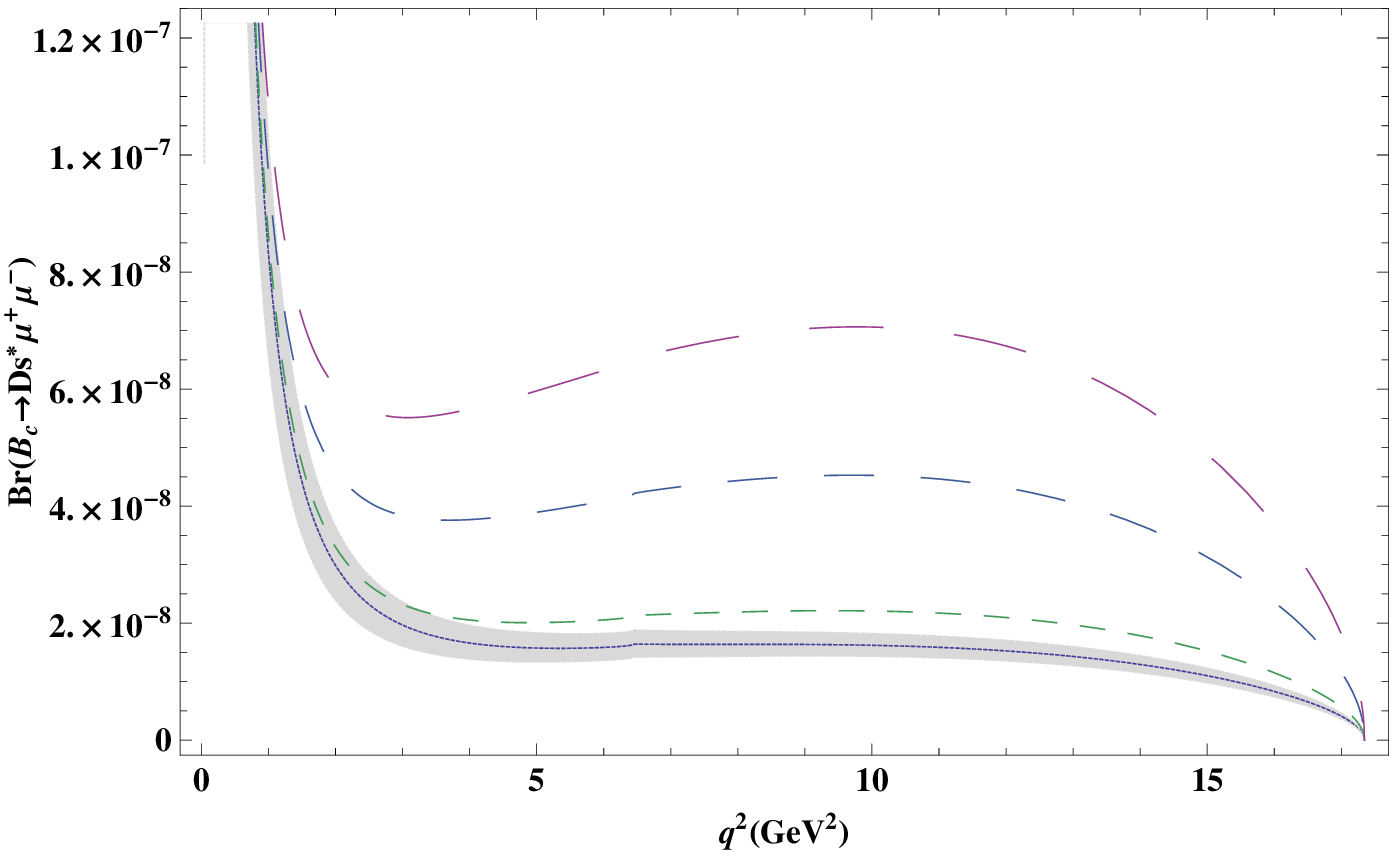}
\put (-350,240){(a)} \put (-100,240){(b)} \put (-350,0.4){(c)}
\put(-100,0.4){(d)} \vspace{-0.5cm} &  &
\end{tabular}%
\end{center}
\caption{The dependence of branching ratio of $B_{c}\rightarrow D_{s}^{\ast
}\protect\mu ^{+}\protect\mu ^{-}$ on $q^{2}$ for different values of $%
m_{t^{\prime }}$ and $\left\vert V_{t^{\prime }b}^{\ast }V_{t^{\prime
}s}\right\vert $. $\left\vert V_{t^{\prime }b}^{\ast }V_{t^{\prime
}s}\right\vert $ $=$ $0.003$, $0.006$, $0.009$ and $0.012$ in (a), (b), (c)
and (d) respectively. In all the graphs, the solid line corresponds to the
SM, dashed, medium dashed and long dashed lines are for $%
m_{t^{\prime }}$ $=$ $300$ GeV , $500$ GeV and $600$ GeV
respectively.}
\label{Branching ratio for muons}
\end{figure}

\begin{figure}[tbp]
\begin{center}
\begin{tabular}{ccc}
\vspace{-0.3cm} \includegraphics[scale=0.6]{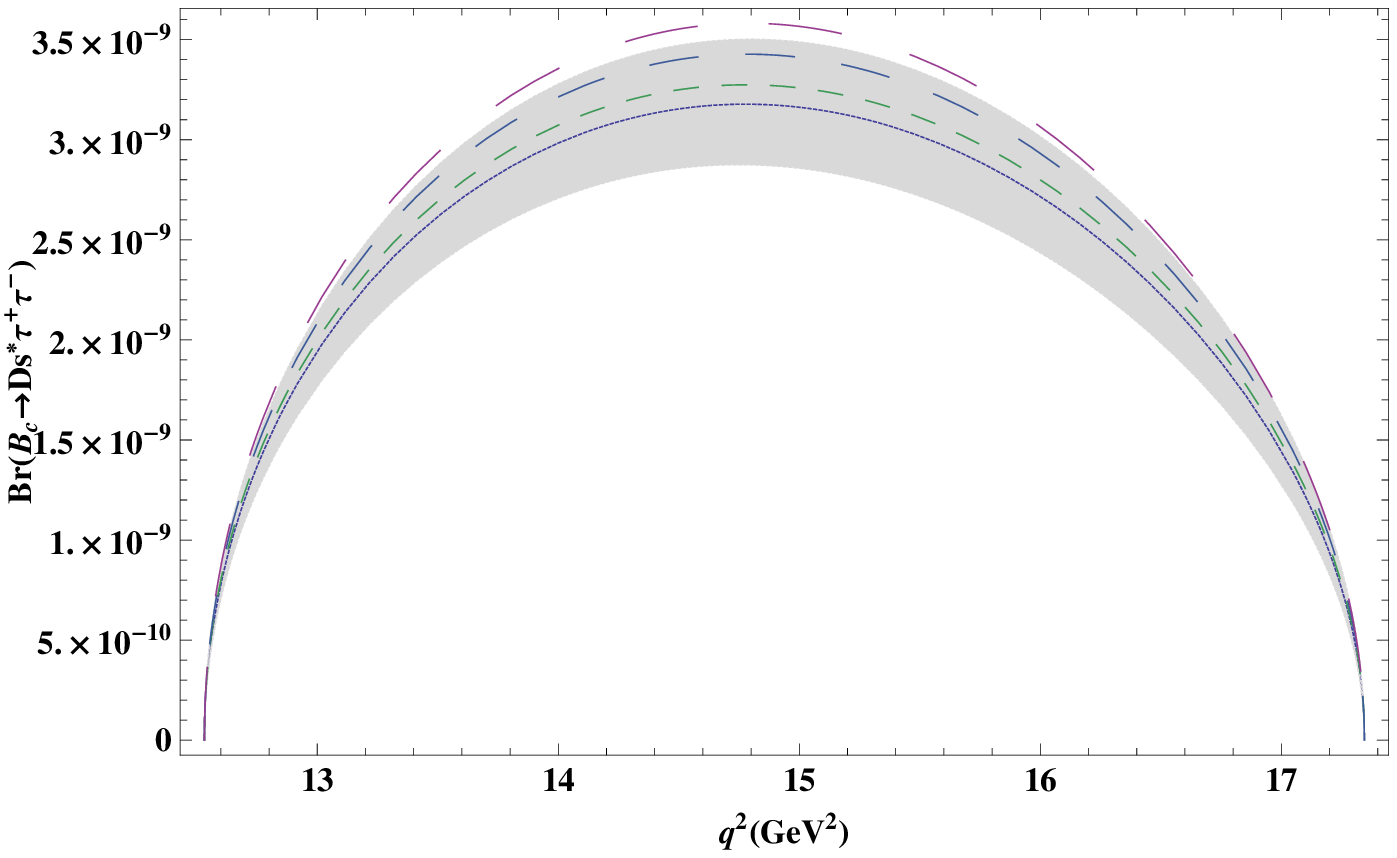} %
\includegraphics[scale=0.6]{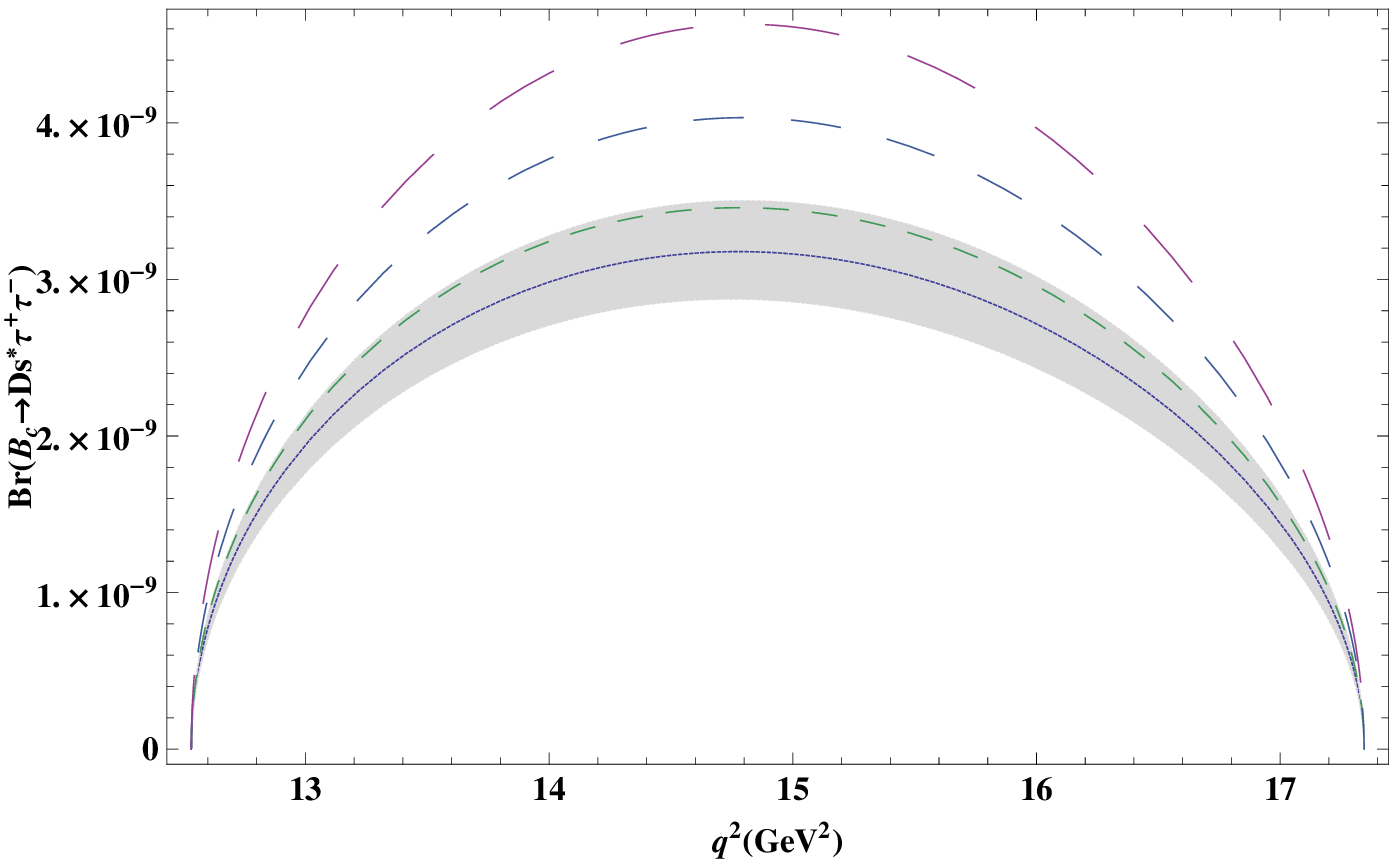} &  &  \\
\includegraphics[scale=0.6]{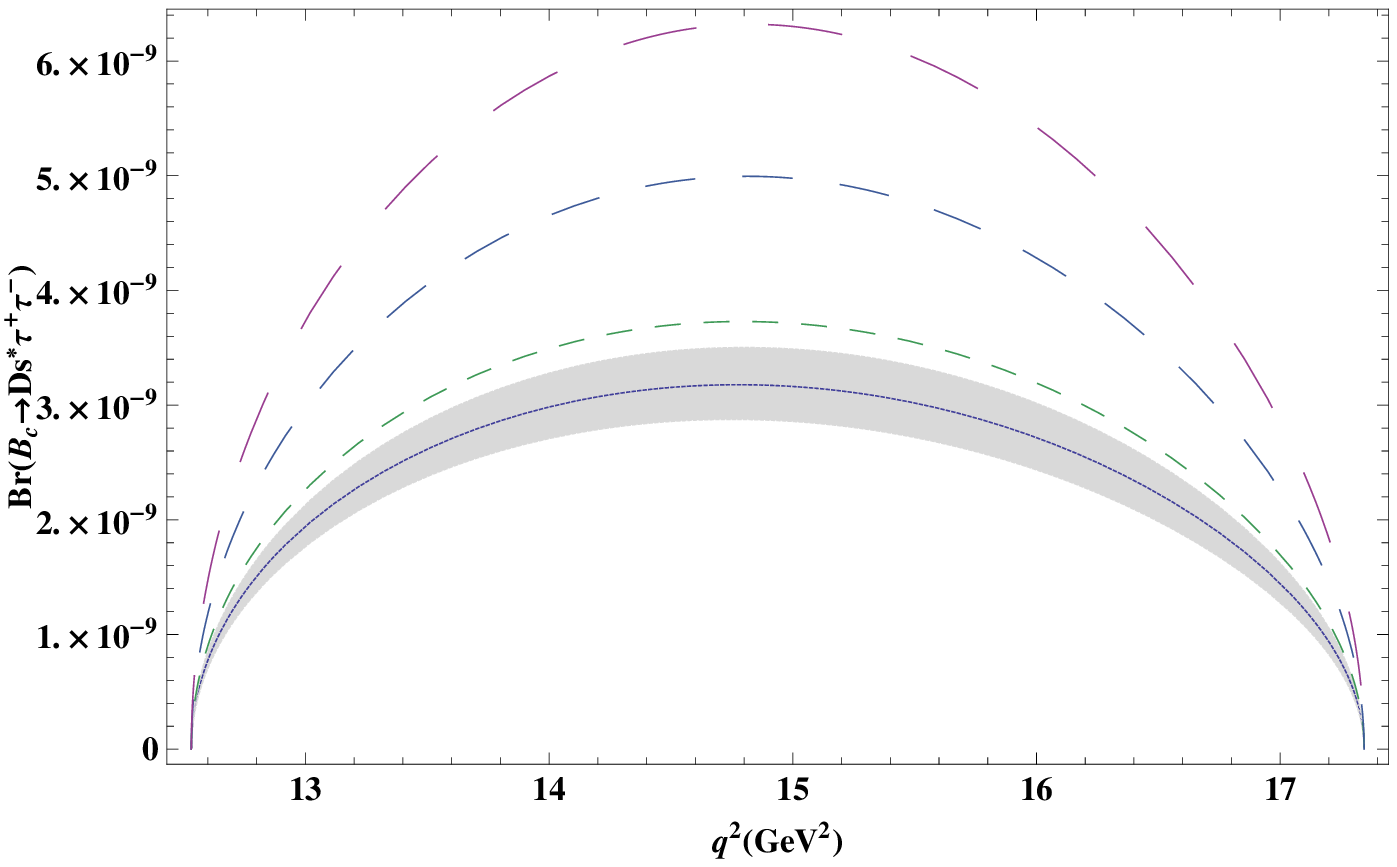} \includegraphics[scale=0.6]{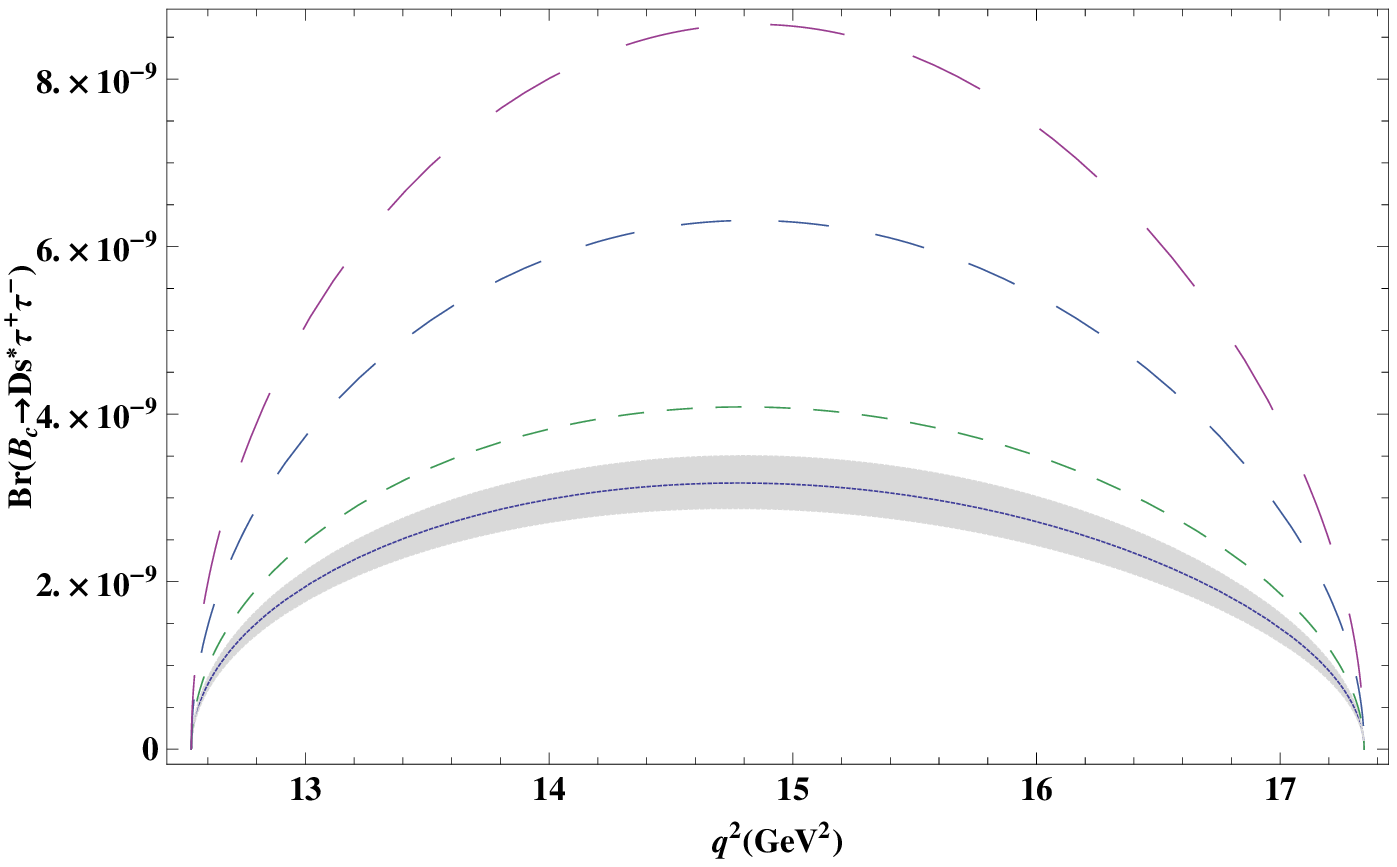}
\put (-350,240){(a)} \put (-100,240){(b)} \put (-350,0.2){(c)}
\put(-100,0.2){(d)} \vspace{-0.5cm} &  &
\end{tabular}%
\end{center}
\caption{The dependence of branching ratio of $B_{c}\rightarrow D_{s}^{\ast
}\protect\tau ^{+}\protect\tau ^{-}$ on $q^{2}$ for different values of $%
m_{t^{\prime }}$ and $\left\vert V_{t^{\prime }b}^{\ast }V_{t^{\prime
}s}\right\vert $. The values of fourth generation parameters and the legends
are same as in Fig.1}
\label{Branching ratio for tau}
\end{figure}

Figs. \ref{Branching ratio for muons} and \ref{Branching ratio for tau}
depict the variation of the differential branching ratio of $B_{c}\rightarrow
D_{s}^{\ast}\mu^{+}\mu^{-}(\tau^{+}\tau^{-})$ decays with $q^2$ both
in the SM and in the sequential fourth generation model (SM4). These figures
indicate that the values of the differential branching ratios are enhanced sizably with increase
in the values of fourth generation parameters $m_{t^{\prime }}$ and $%
\left\vert V_{t^{\prime }b}^{\ast }V_{t^{\prime }s}\right\vert $.
These new physics effects are prominent in whole $q^2$ region both for the
$\mu$ and $\tau$ as the final state leptons which is due to the fact that
at small value of $q^2$
the dominant contribution comes from $C_{7}^{tot}(\mu)$
whereas for the large value of $q^{2}$ the major contribution is from the $%
Z$ exchange i.e., $C_{10}^{tot}$, which is sensitive to the mass of the
fourth generation quark $(m_{t^{\prime }})$.

As an exclusive decay, the new physics effects in the branching ratios
are usually masked by the uncertainties involved in different input parameters
where form factors are the major contributors. However, for the present case
the new physics effects are very prominent and lies well separated from the
SM values even in the error bounds. However there exists some other obseverables
which have very mild dependence on the choice of the form factors.
Among them the zero position of the
forward-backward asymmetry, helicity fractions of final state $D_{s}^{\ast}$ meson and
the different lepton polarization asymmetries,
are almost free from the hadronic uncertainties, in particular at low
$q^2$ region, and hence serve as an
important tool to study NP.

\begin{figure}[tbp]
\begin{center}
\begin{tabular}{ccc}
\vspace{-0.3cm} \includegraphics[scale=0.6]{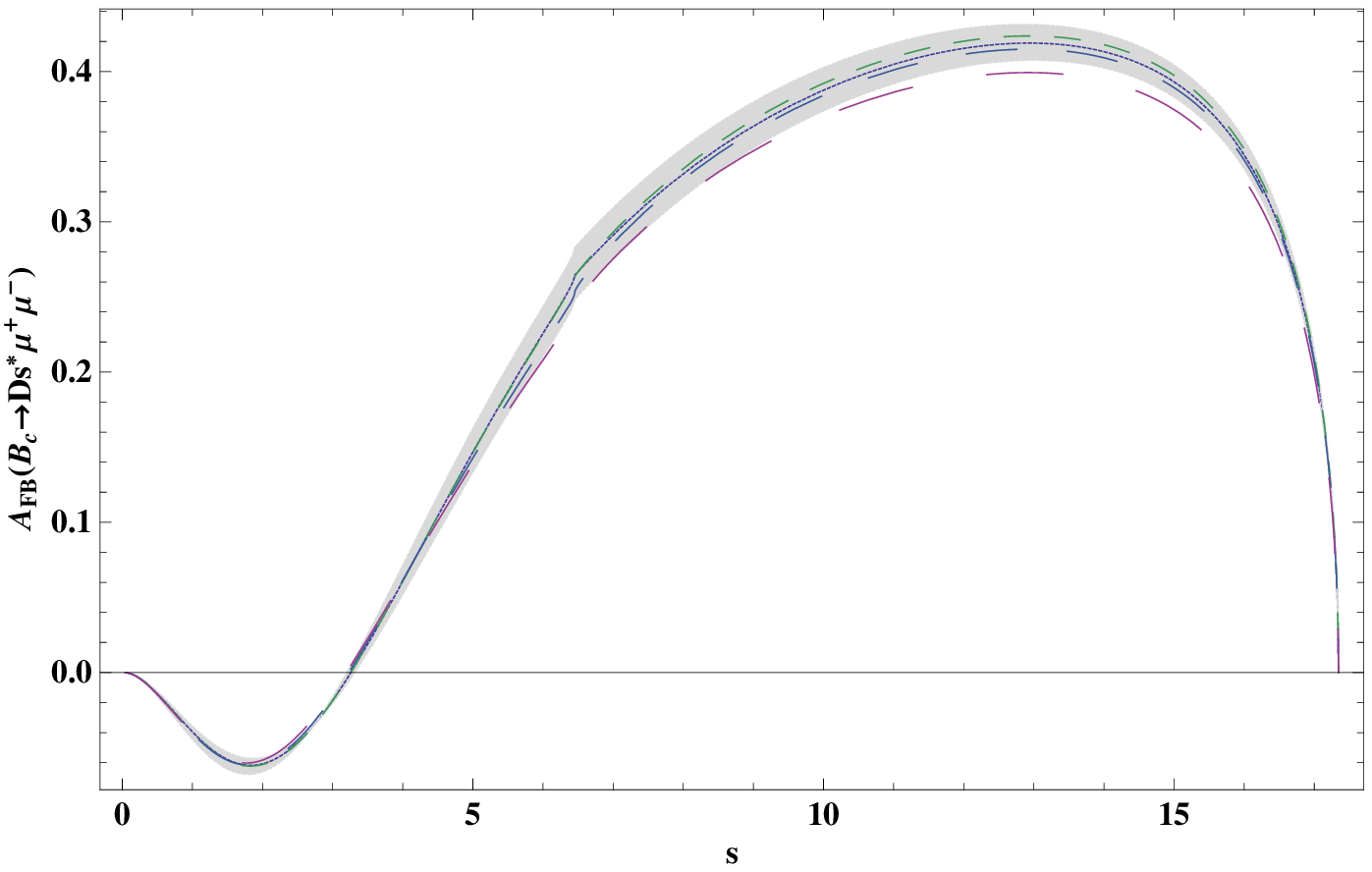} %
\includegraphics[scale=0.6]{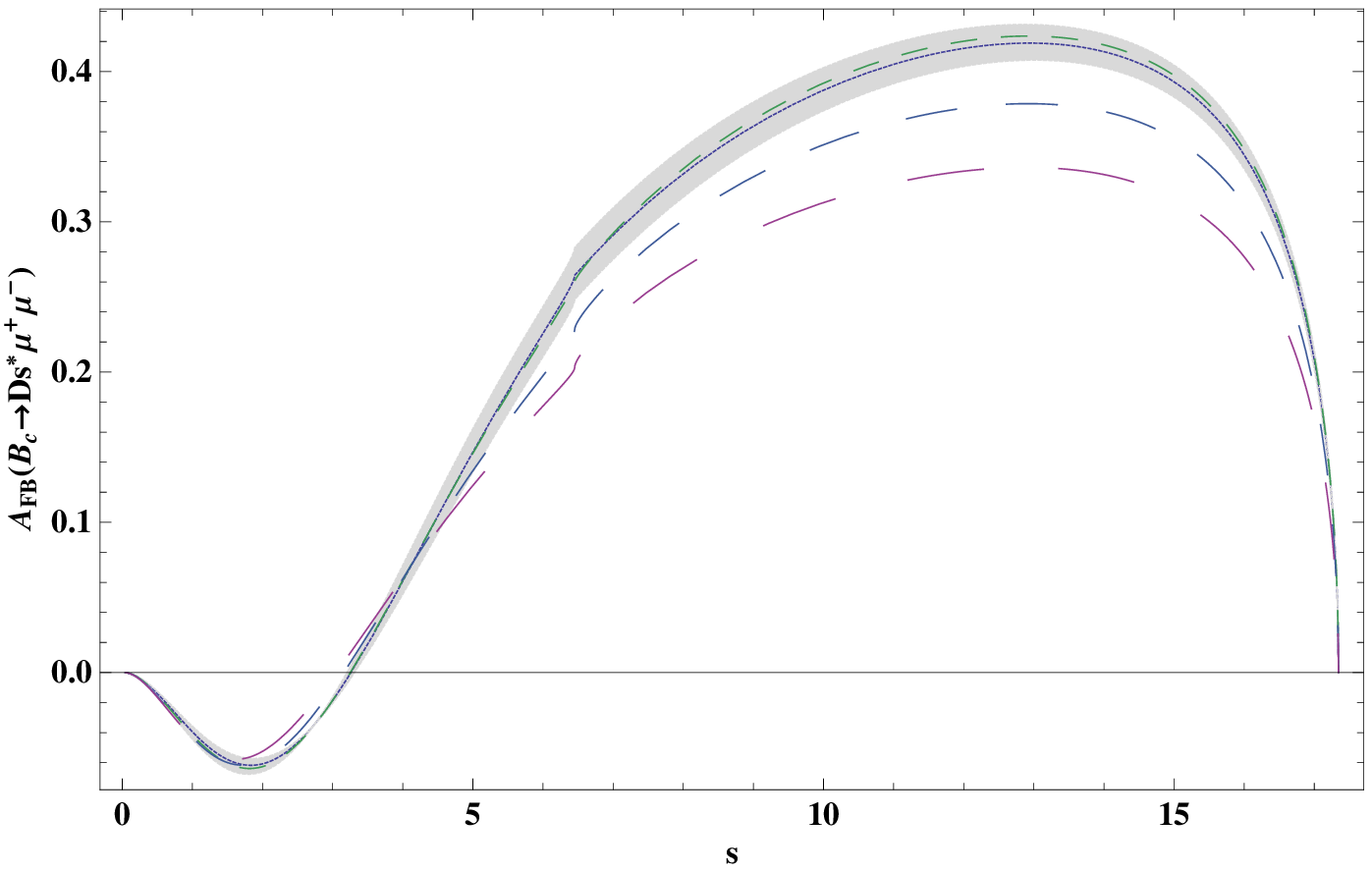} &  &  \\
\includegraphics[scale=0.6]{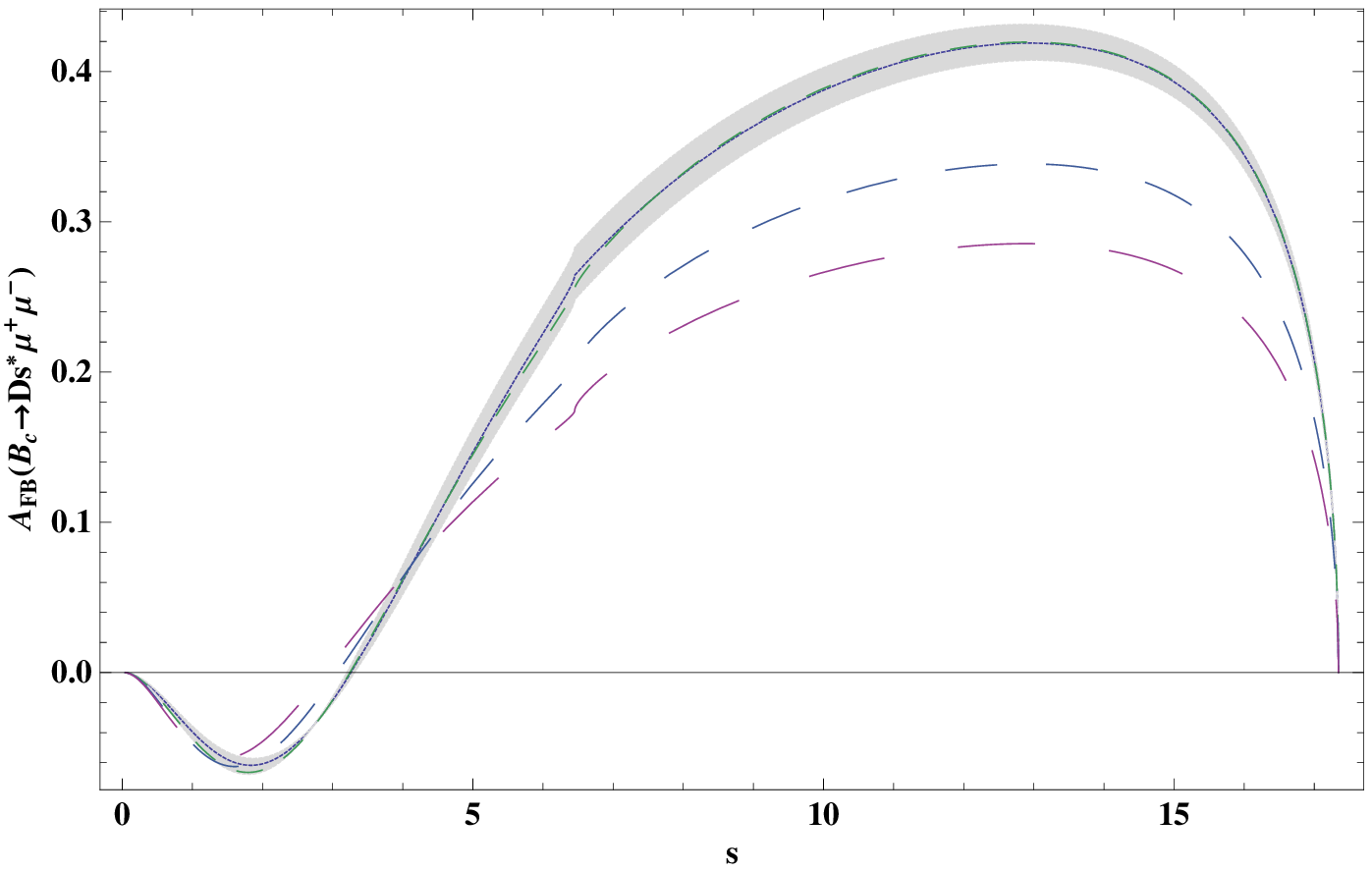} \includegraphics[scale=0.6]{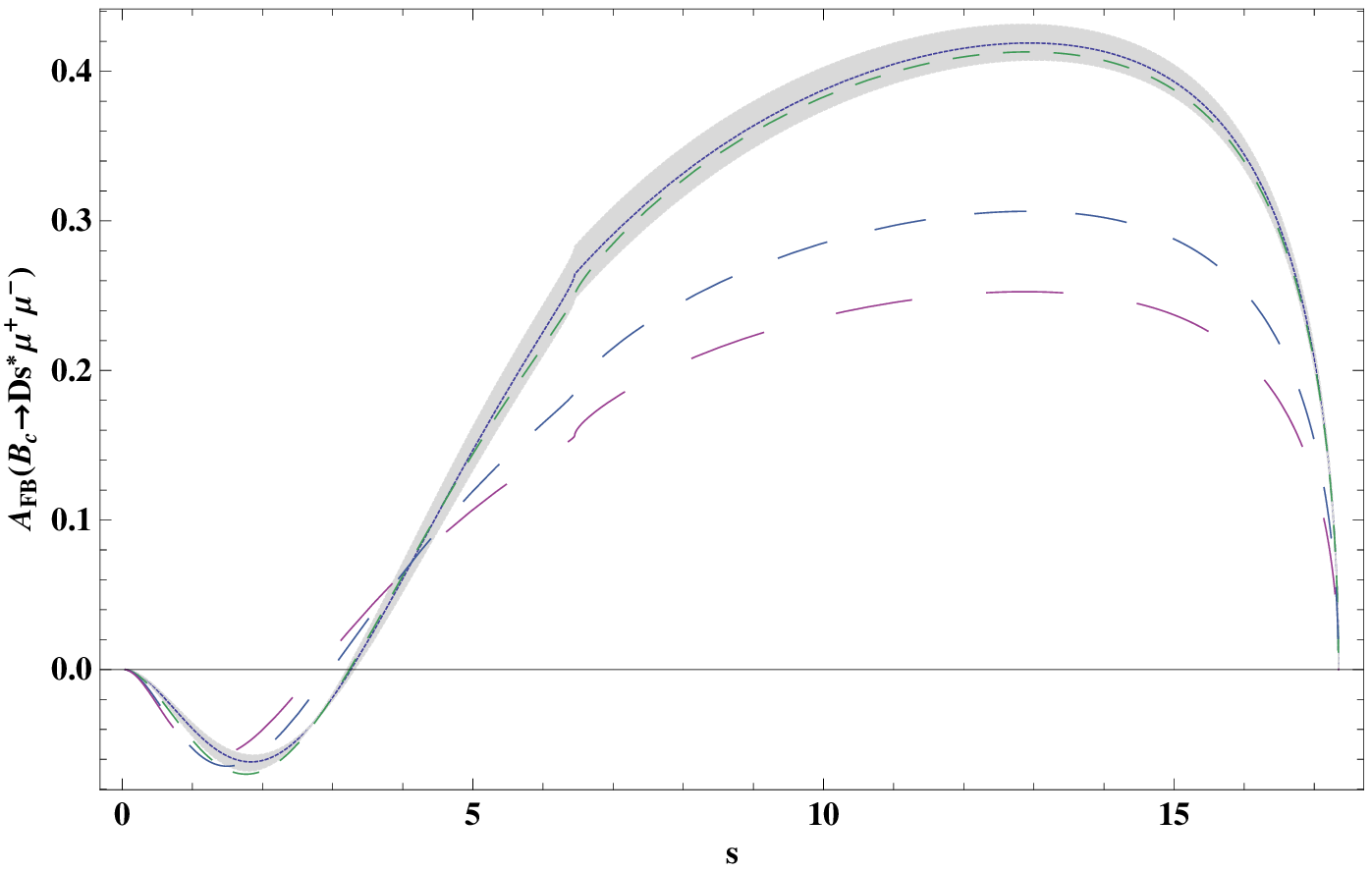}
\put (-350,240){(a)} \put (-100,240){(b)} \put (-350,0.2){(c)}
\put(-100,0.2){(d)} \vspace{-0.5cm} &  &
\end{tabular}%
\end{center}
\caption{The dependence of forward-backward asymmetry of $B_{c}\rightarrow D_{s}^{\ast
}\protect\mu ^{+}\protect\mu ^{-}$ on $q^{2}$ for different values of $%
m_{t^{\prime }}$ and $\left\vert V_{t^{\prime }b}^{\ast }V_{t^{\prime
}s}\right\vert $. The values of fourth generation parameters and the legends
are same as in Fig.1.}
\label{Forward-Backward asymmetry for muons}
\end{figure}

Fig. \ref{Forward-Backward asymmetry for muons} describes the behavior of the
forward-backward asymmetry of $B_{c}\rightarrow D_{s}^{\ast} \mu ^{+}\mu ^{-}$ with $q^2$. Here one can see that the value of the
forward-backward asymmetry passes from the zero at a particular value of
$q^2$ both in the SM as well as in the SM4.
This is because of the destructive interference between the
photon penguin ($C_{7}^{eff}$) and the $Z$
penguin ($C_{9}^{eff}$). It has already been mentioned that the SM4 effects display themselves
in the Wilson coefficients, therefore, one expects that both the zero crossing of
FBA and its magnitude will be different from SM. This fact is illustrated
in \ref{Forward-Backward asymmetry for muons}. We can see that the value of
the forward-backward asymmetry decreases from the SM value but the
position of zero crossing remains the same for the low value of SM4
parameters $(m_{t^{\prime}}, \left\vert V_{t^{\prime }b}^{\ast
}V_{t^{\prime }s}\right\vert)$ (c.f. Fig. 3(a,b)).
However, at the large value of the CKM4 matrix elements and the mass $%
m_{t^{\prime }}$ the zero position is shifted to the
left which
makes it an important candidate for the search of SM4 effects.

\begin{figure}[tbp]
\begin{center}
\begin{tabular}{ccc}
\vspace{-0.3cm} \includegraphics[scale=0.6]{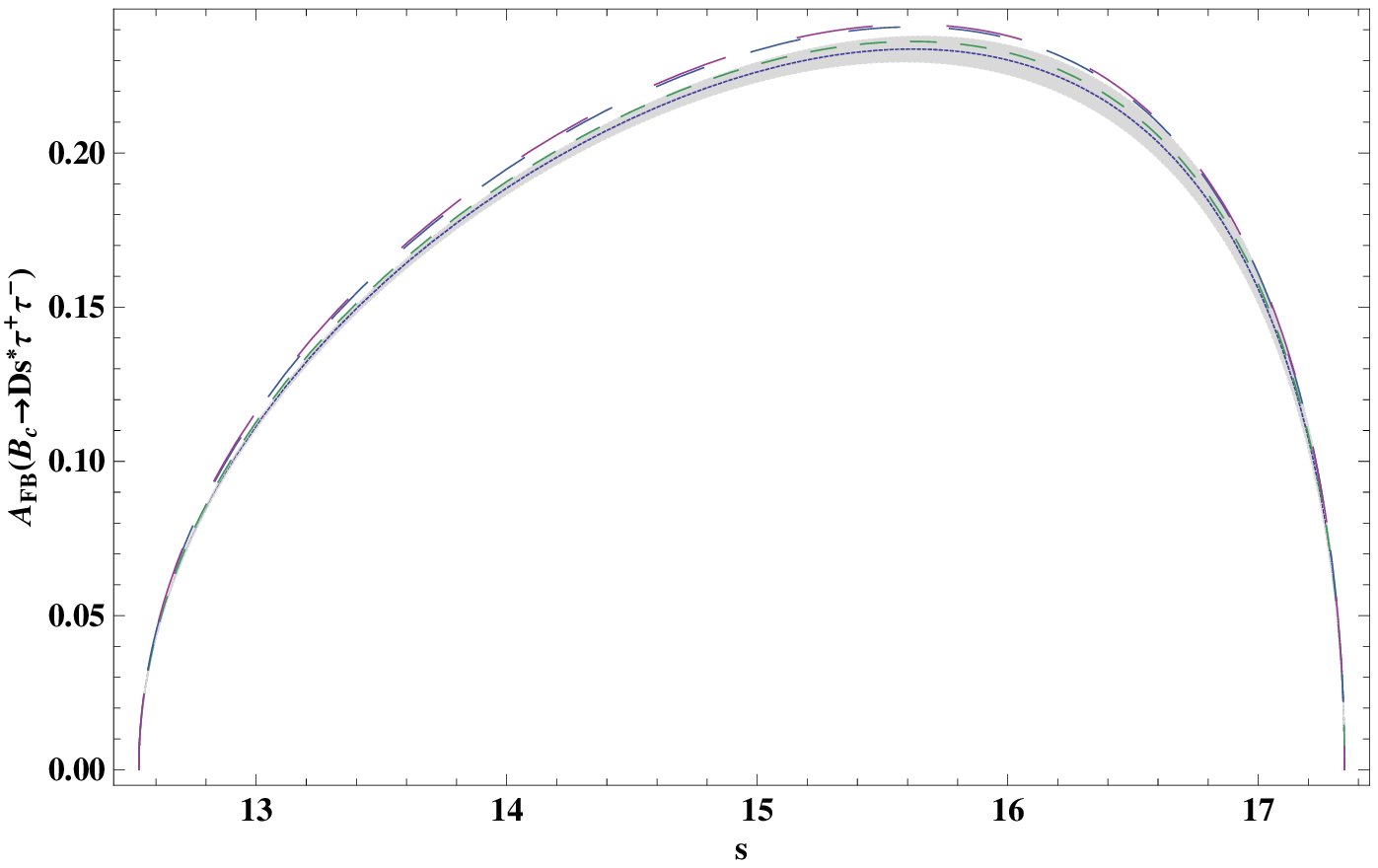} %
\includegraphics[scale=0.6]{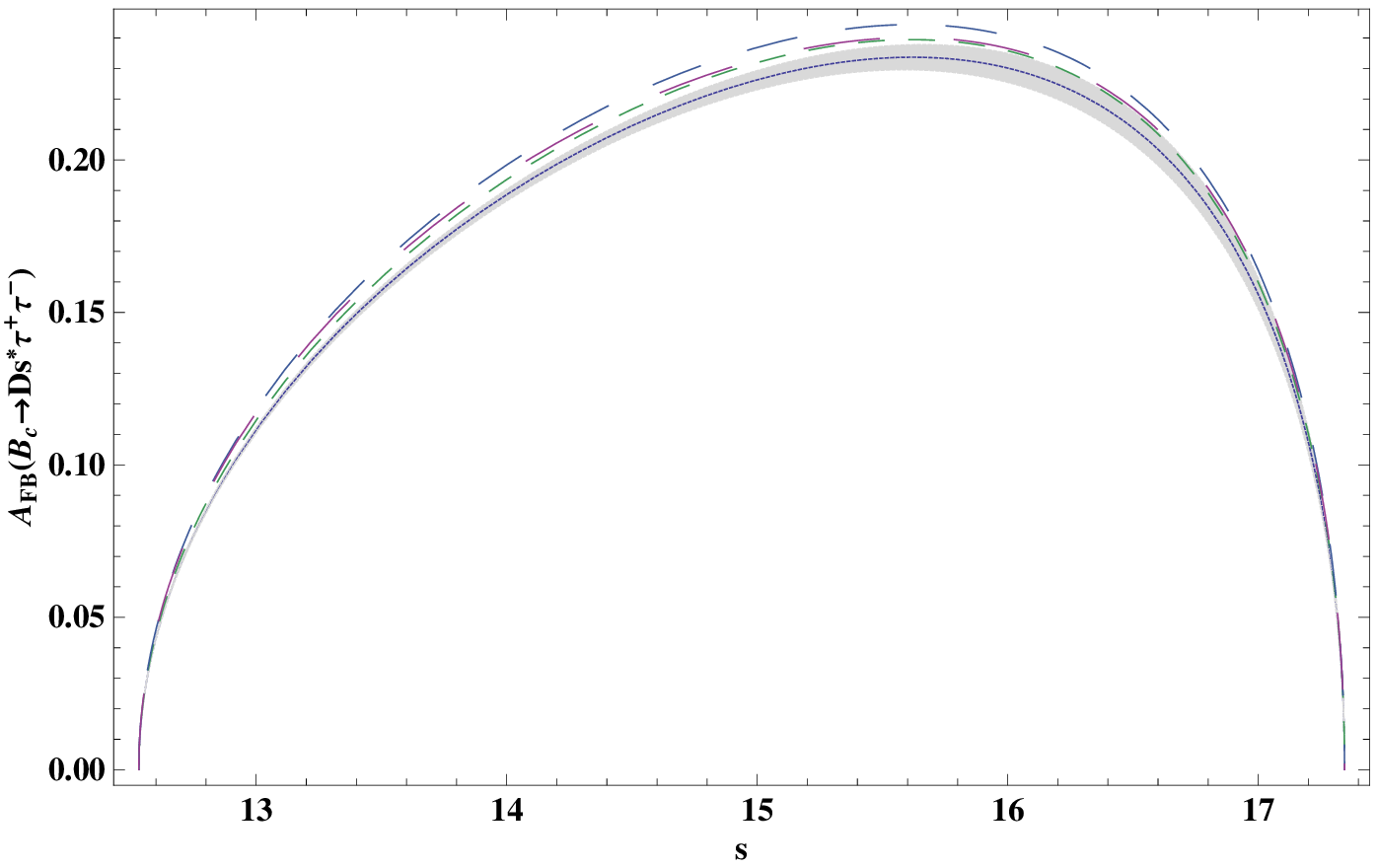} &  &  \\
\includegraphics[scale=0.6]{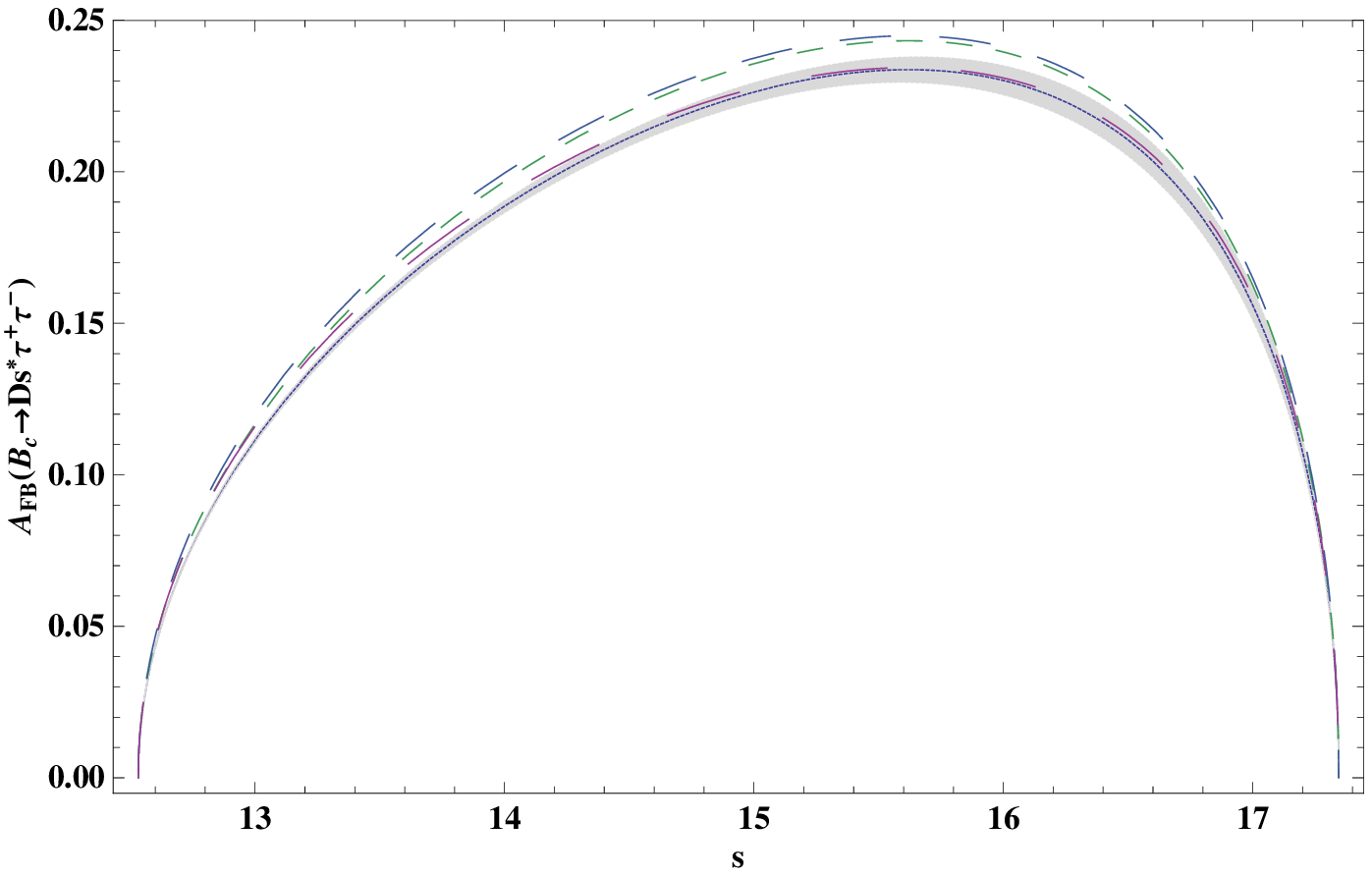} \includegraphics[scale=0.6]{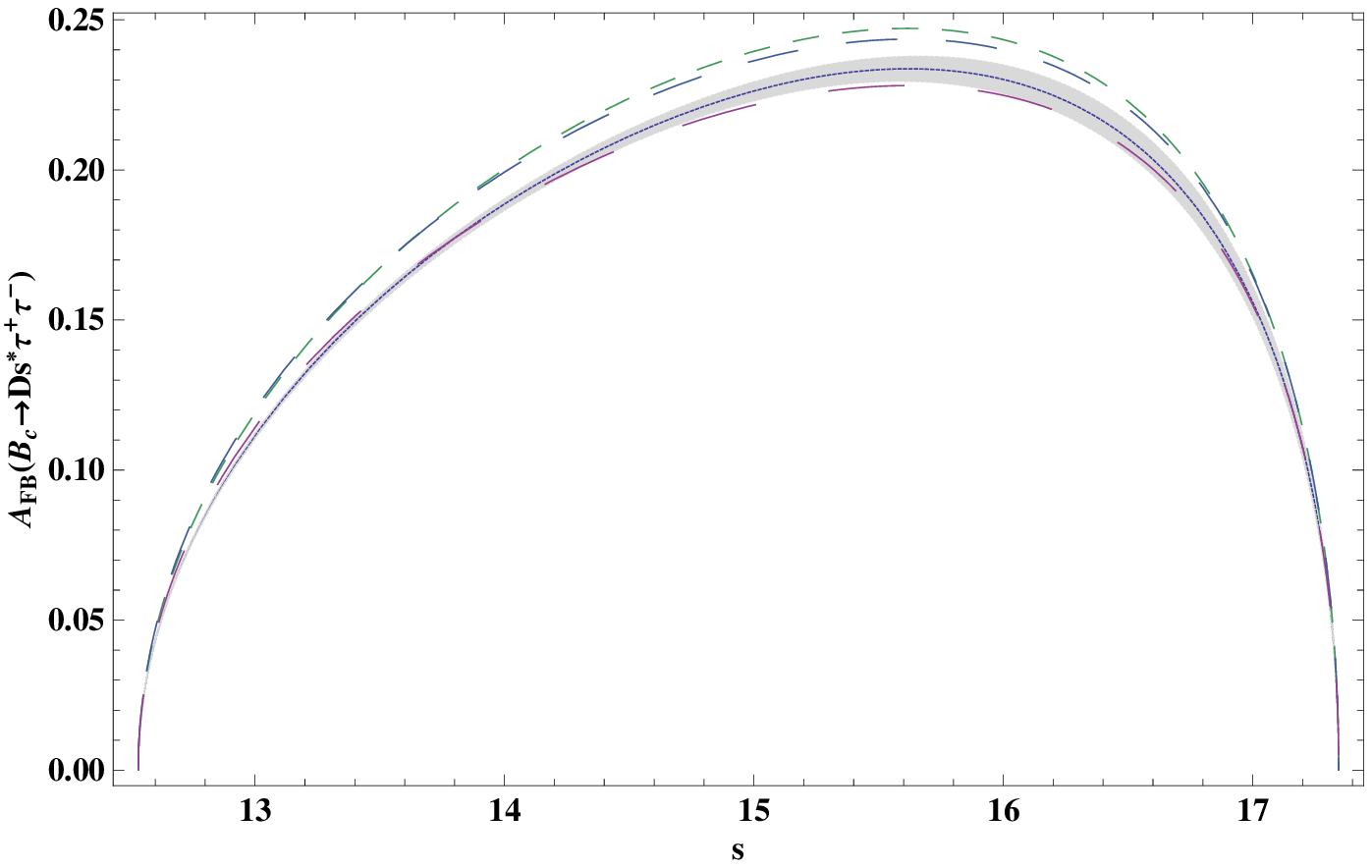}
\put (-350,240){(a)} \put (-100,240){(b)} \put (-350,0.2){(c)}
\put(-100,0.2){(d)} \vspace{-0.5cm} &  &
\end{tabular}%
\end{center}
\caption{The dependence of forward-backward asymmetry of $B_{c}\rightarrow D_{s}^{\ast
}\protect\tau ^{+}\protect\tau ^{-}$ on $q^{2}$ for different values of $%
m_{t^{\prime }}$ and $\left\vert V_{t^{\prime }b}^{\ast }V_{t^{\prime
}s}\right\vert $. The values of fourth generation parameters and the legends
are same as in Fig.1. }
\label{Forward-Backward asymmetry for tau}
\end{figure}

Now for the $%
B_{c}\rightarrow D_{s}^{\ast} \tau ^{+}\tau ^{-}$ decay the
FBA is presented in Fig. \ref{Forward-Backward asymmetry for tau}.
In this case the crossing of the FBA is absent both in the SM and in the SM4, however, there is
a significant deviation in its magnitude for large values of the SM4 parameters. As the magnitude of the forward-backward
asymmetry is also an important tool to establish the NP, therefore, the experimental study of
it will give us some distinguishing effects of SM4.

Another handy tool to explore the NP is the study of helicity fractions of the final state meson
which are associated with its spin. In literature, there exist some studies on the helicity fractions
for the case of vector $(K^{\ast}, D_{s}^{\ast})$ and $K_1(1270,1400)$ mesons both in the SM and in some NP
scenarios, whereas for the $K^{\ast}$ there exists some experimental observations too \cite{51, colangelo}.
It is therefore legitimate to study SM4 contributions to the helicity fractions of $D_{s}^{\ast}$ meson
in $B_{c}\rightarrow D_{s}^{\ast}\ell^{+}\ell ^{-}$ decays.

\begin{figure}[tbp]
\begin{center}
\begin{tabular}{ccc}
\vspace{-0.3cm} \includegraphics[scale=0.6]{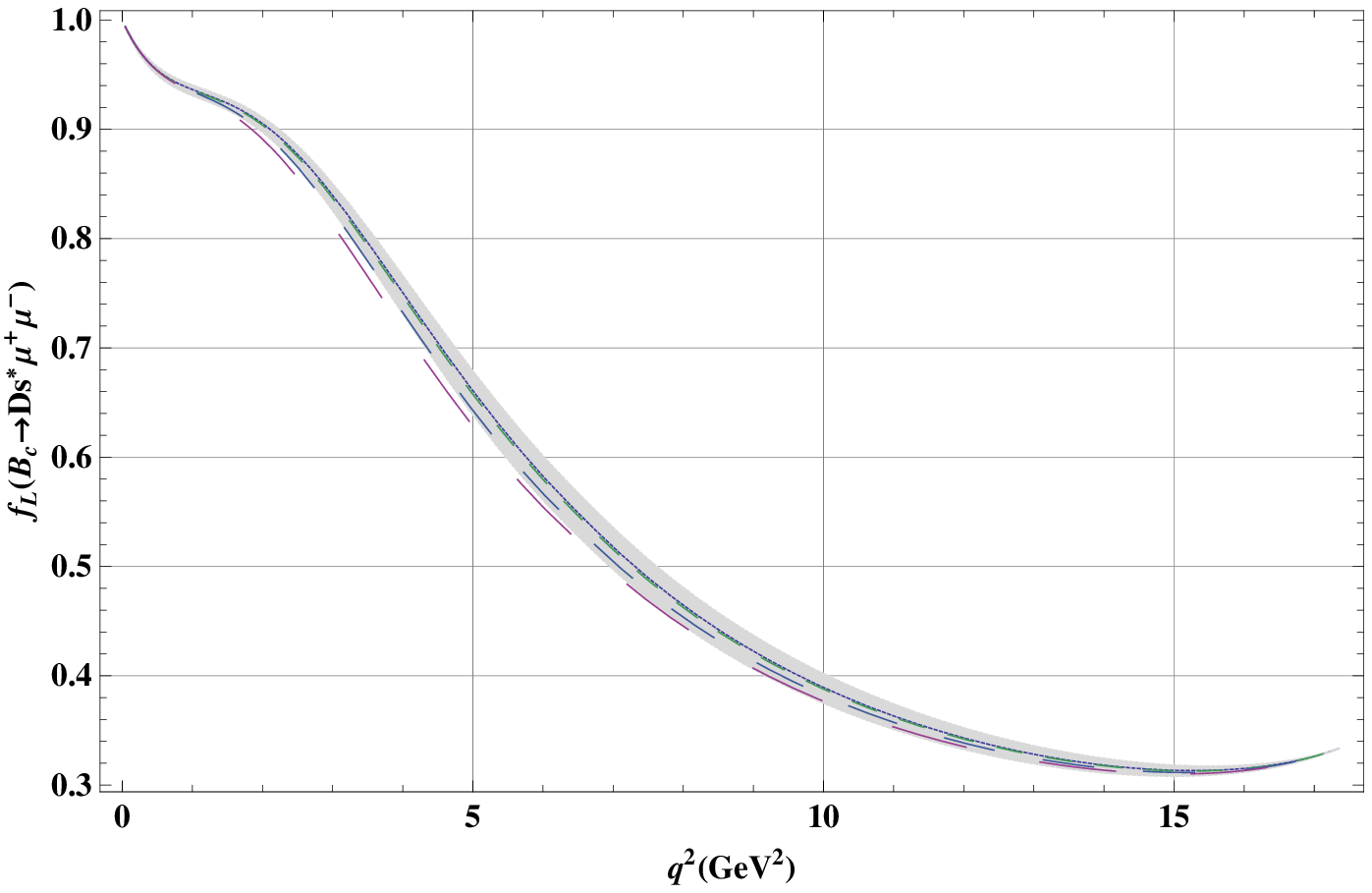} %
\includegraphics[scale=0.6]{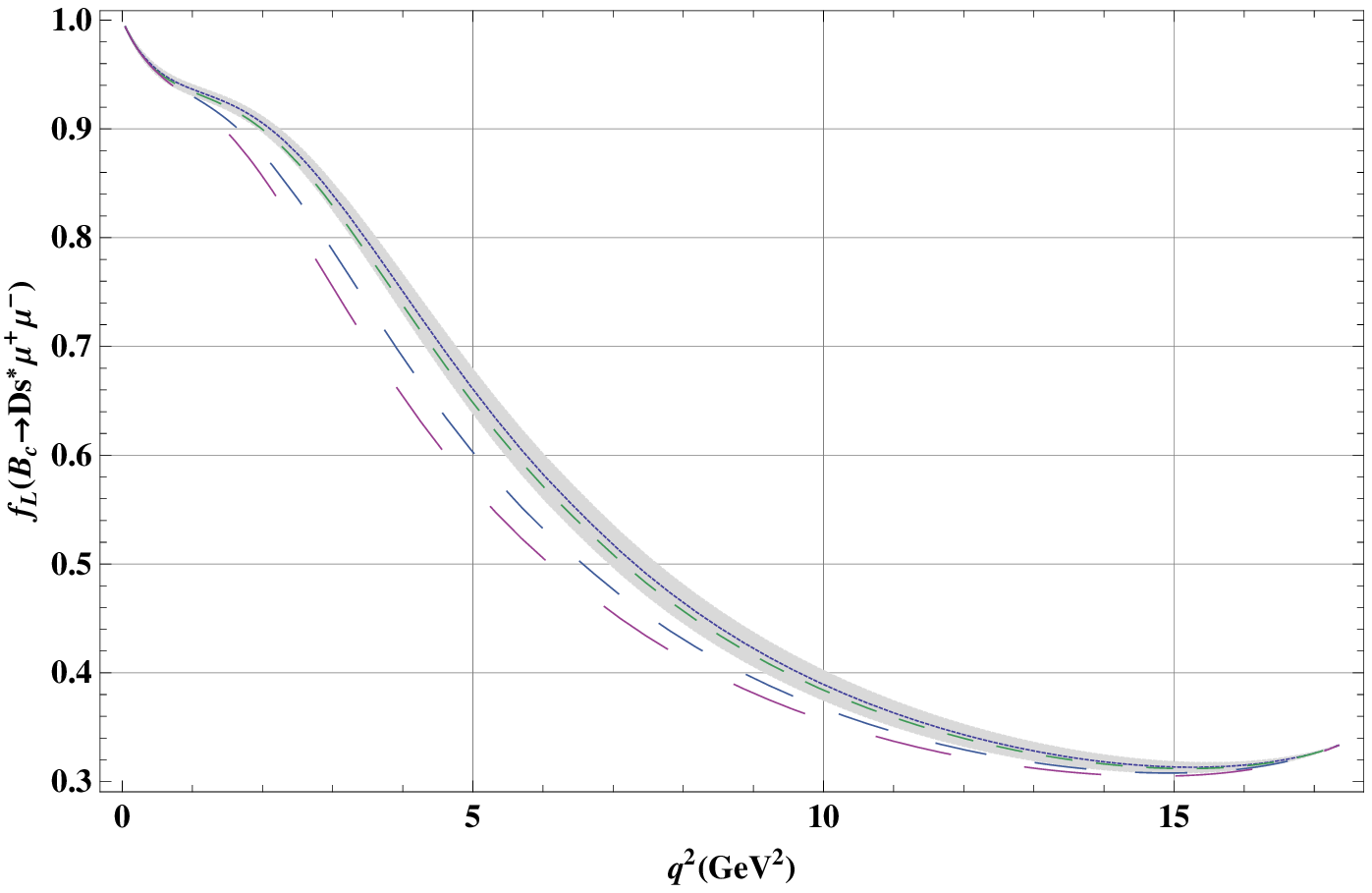} &  &  \\
\includegraphics[scale=0.6]{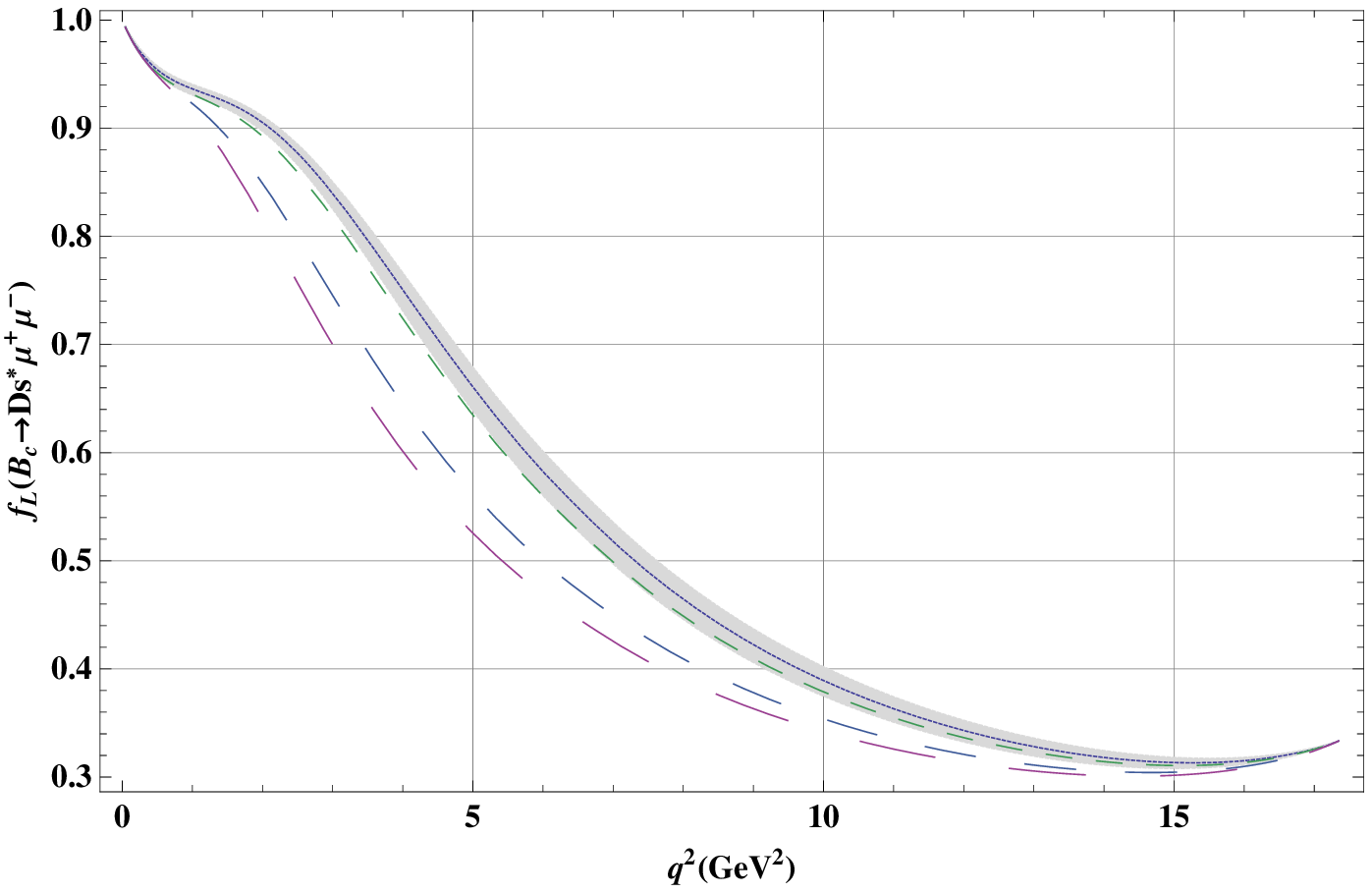} \includegraphics[scale=0.6]{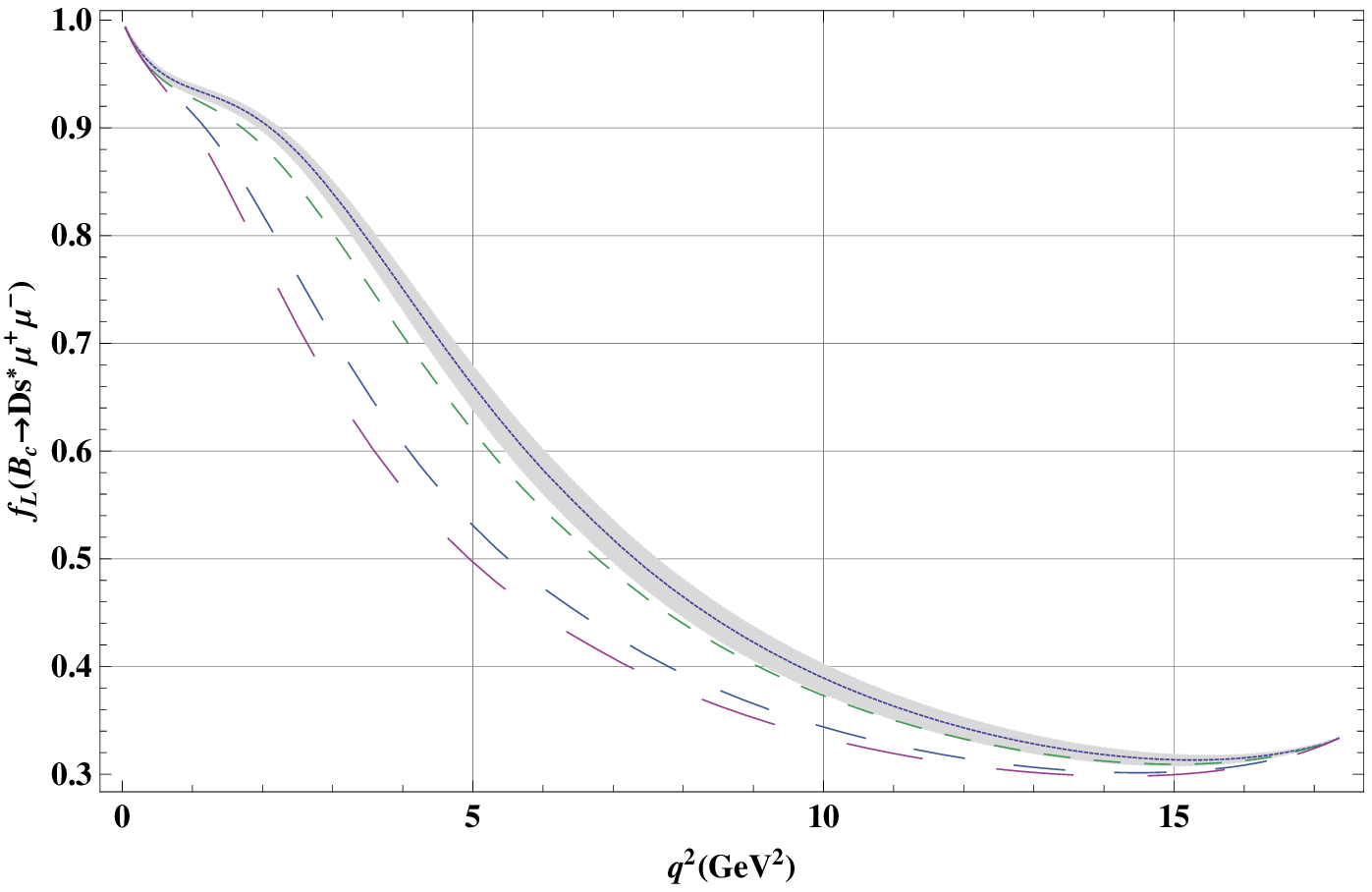}
\put (-350,240){(a)} \put (-100,240){(b)} \put (-350,0.2){(c)}
\put(-100,0.2){(d)} \vspace{-0.5cm} &  &
\end{tabular}%
\end{center}
\caption{The dependence of longitudinal helicity fractions of $B_{c}\rightarrow D_{s}^{\ast
}\protect\mu ^{+}\protect\mu ^{-}$ on $q^{2}$ for different values of $%
m_{t^{\prime }}$ and $\left\vert V_{t^{\prime }b}^{\ast }V_{t^{\prime
}s}\right\vert $. The values of fourth generation parameters and the legends
are same as in Fig.1.}
\label{LHF for muons}
\end{figure}
\begin{figure}[tbp]
\begin{center}
\begin{tabular}{ccc}
\vspace{-0.3cm} \includegraphics[scale=0.6]{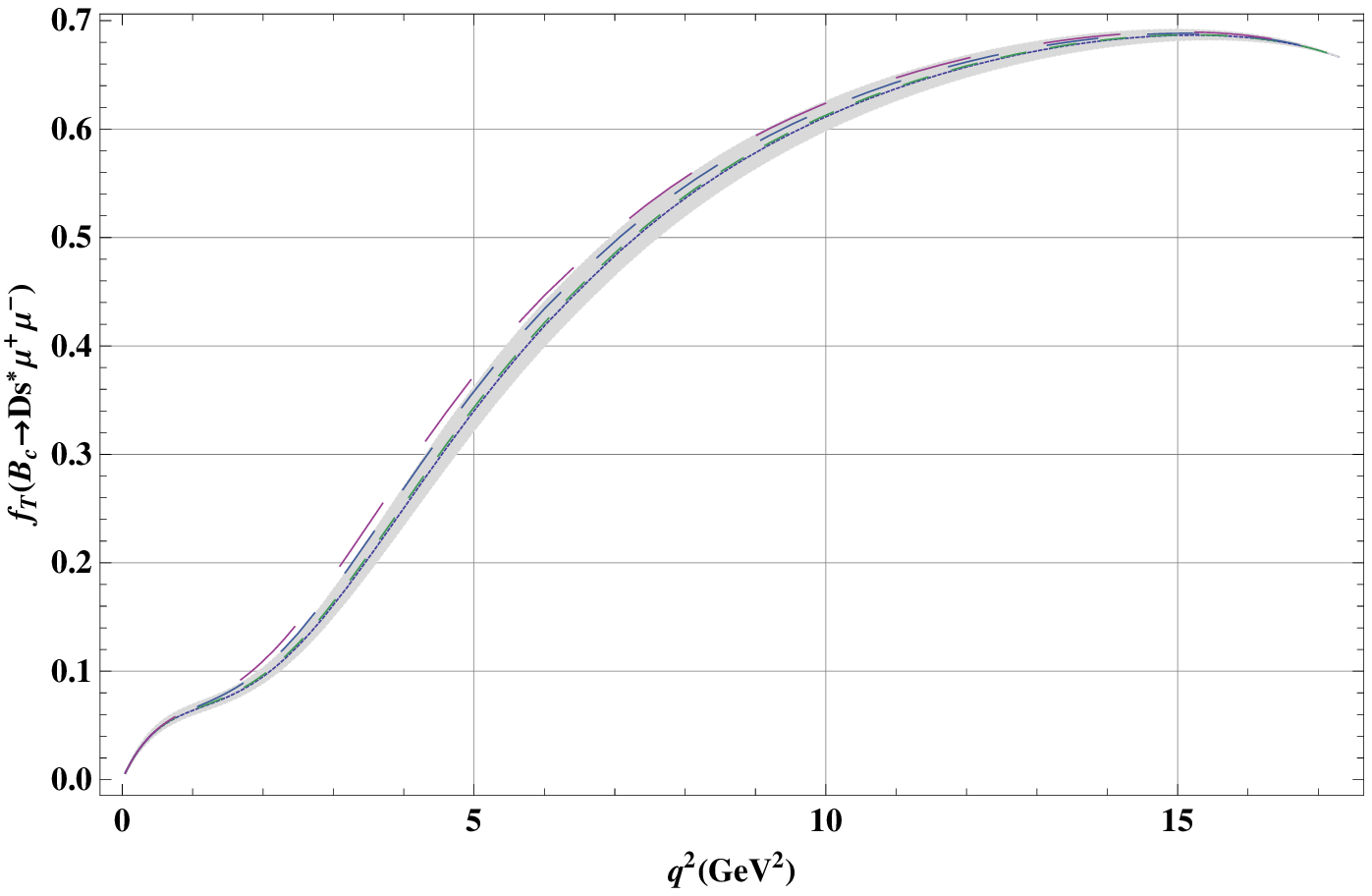} %
\includegraphics[scale=0.6]{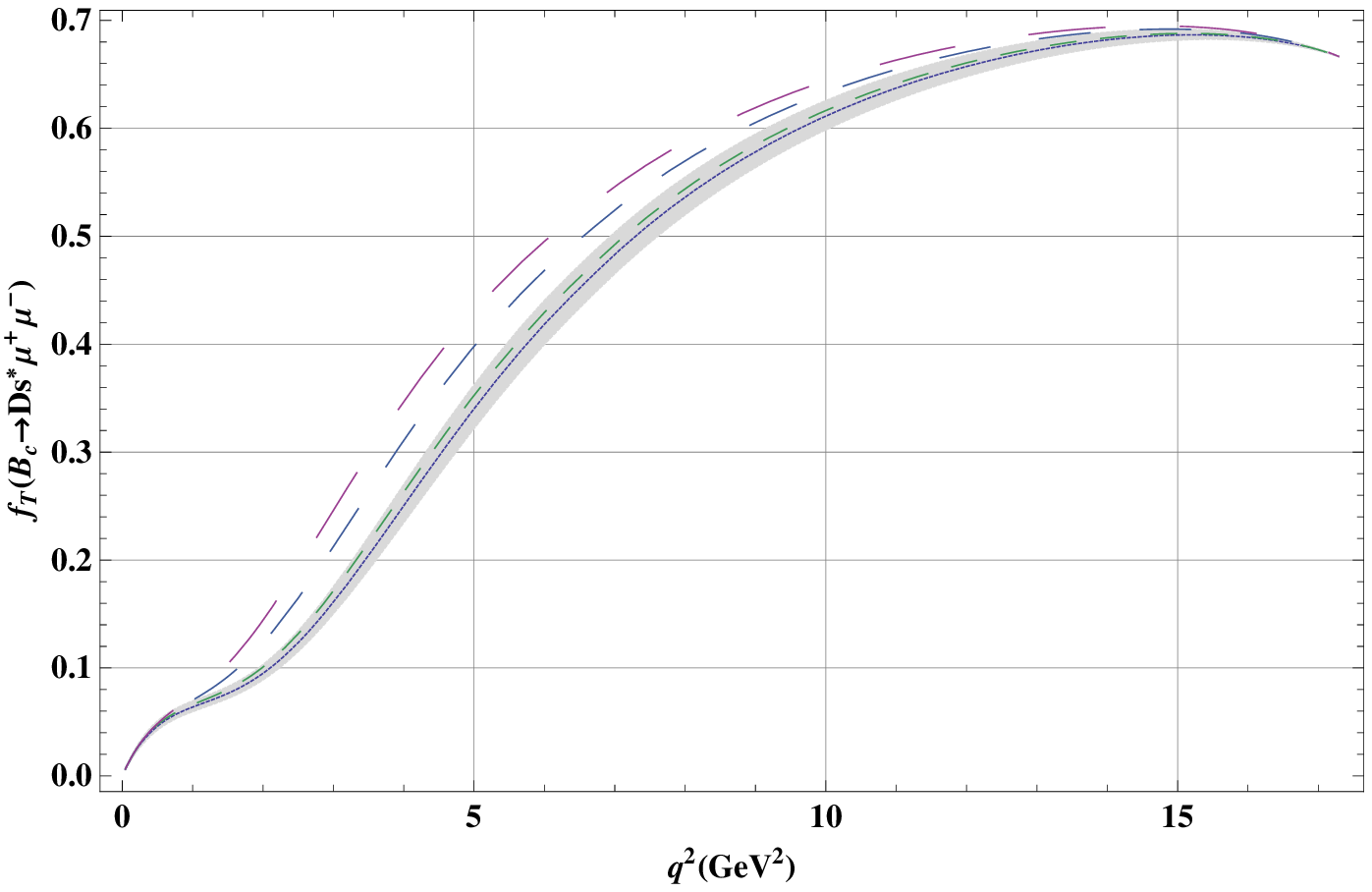} &  &  \\
\includegraphics[scale=0.6]{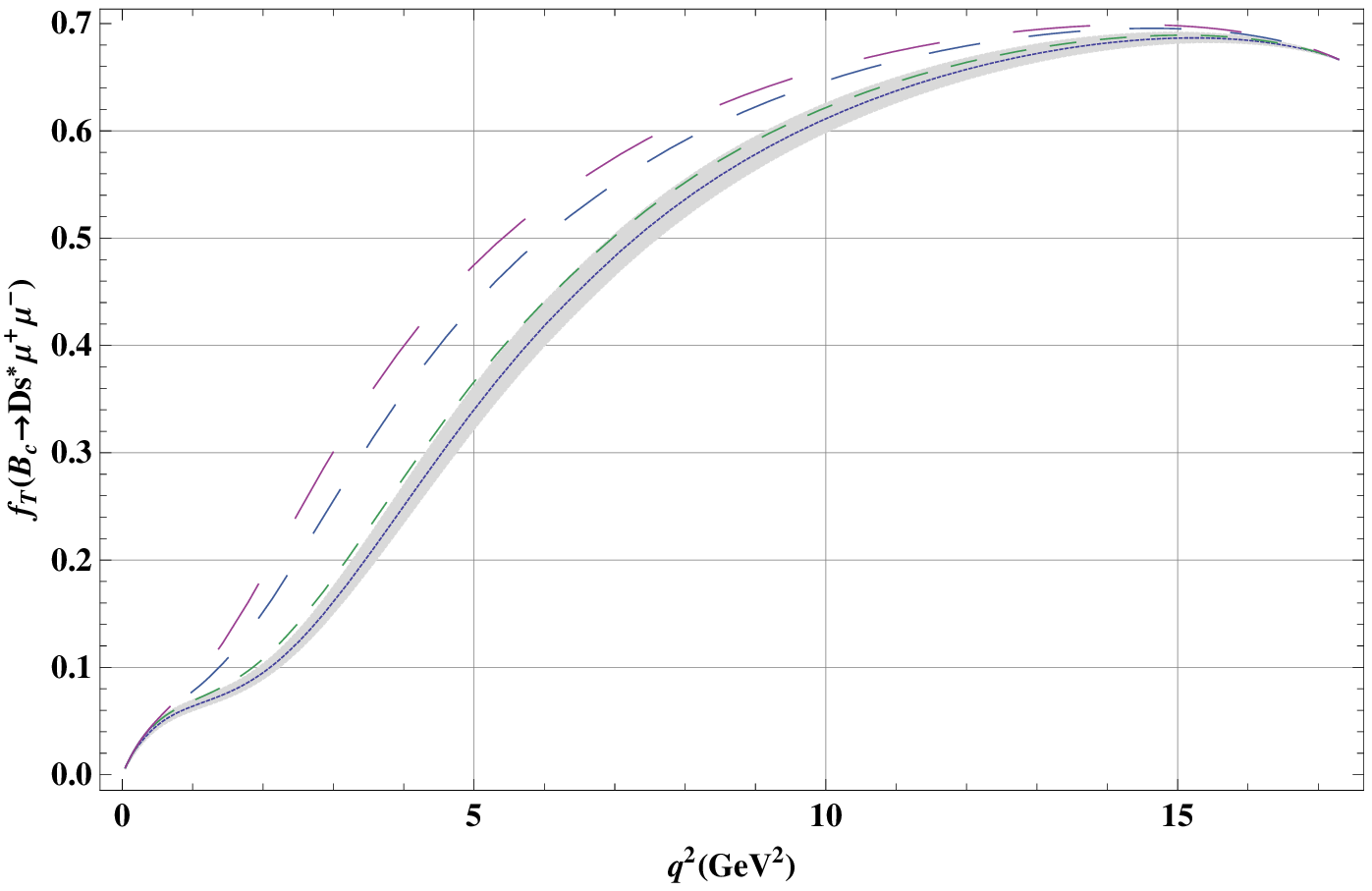} \includegraphics[scale=0.6]{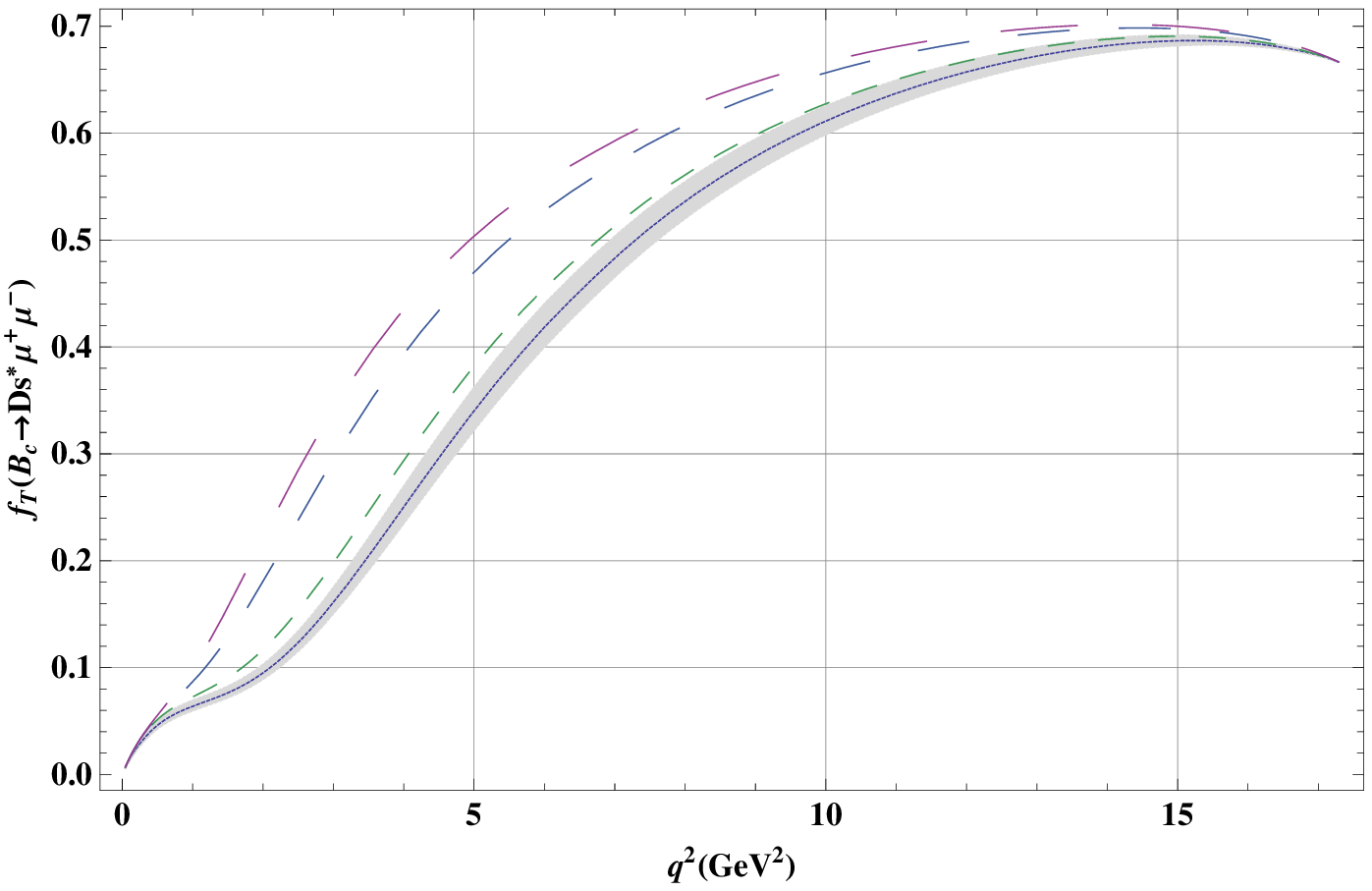}
\put (-350,240){(a)} \put (-100,240){(b)} \put (-350,0.2){(c)}
\put(-100,0.2){(d)} \vspace{-0.5cm} &  &
\end{tabular}%
\end{center}
\caption{The dependence of transverse helicity fractions of $B_{c}\rightarrow D_{s}^{\ast
}\protect\mu ^{+}\protect\mu ^{-}$ on $q^{2}$ for different values of $%
m_{t^{\prime }}$ and $\left\vert V_{t^{\prime }b}^{\ast }V_{t^{\prime
}s}\right\vert $. The values of fourth generation parameters and the legends
are same as in Fig.1.}
\label{THF for muons}
\end{figure}

In Figs. \ref{LHF for muons} and \ref{THF for muons} we have shown the dependence of longitudinal and transverse helicity
fractions of $D_{s}^{\ast}$ meson in $B_{c}\rightarrow D_{s}^{\ast} \mu^{+} \mu ^{-}$ decay on $q^2$.
We can establish from Fig. \ref{LHF for muons} (\ref{THF for muons}) that the effects of the
fourth generation on the longitudinal (transverse) helicity
fractions of $D_{s}^{\ast}$ are marked up in the
$2.5<q^{2}\leq13.5$GeV$^{2}$ region. Here one can see that the shift in
the value of longitudinal and transverse helicity fractions from that of the SM value
is small for the lower values of the fourth generation parameters $(m_{t^{\prime}}$, $|V_{t^{\prime}b}V_{t^{\prime}s}|)$
which however becomes prominent when the values of these input parameters becomes maximum.
The value of the longitudinal (transverse)
helicity fraction of $D_{s}^{\ast}$ meson decreases (increase) with the
increment in the values of fourth generation input parameters
$m_{t^{\prime}}$ and $|V_{t^{\prime}b}V_{t^{\prime}s}|$.
\textit{In contrast to analysis of same observable made in the UED model \cite{51} where UED effects were mitigated
by the WA contributions, we can
see that in the present scenario the NP effects are very promising. Therefore, we expect that the experimental study of helicity
fraction of $D_{s}^{\ast}$ meson will serve as a handy tool to distinguish
different NP scenarios.}

\begin{figure}[tbp]
\begin{center}
\begin{tabular}{ccc}
\vspace{-0.3cm} \includegraphics[scale=0.6]{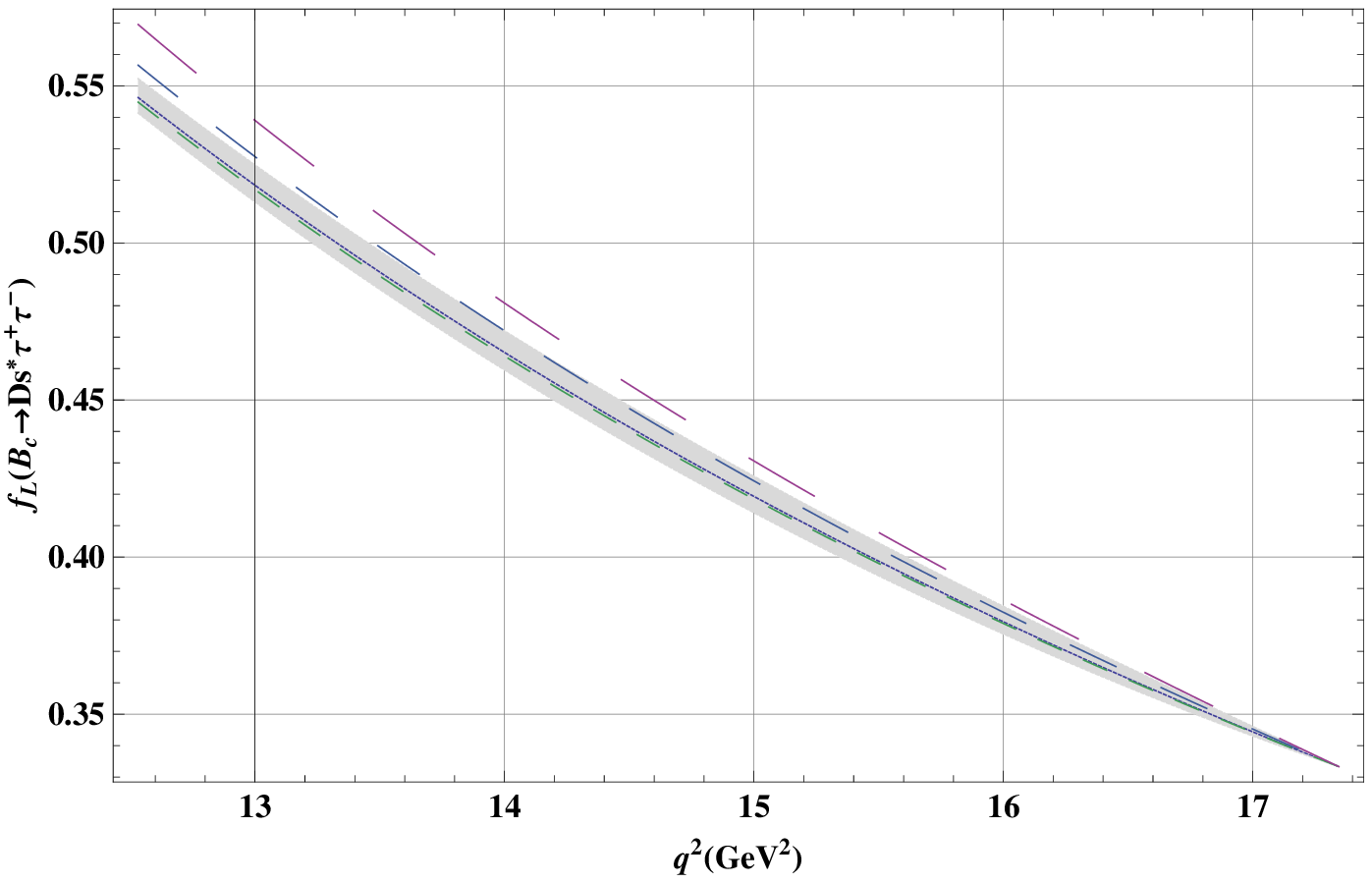} %
\includegraphics[scale=0.6]{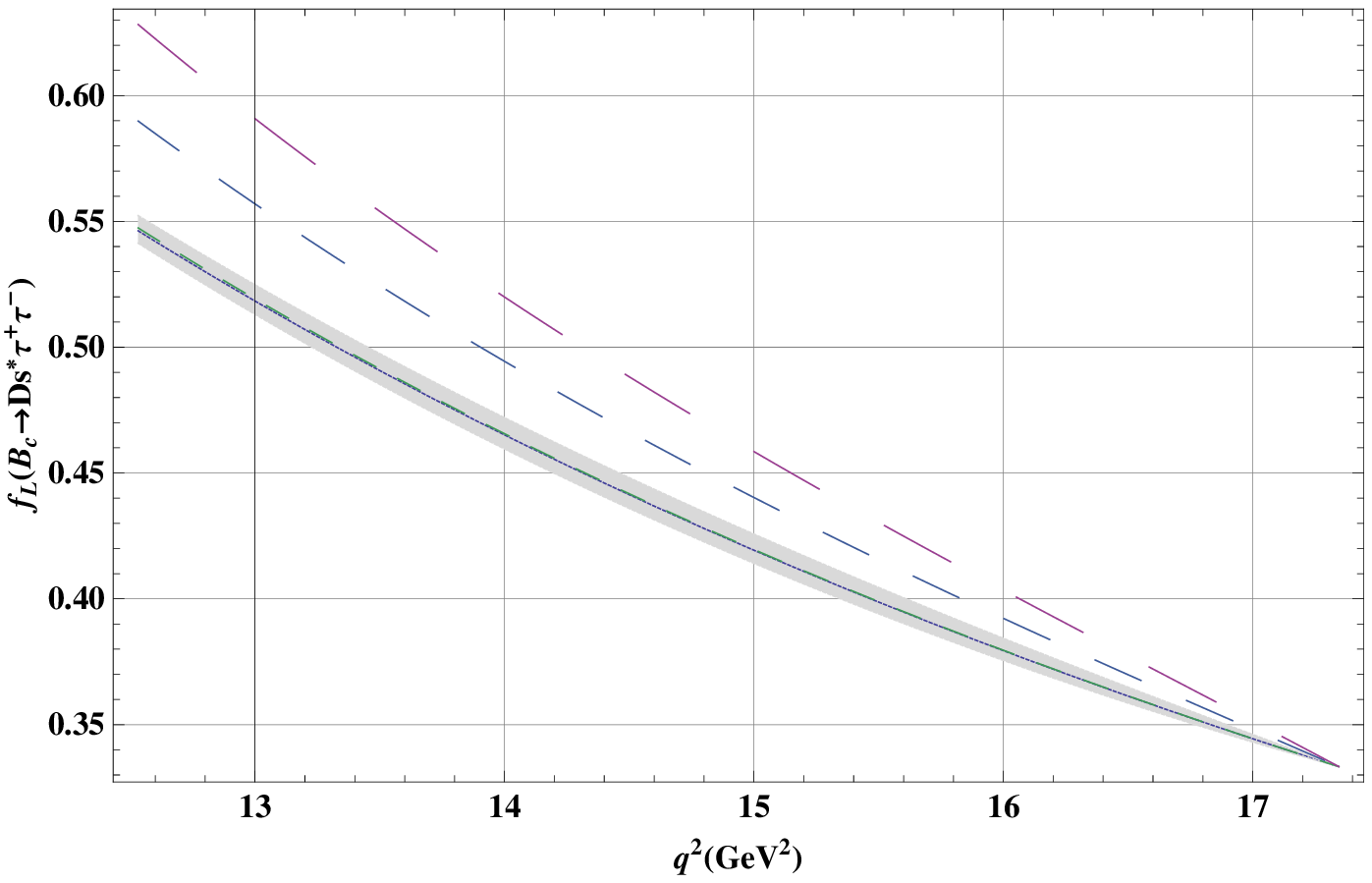} &  &  \\
\includegraphics[scale=0.6]{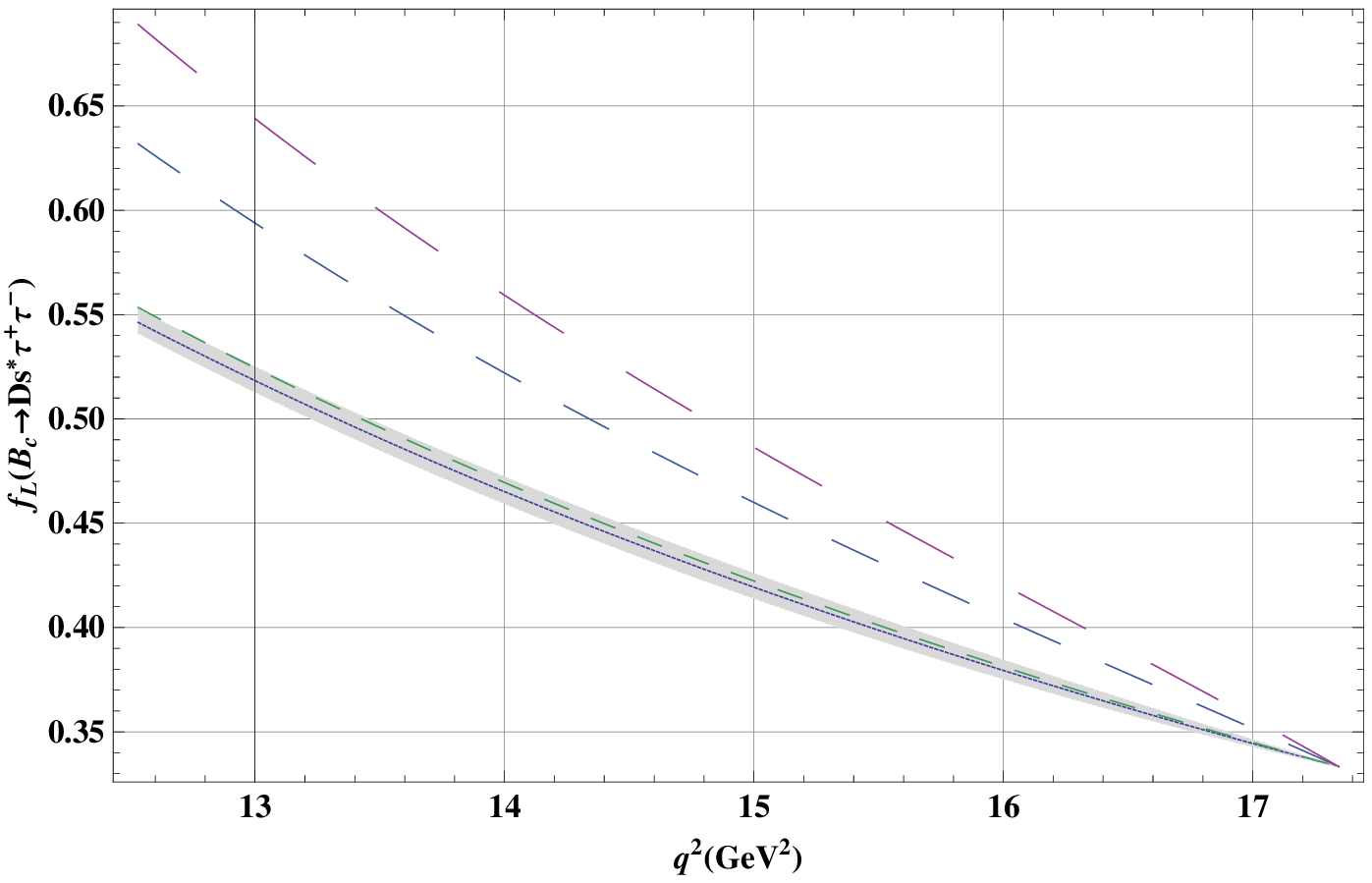} \includegraphics[scale=0.6]{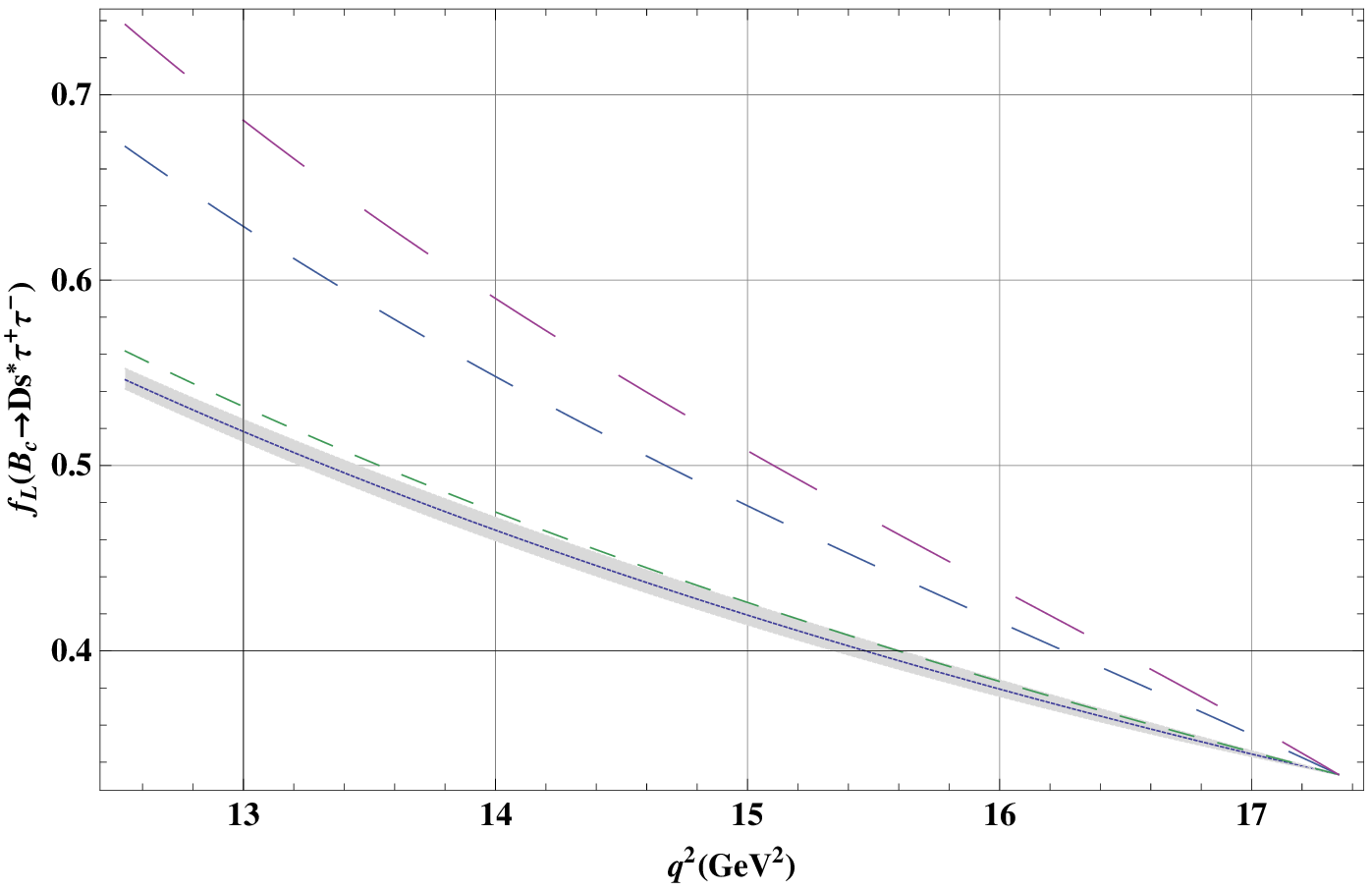}
\put (-350,240){(a)} \put (-100,240){(b)} \put (-350,0.2){(c)}
\put(-100,0.2){(d)} \vspace{-0.5cm} &  &
\end{tabular}%
\end{center}
\caption{The dependence of longitudinal helicity fractions of $B_{c}\rightarrow D_{s}^{\ast
}\protect\tau ^{+}\protect\tau ^{-}$ on $q^{2}$ for different values of $%
m_{t^{\prime }}$ and $\left\vert V_{t^{\prime }b}^{\ast }V_{t^{\prime
}s}\right\vert $. The values of fourth generation parameters and the legends
are same as in Fig.1.}
\label{LHF for tau.}
\end{figure}

\begin{figure}[tbp]
\begin{center}
\begin{tabular}{ccc}
\vspace{-0.3cm} \includegraphics[scale=0.6]{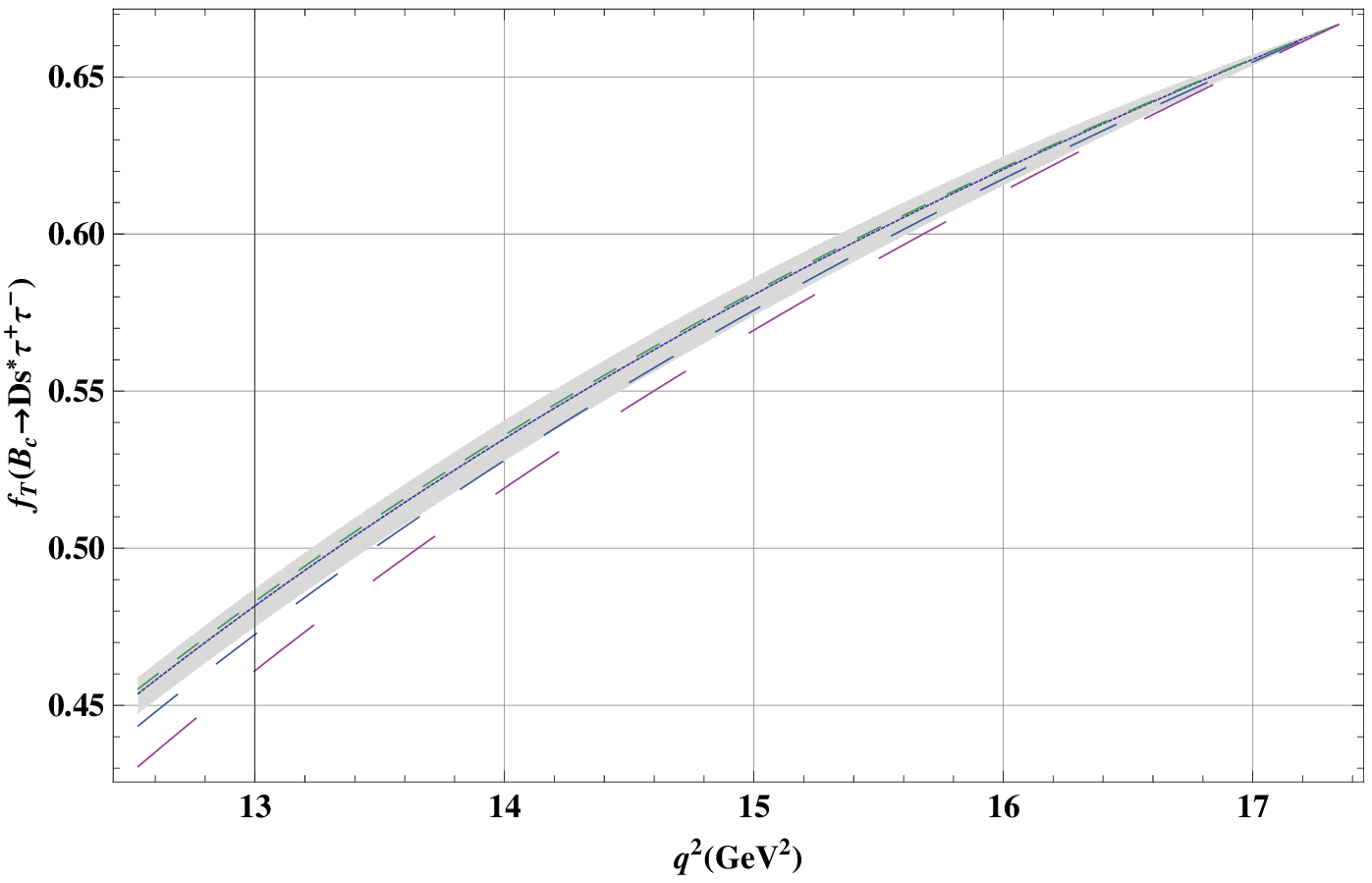} %
\includegraphics[scale=0.6]{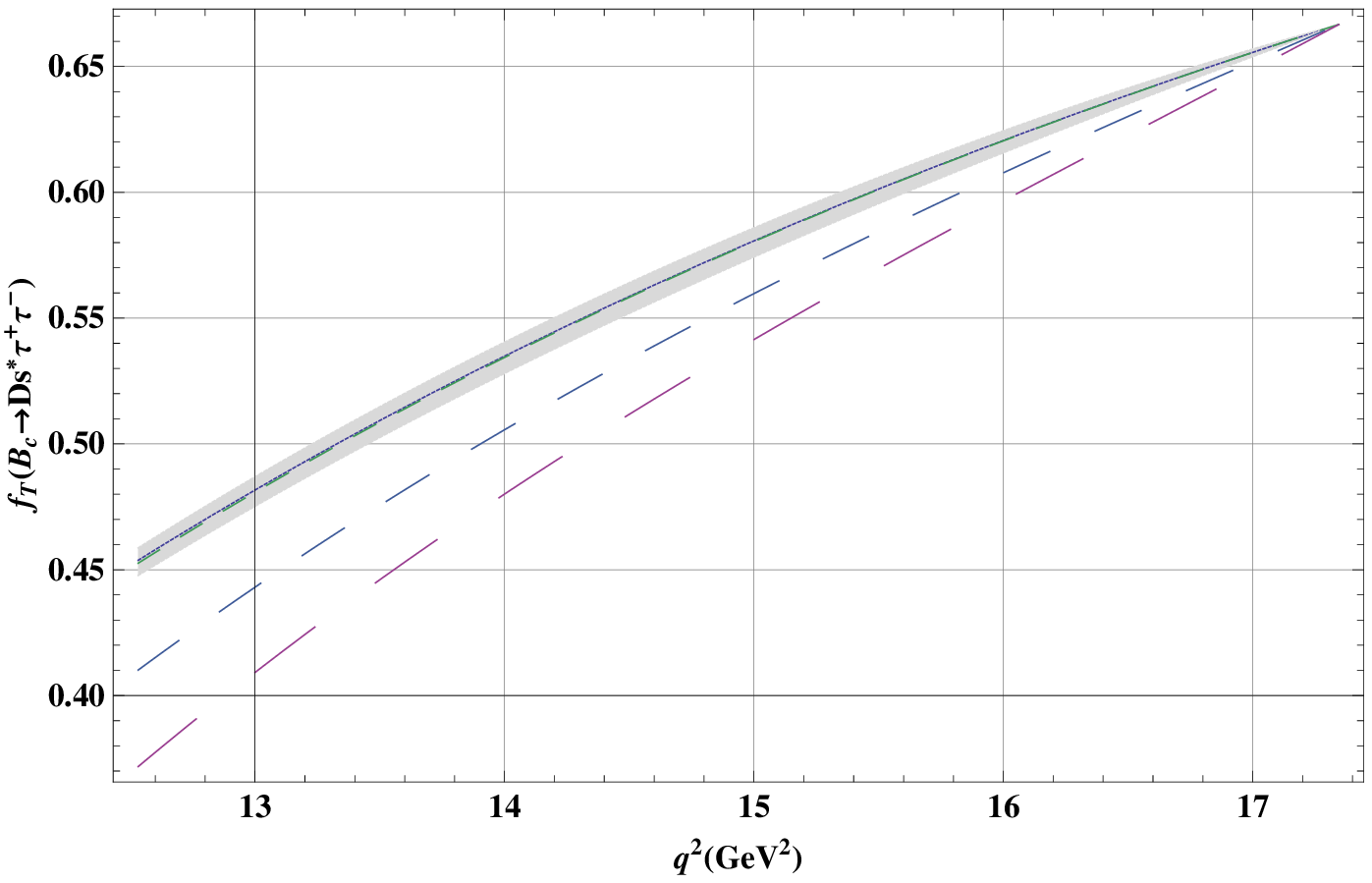} &  &  \\
\includegraphics[scale=0.6]{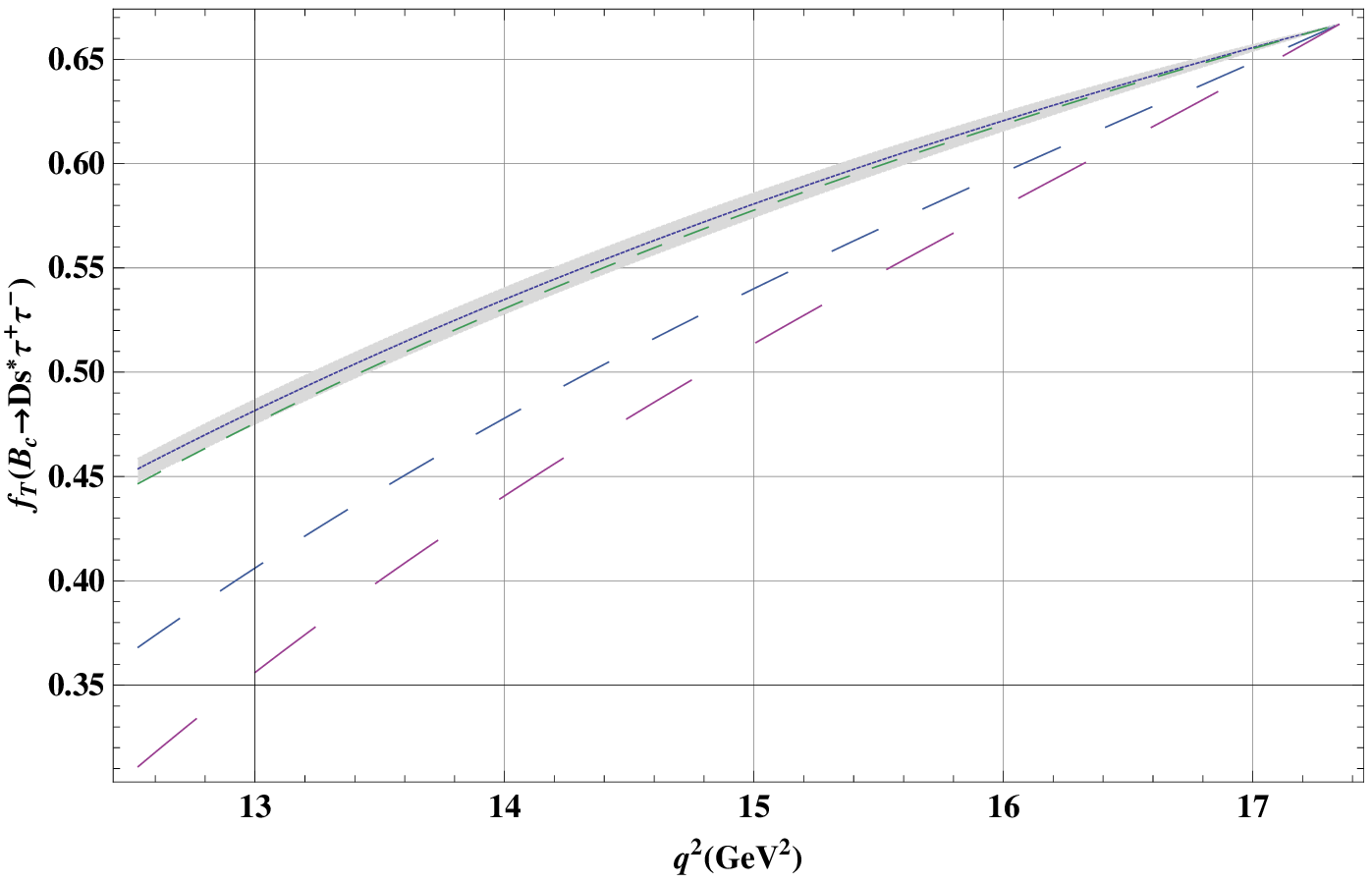} \includegraphics[scale=0.6]{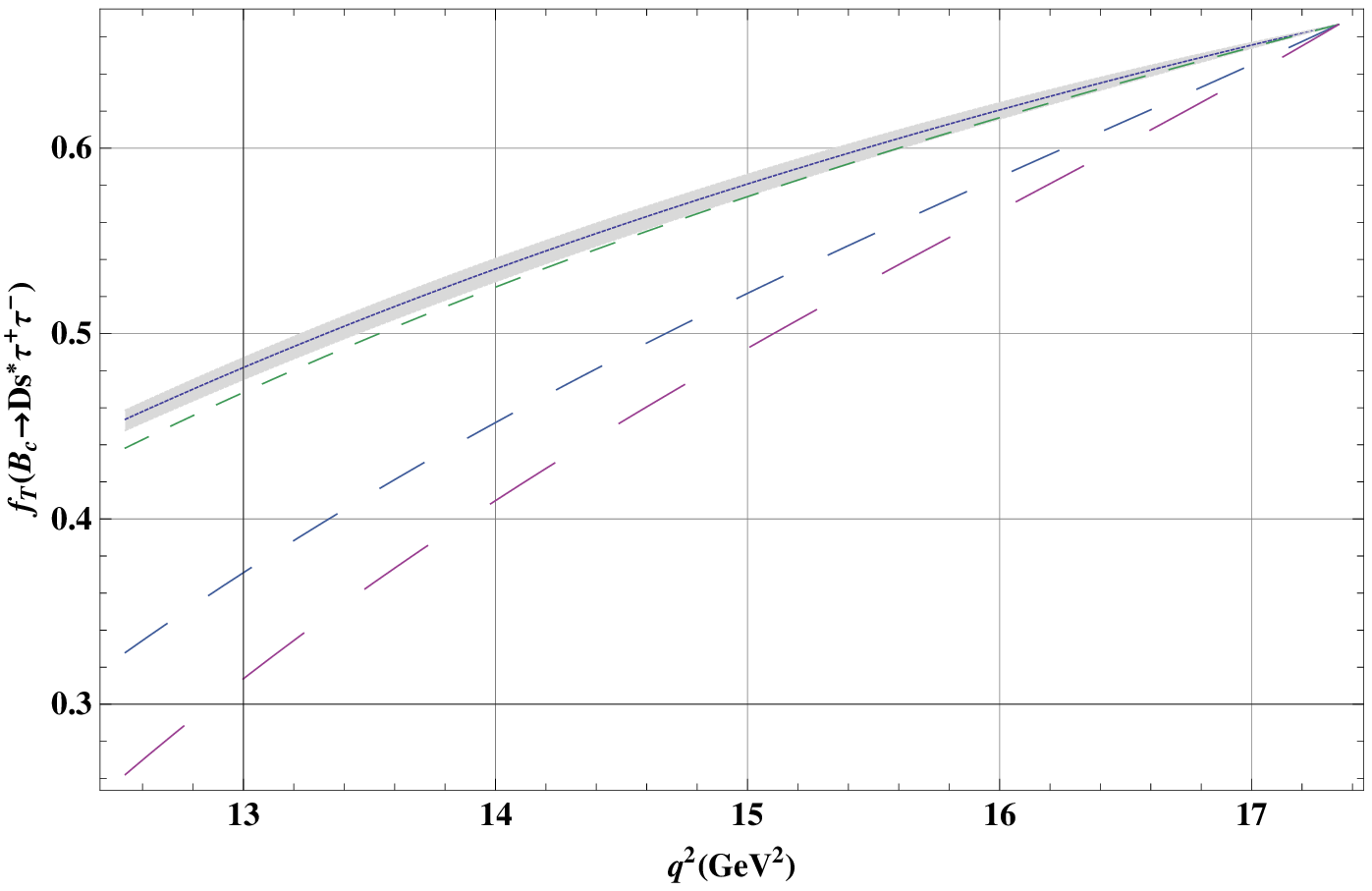}
\put (-350,240){(a)} \put (-100,240){(b)} \put (-350,0.2){(c)}
\put(-100,0.2){(d)} \vspace{-0.5cm} &  &
\end{tabular}%
\end{center}
\caption{The dependence of transverse helicity fractions of $B_{c}\rightarrow D_{s}^{\ast
}\protect\tau ^{+}\protect\tau ^{-}$ on $q^{2}$ for different values of $%
m_{t^{\prime }}$ and $\left\vert V_{t^{\prime }b}^{\ast }V_{t^{\prime
}s}\right\vert $. The values of fourth generation parameters and the legends
are same as in Fig.1. }
\label{THF for tau.}
\end{figure}

Performing the similar study when tau's are the final state leptons
we have shown the dependence of longitudinal and transverse helicity fractions of the $D_{s}^{\ast}$ meson
in Figs. \ref{LHF for tau.} and \ref{THF for tau.} against the square of the momentum transfer $(q^2)$.
Compared to the muon in the final state, the SM4 effects in this case are somewhat dim but still visible. Therefore, we expect that the experimental study of
the helicity fraction will shed some light on the NP searches especially in $B_{c}\rightarrow D_{s}^{\ast} \mu^{+} \mu ^{-}$ decays.

\begin{figure}[tbp]
\begin{center}
\begin{tabular}{ccc}
\vspace{-0.3cm} \includegraphics[scale=0.6]{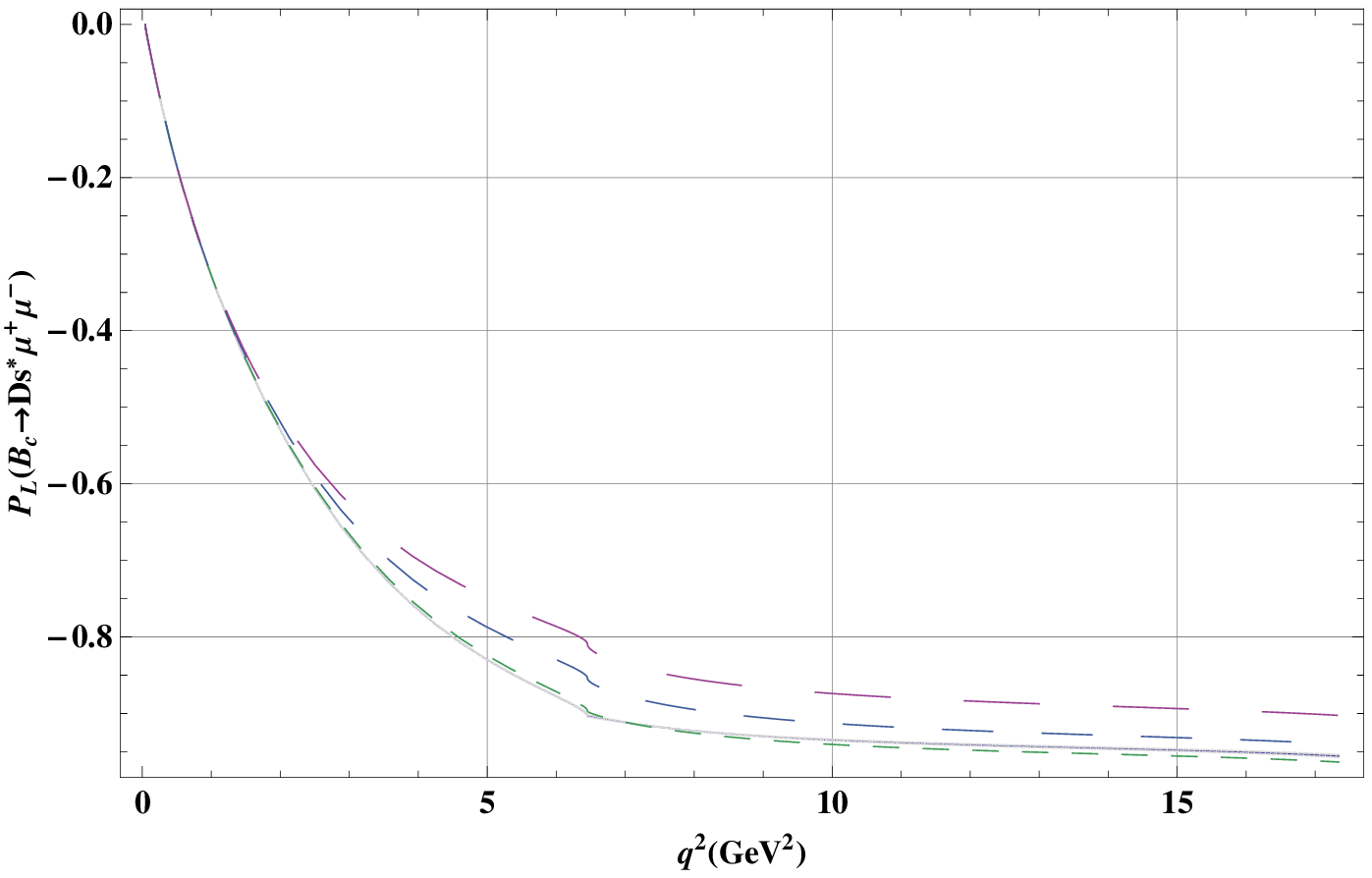} %
\includegraphics[scale=0.6]{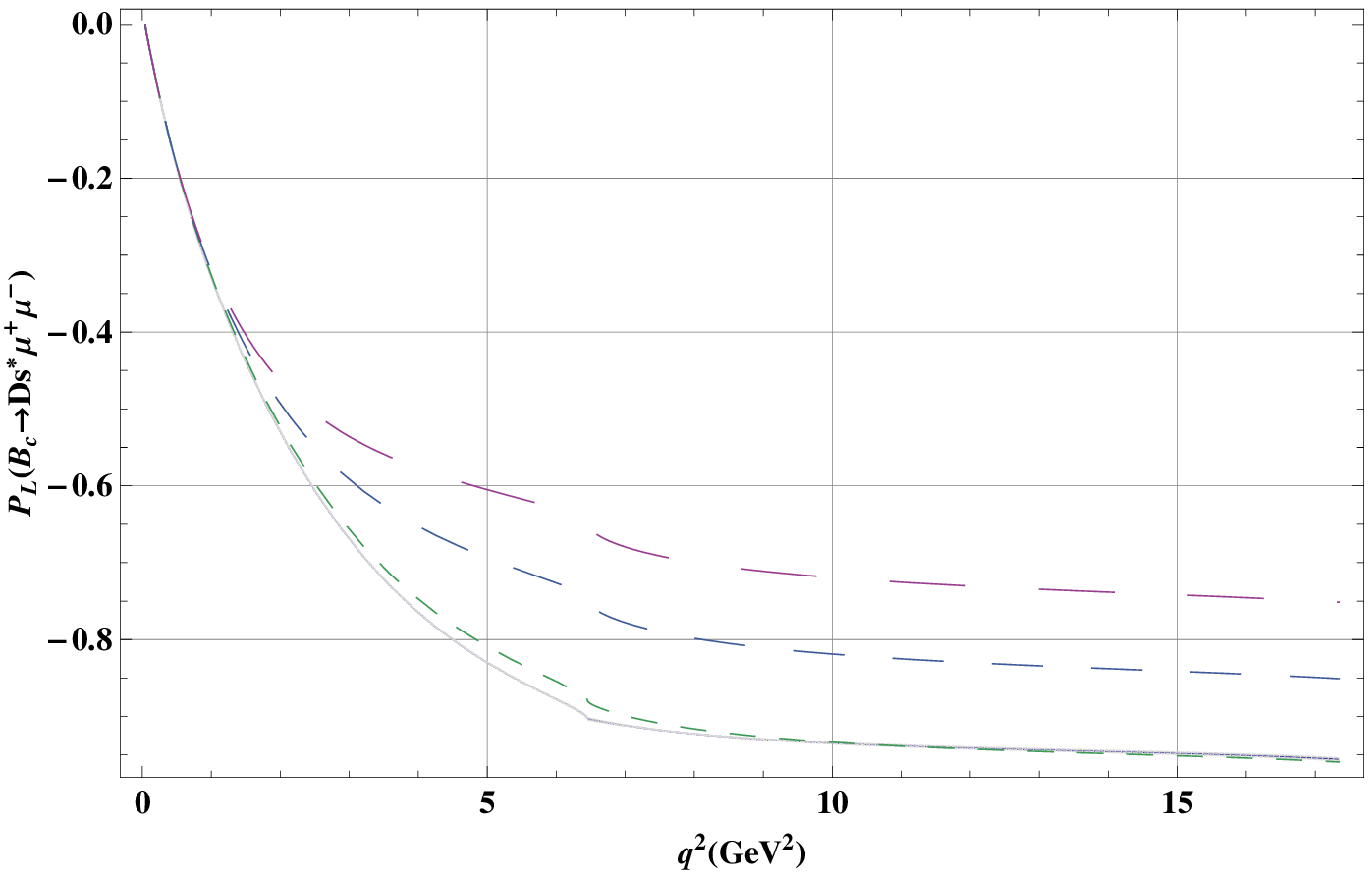} &  &  \\
\includegraphics[scale=0.6]{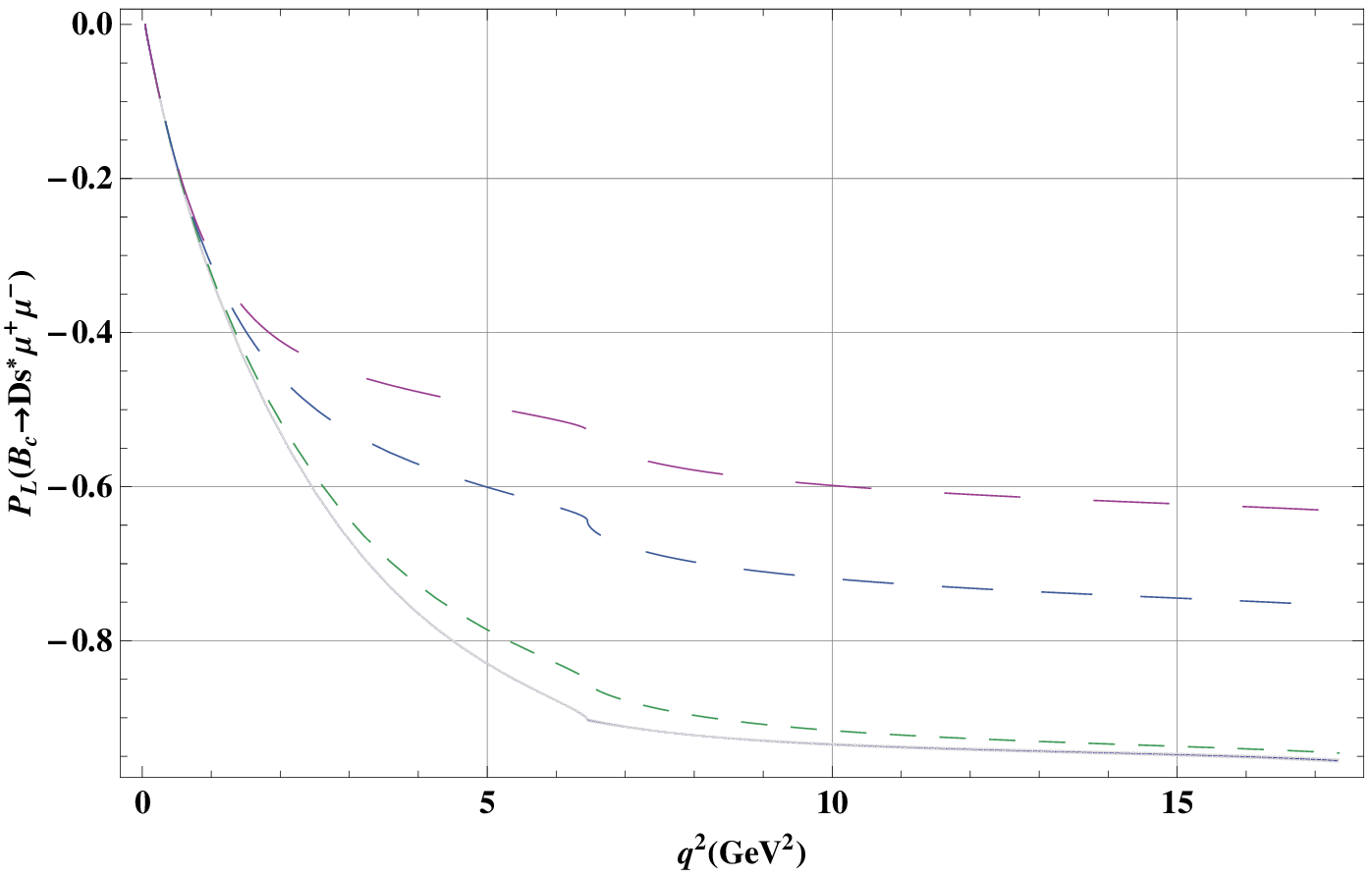} \includegraphics[scale=0.6]{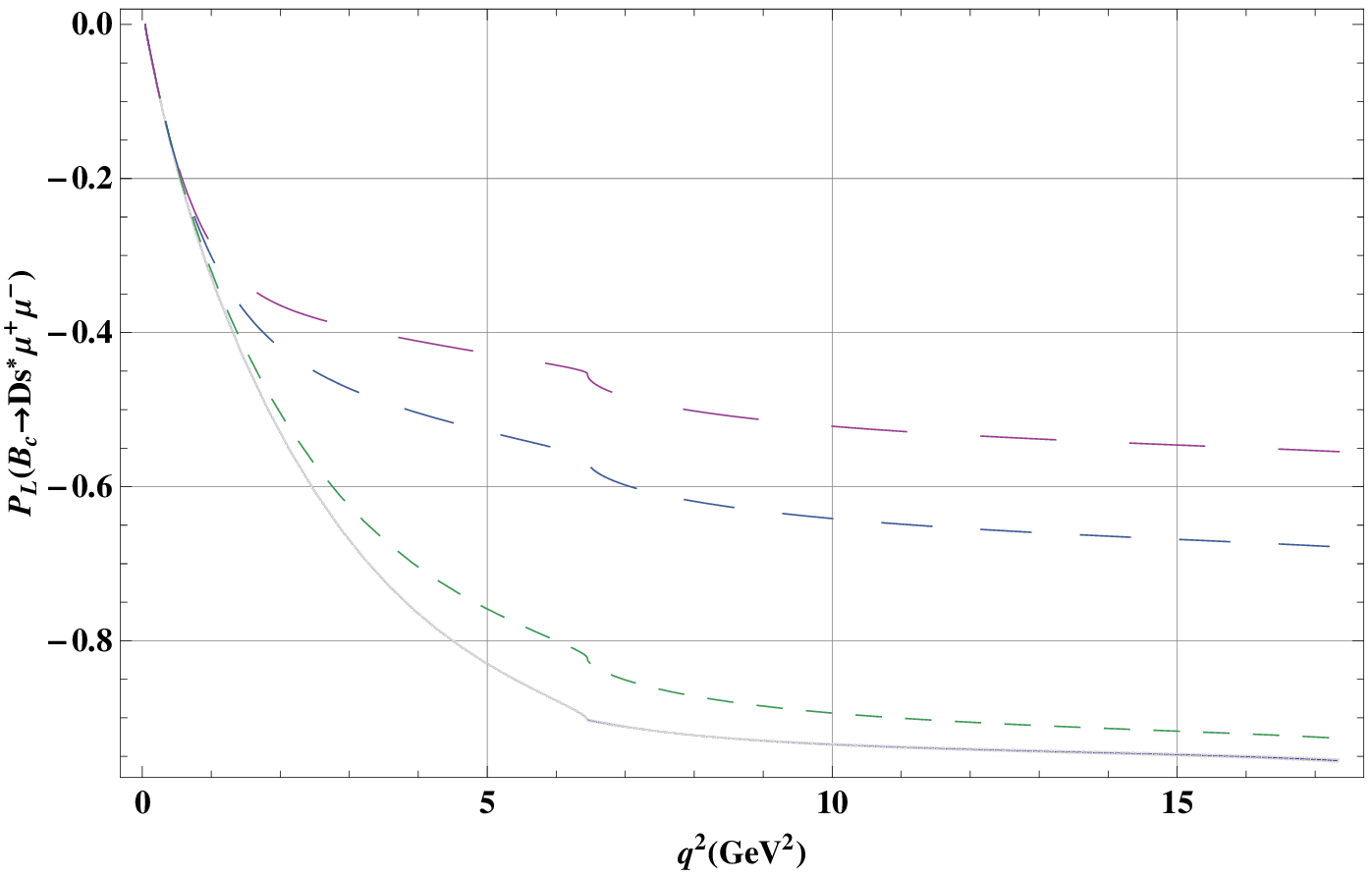}
\put (-350,240){(a)} \put (-100,240){(b)} \put (-350,0.2){(c)}
\put(-100,0.2){(d)} \vspace{-0.5cm} &  &
\end{tabular}%
\end{center}
\caption{The dependence of longitudinal lepton polarization of $B_{c}\rightarrow D_{s}^{\ast
}\protect\mu ^{+}\protect\mu ^{-}$ on $q^{2}$ for different values of $%
m_{t^{\prime }}$ and $\left\vert V_{t^{\prime }b}^{\ast }V_{t^{\prime
}s}\right\vert $. The values of fourth generation parameters and the legends
are same as in Fig.1.}
\label{LP-muon}
\end{figure}

\begin{figure}[tbp]
\begin{center}
\begin{tabular}{ccc}
\vspace{-0.3cm} \includegraphics[scale=0.6]{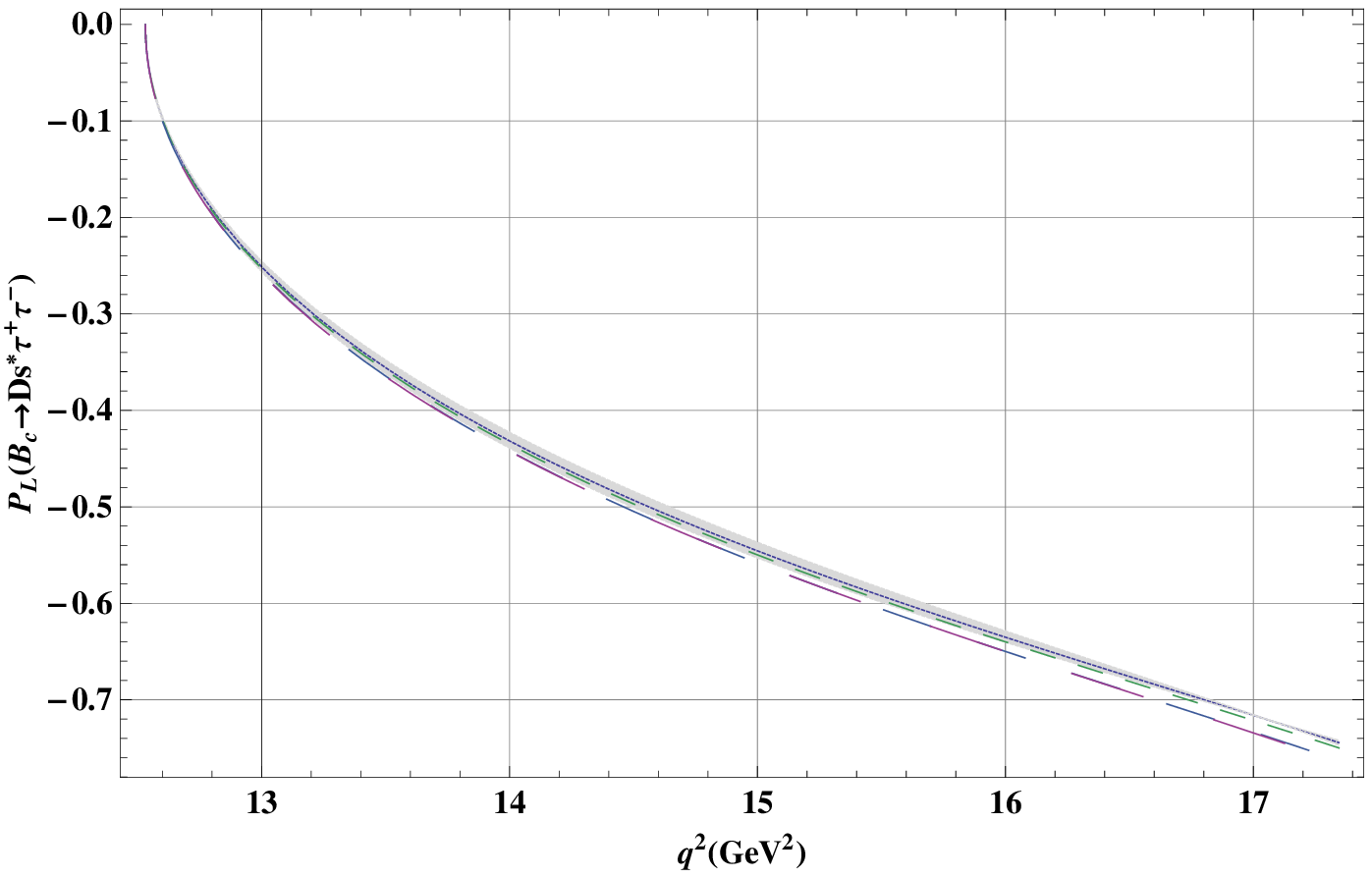} %
\includegraphics[scale=0.6]{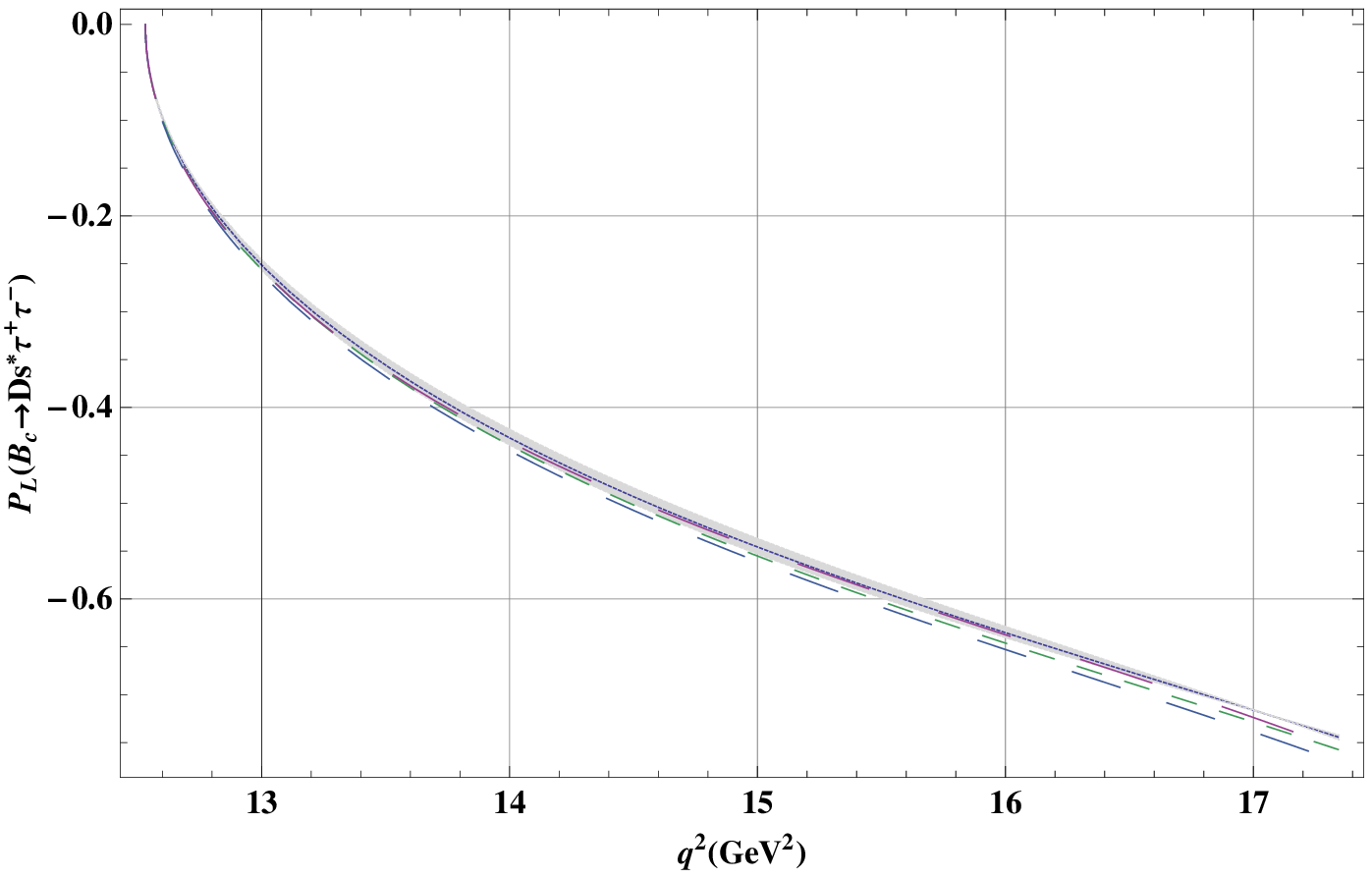} &  &  \\
\includegraphics[scale=0.6]{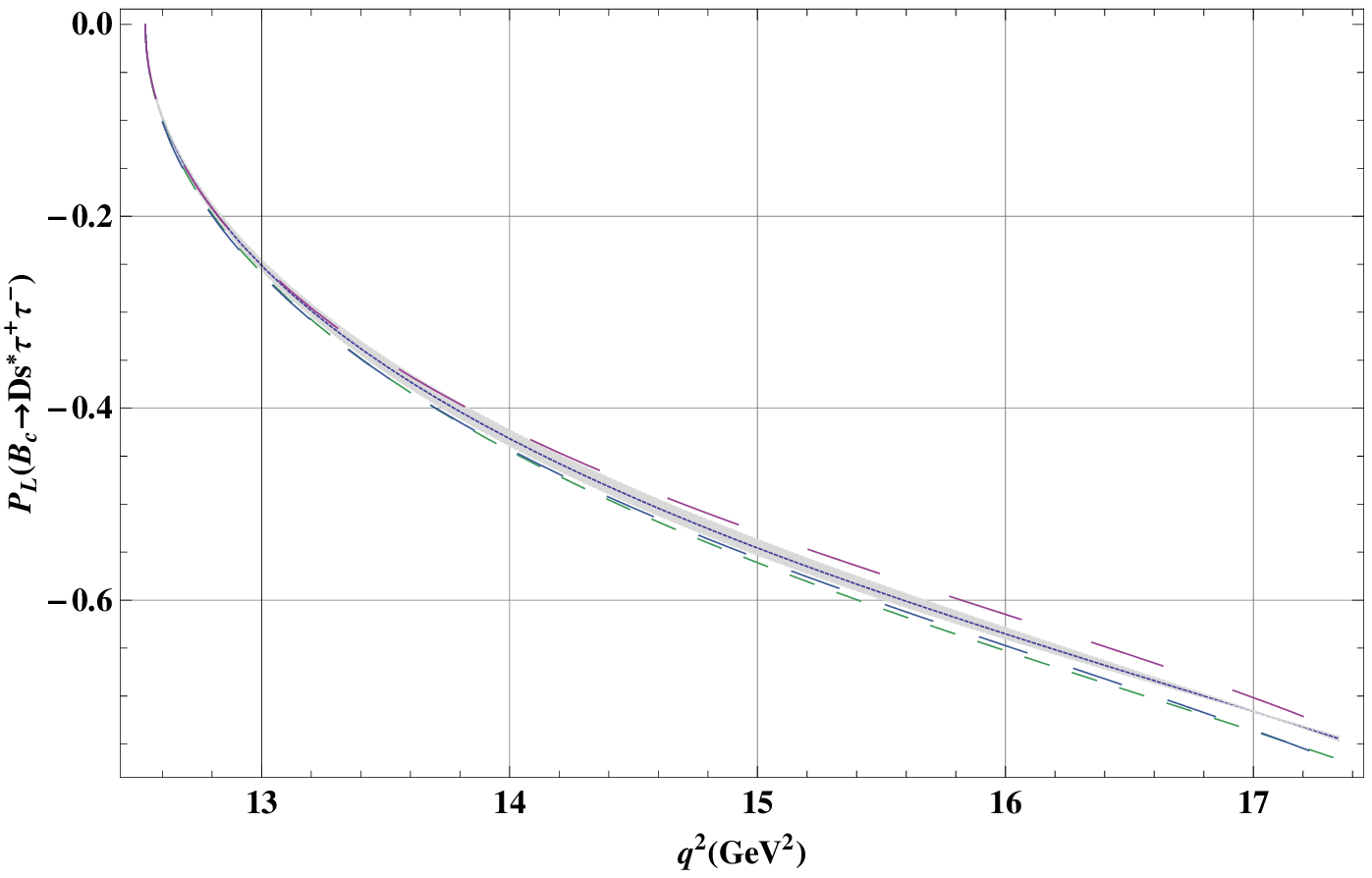} \includegraphics[scale=0.6]{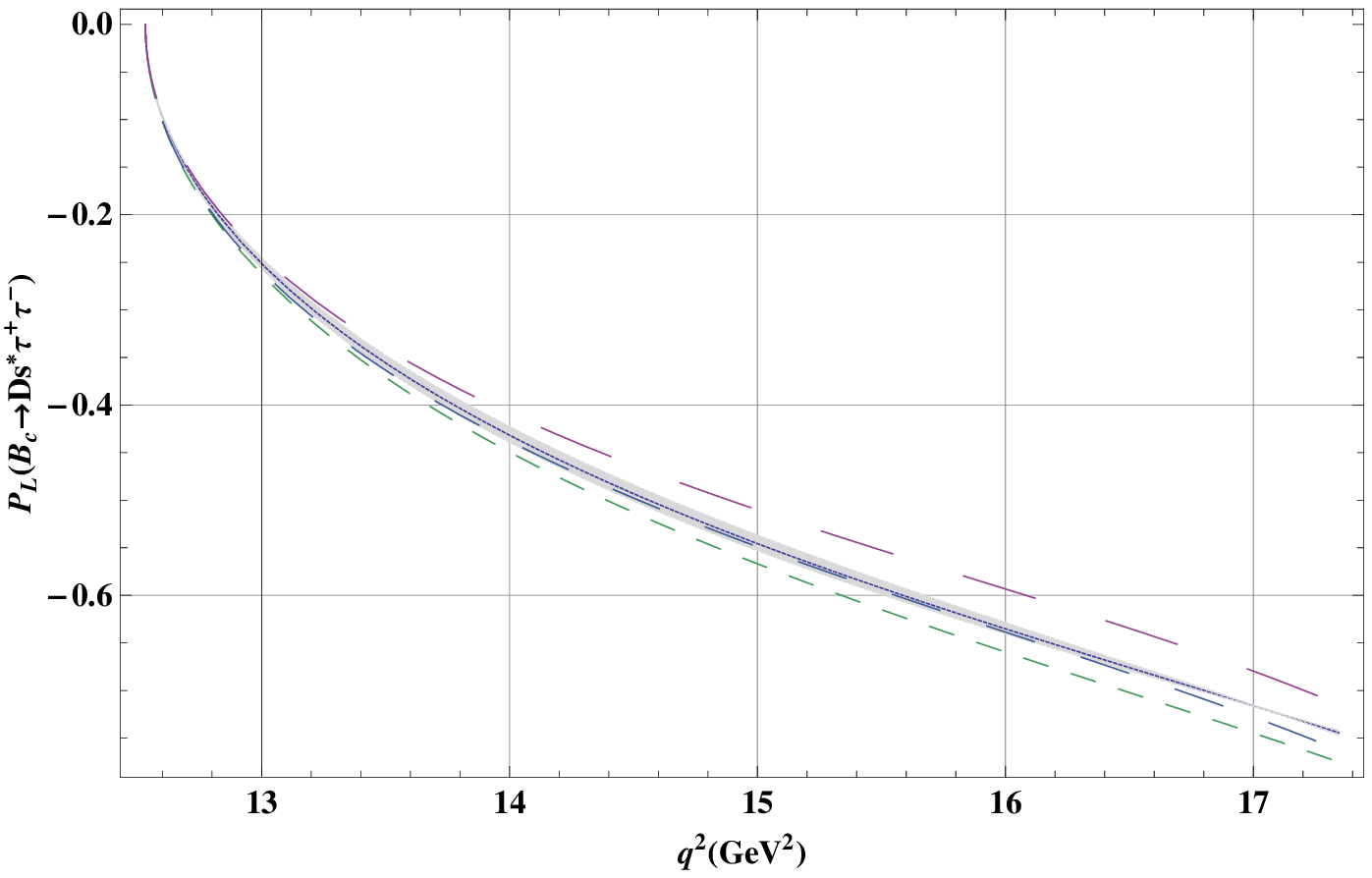}
\put (-350,240){(a)} \put (-100,240){(b)} \put (-350,0.2){(c)}
\put(-100,0.2){(d)} \vspace{-0.5cm} &  &
\end{tabular}%
\end{center}
\caption{The dependence of longitudinal lepton polarization of $B_{c}\rightarrow D_{s}^{\ast
}\protect\tau ^{+}\protect\tau ^{-}$ on $q^{2}$ for different values of $%
m_{t^{\prime }}$ and $\left\vert V_{t^{\prime }b}^{\ast }V_{t^{\prime
}s}\right\vert $. The values of fourth generation parameters and the legends
are same as in Fig.1. }
\label{LP tauon}
\end{figure}

Fig. \ref{LP-muon} shows the dependence of longitudinal lepton polarization
asymmetry for the $B_c\rightarrow D_{s}^{\ast }\mu ^{+}\mu ^{-}$ decay
on $q^2$ for different values of $m_{t^{\prime }}$
and $\left\vert V_{t^{\prime }b}^{\ast }V_{t^{\prime }s}\right\vert $. From Eq. (\ref{long-polarization})
we can see that the WA contributions are canceled out and hence the effects of SM4 will
be distinctively clear in longitudinal lepton polarization asymmetry which is depicted in Fig.
\ref{LP-muon}. The
value of longitudinal lepton polarization for muons is around $-0.9$ in the SM3
and we have significant deviation in this value in SM4. Just in the case of $%
m_{t^{\prime }}=600$ GeV and $\left\vert V_{t^{\prime }b}^{\ast
}V_{t^{\prime }s}\right\vert =1.2\times 10^{-2}$ the value of the
longitudinal lepton polarization becomes $-0.5$ which will help us to see
experimentally the SM4 effects in these decays. Similar effects can
be seen for the final state tauons (c.f. Fig. \ref{LP tauon}). In this case the
shift from the SM value is small compared to the muons in the final state because of the
factor $\left( 1-\frac{%
4m_{l}^{2}}{q^{2}}\right)$ appear in the calculation of the longitudinal lepton polarization asymmetry.

\begin{figure}[tbp]
\begin{center}
\begin{tabular}{ccc}
\vspace{-0.3cm} \includegraphics[scale=0.6]{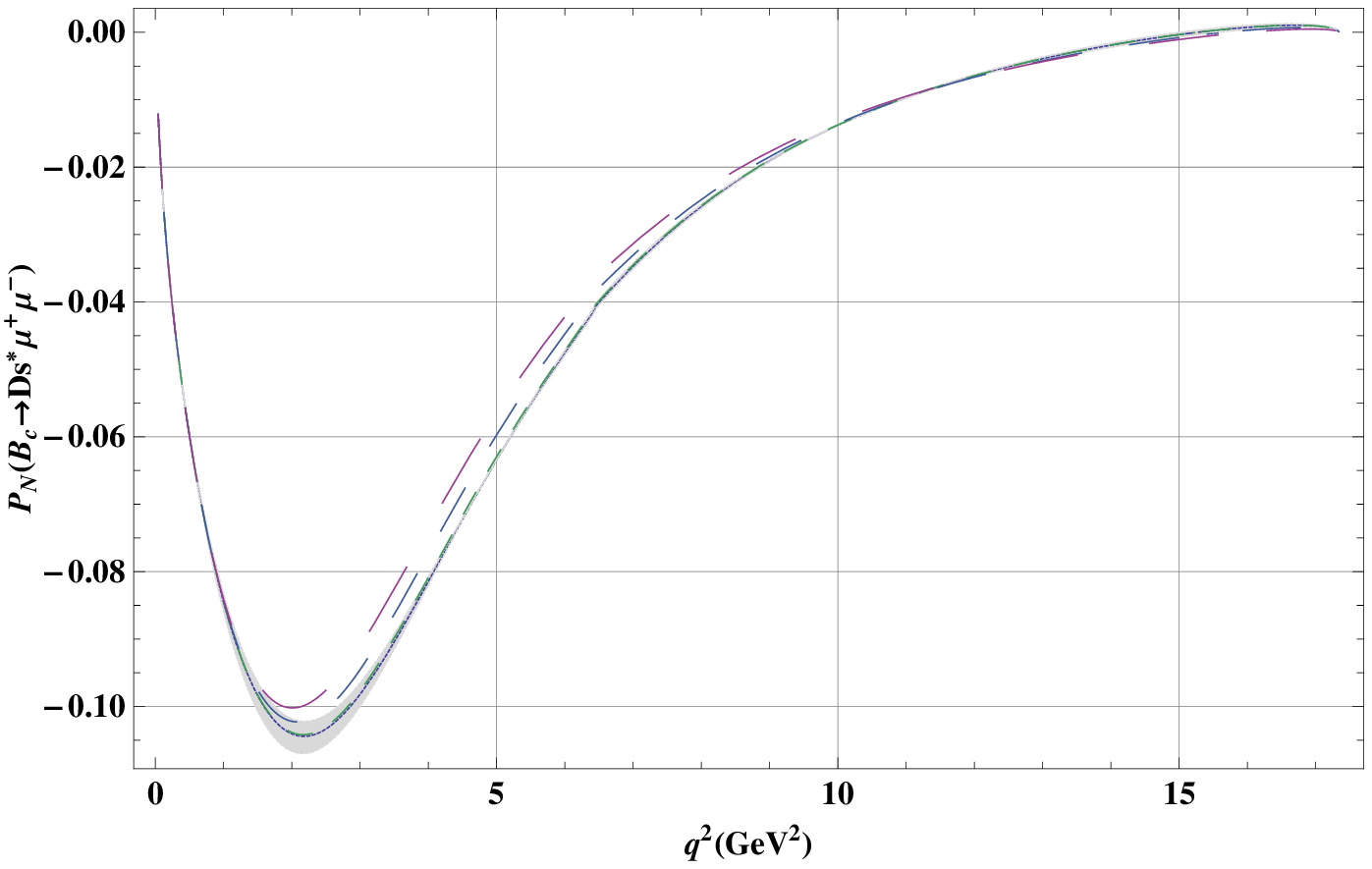} %
\includegraphics[scale=0.6]{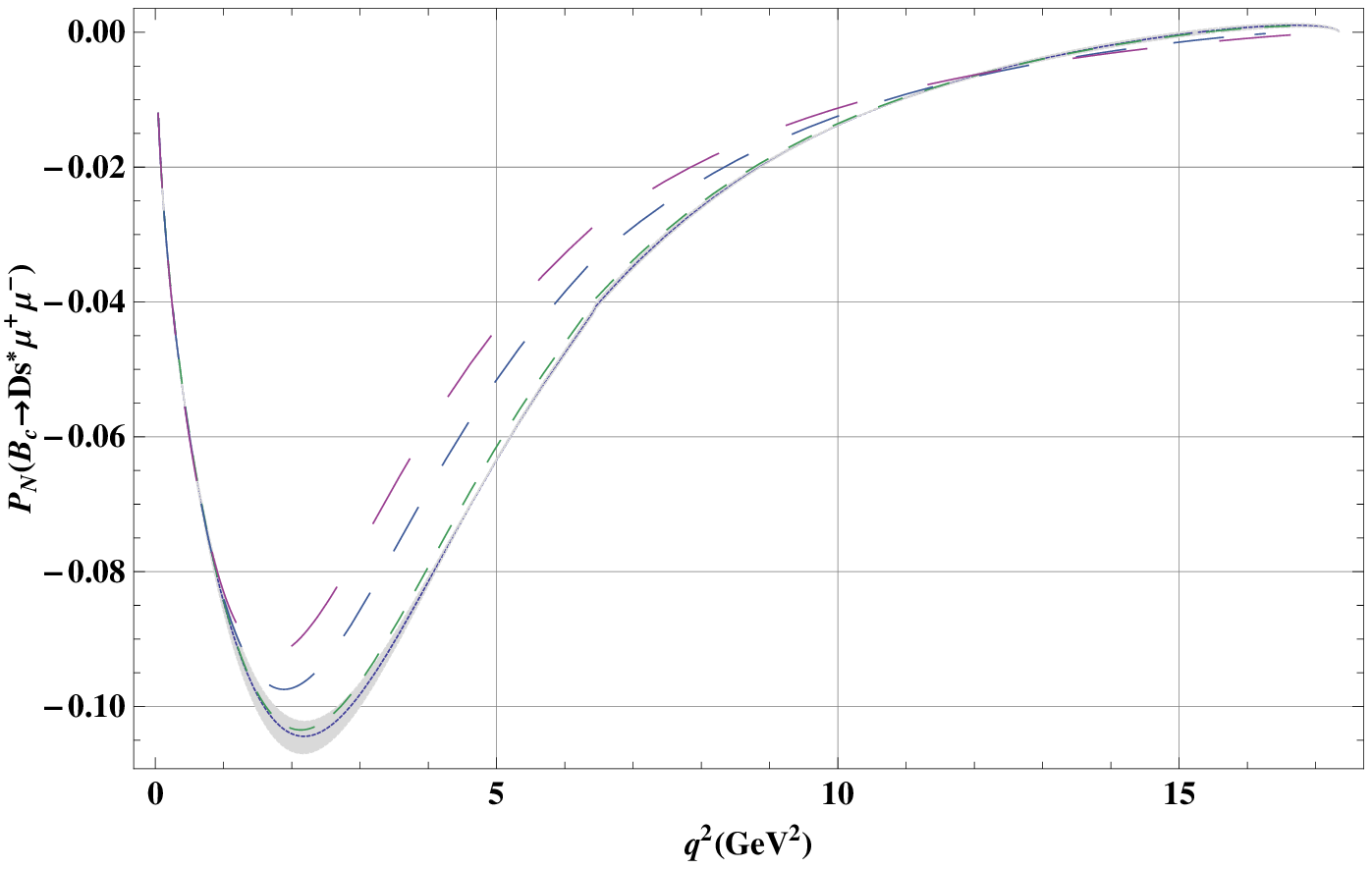} &  &  \\
\includegraphics[scale=0.6]{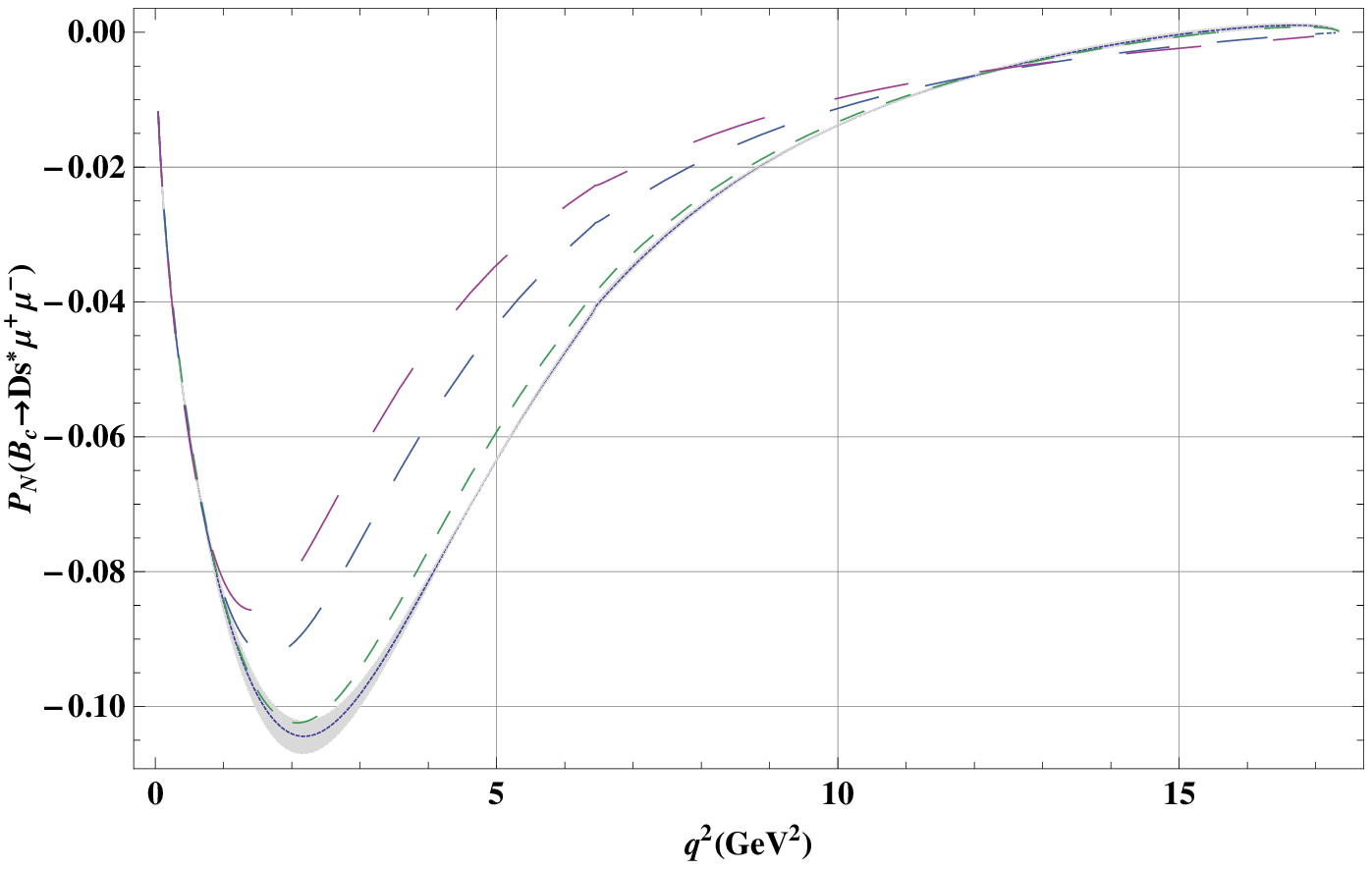} \includegraphics[scale=0.6]{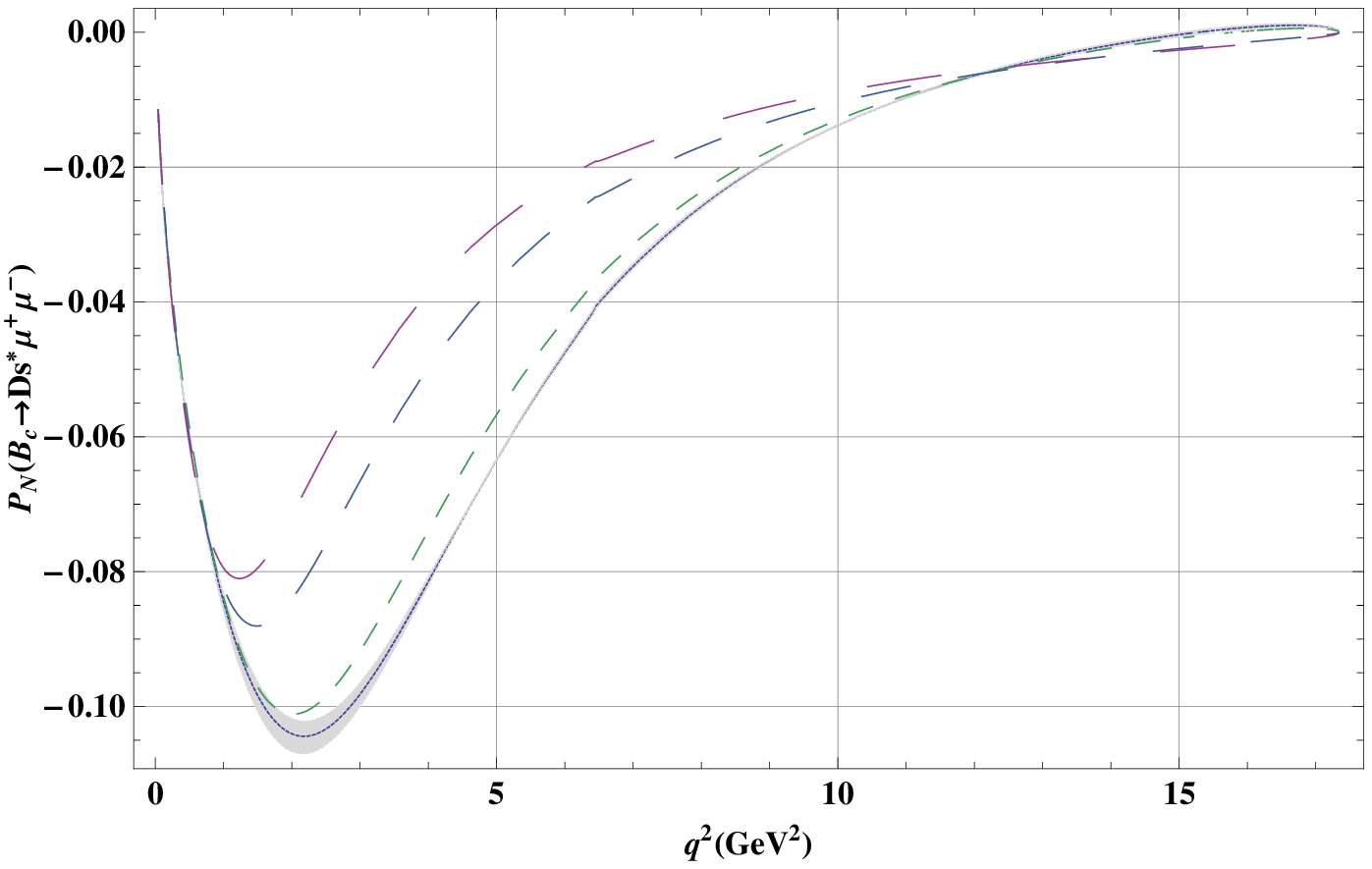}
\put (-350,240){(a)} \put (-100,240){(b)} \put (-350,0.2){(c)}
\put(-100,0.2){(d)} \vspace{-0.5cm} &  &
\end{tabular}%
\end{center}
\caption{The dependence of normal lepton polarization of $B_{c}\rightarrow D_{s}^{\ast
}\protect\mu ^{+}\protect\mu ^{-}$ on $q^{2}$ for different values of $%
m_{t^{\prime }}$ and $\left\vert V_{t^{\prime }b}^{\ast }V_{t^{\prime
}s}\right\vert $. The values of fourth generation parameters and the legends
are same as in Fig.1. }
\label{NP-muon}
\end{figure}

\begin{figure}[tbp]
\begin{center}
\begin{tabular}{ccc}
\vspace{-0.3cm} \includegraphics[scale=0.6]{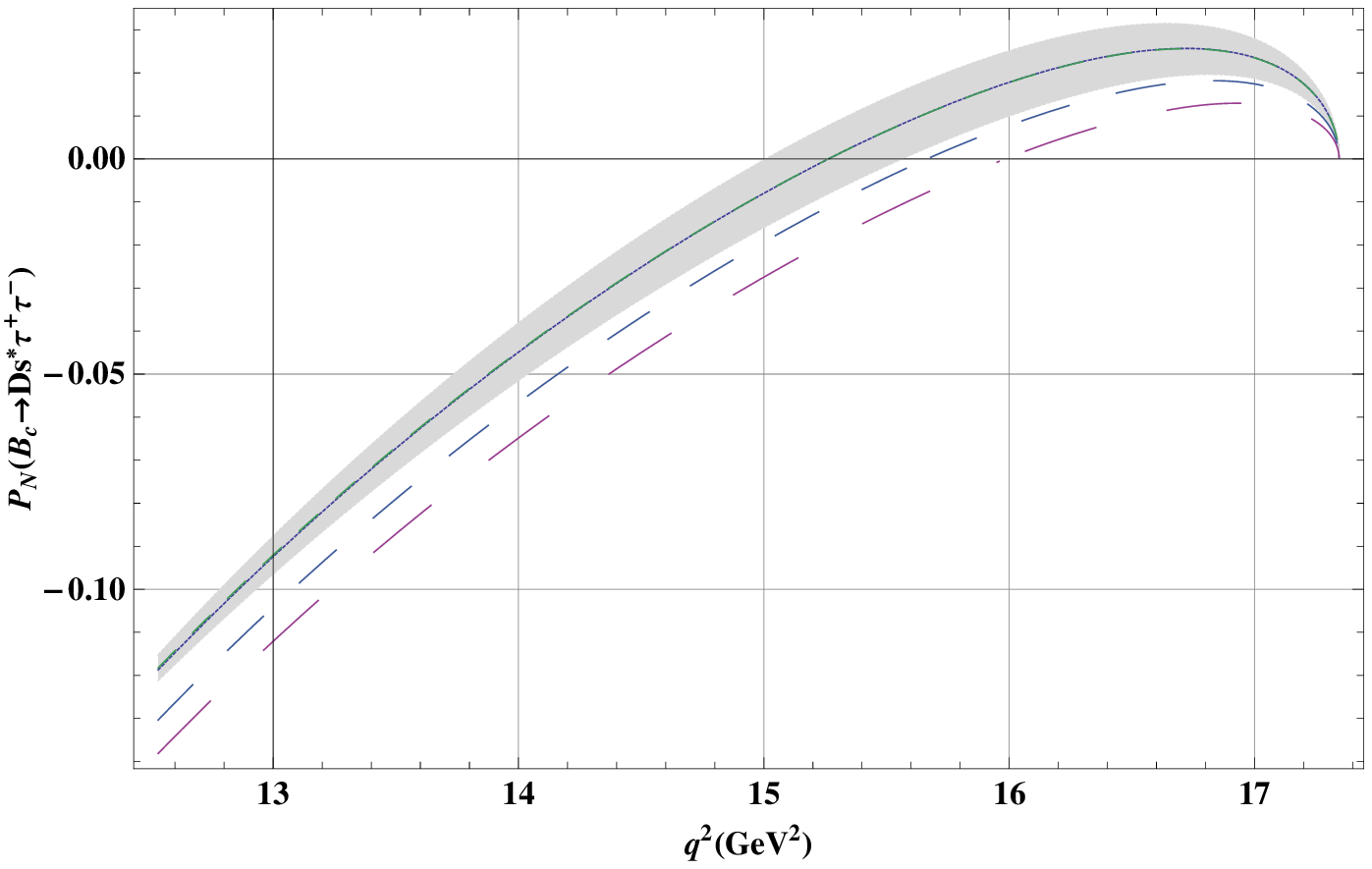} %
\includegraphics[scale=0.6]{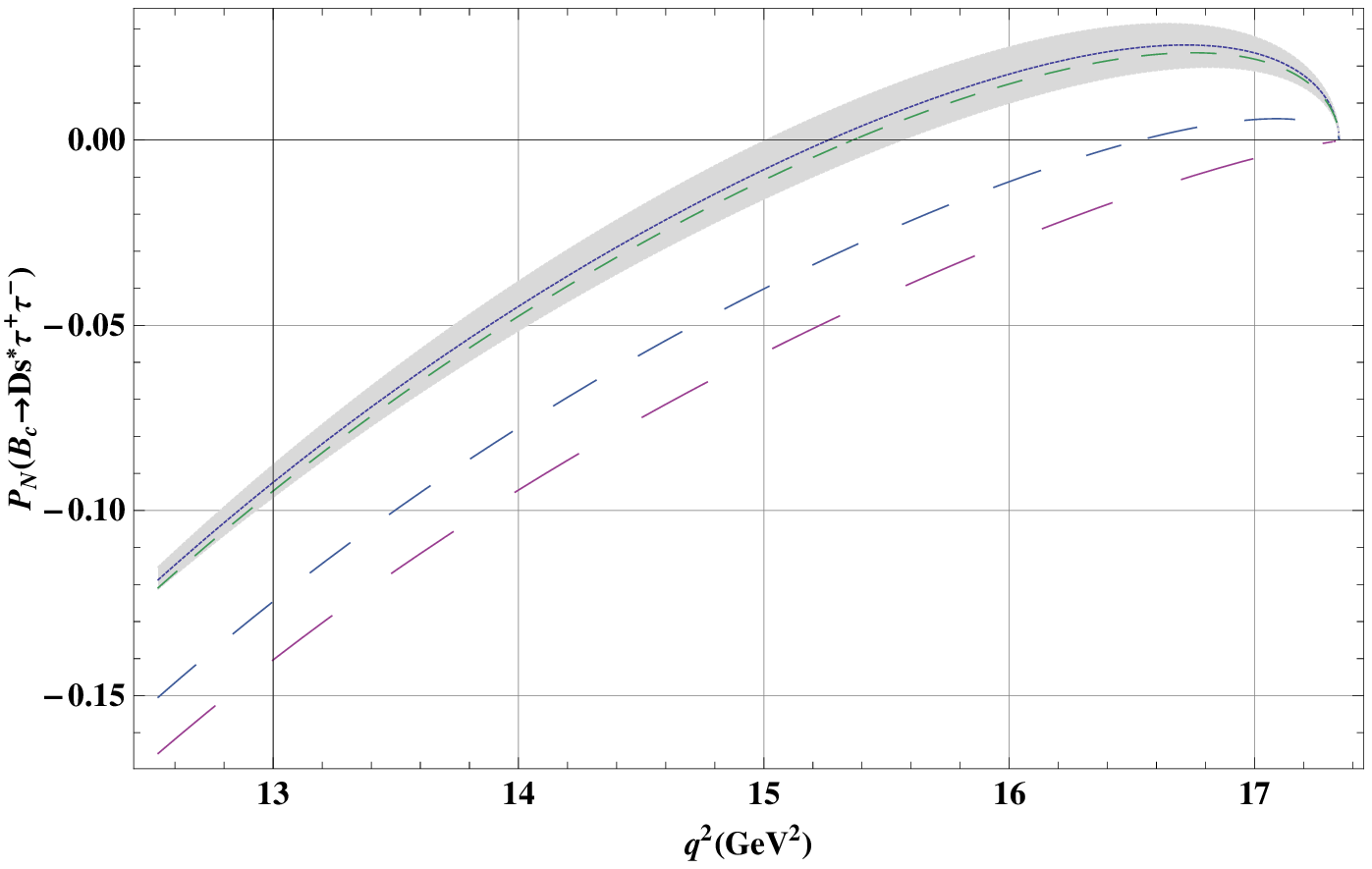} &  &  \\
\includegraphics[scale=0.6]{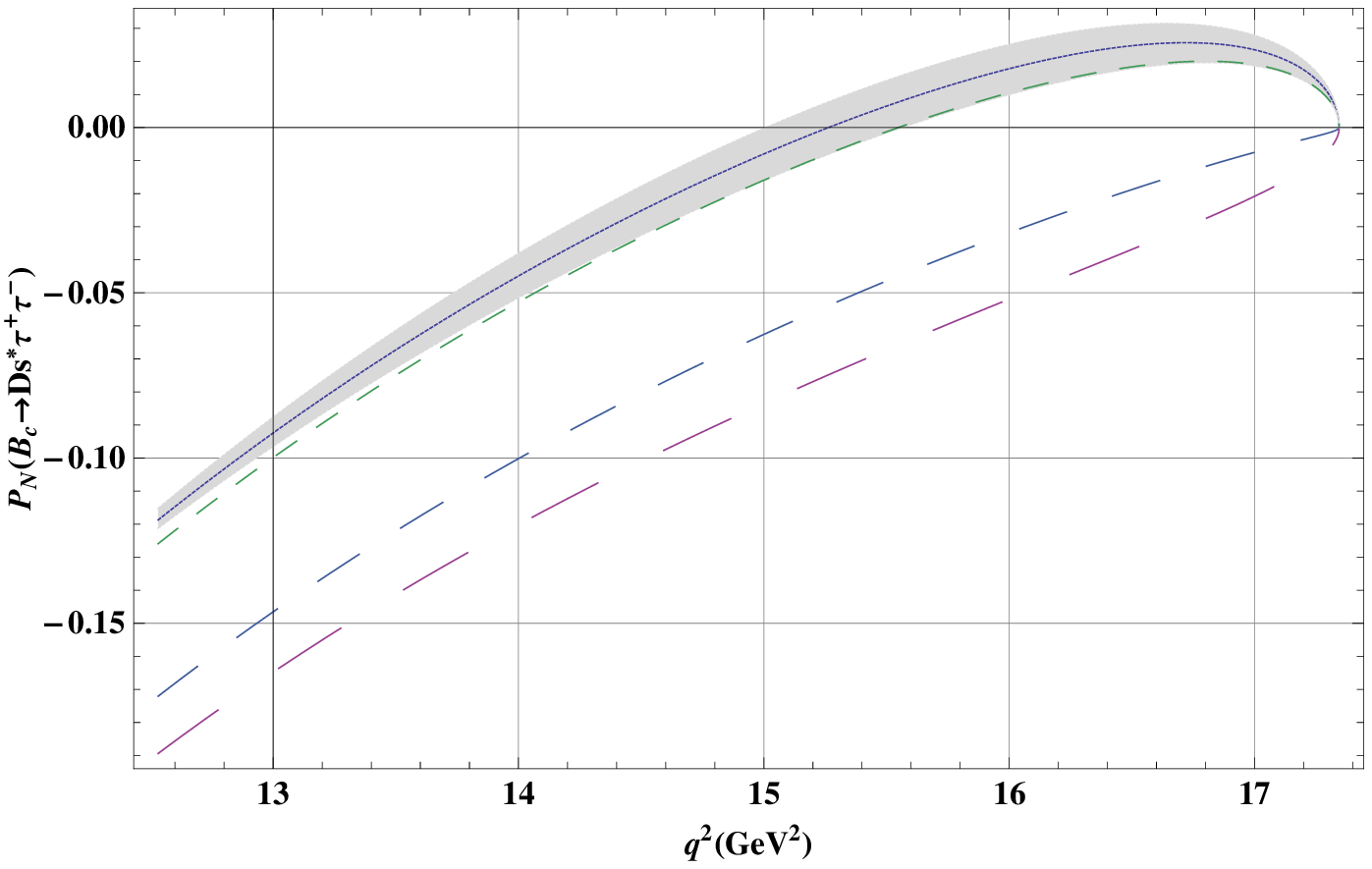} \includegraphics[scale=0.6]{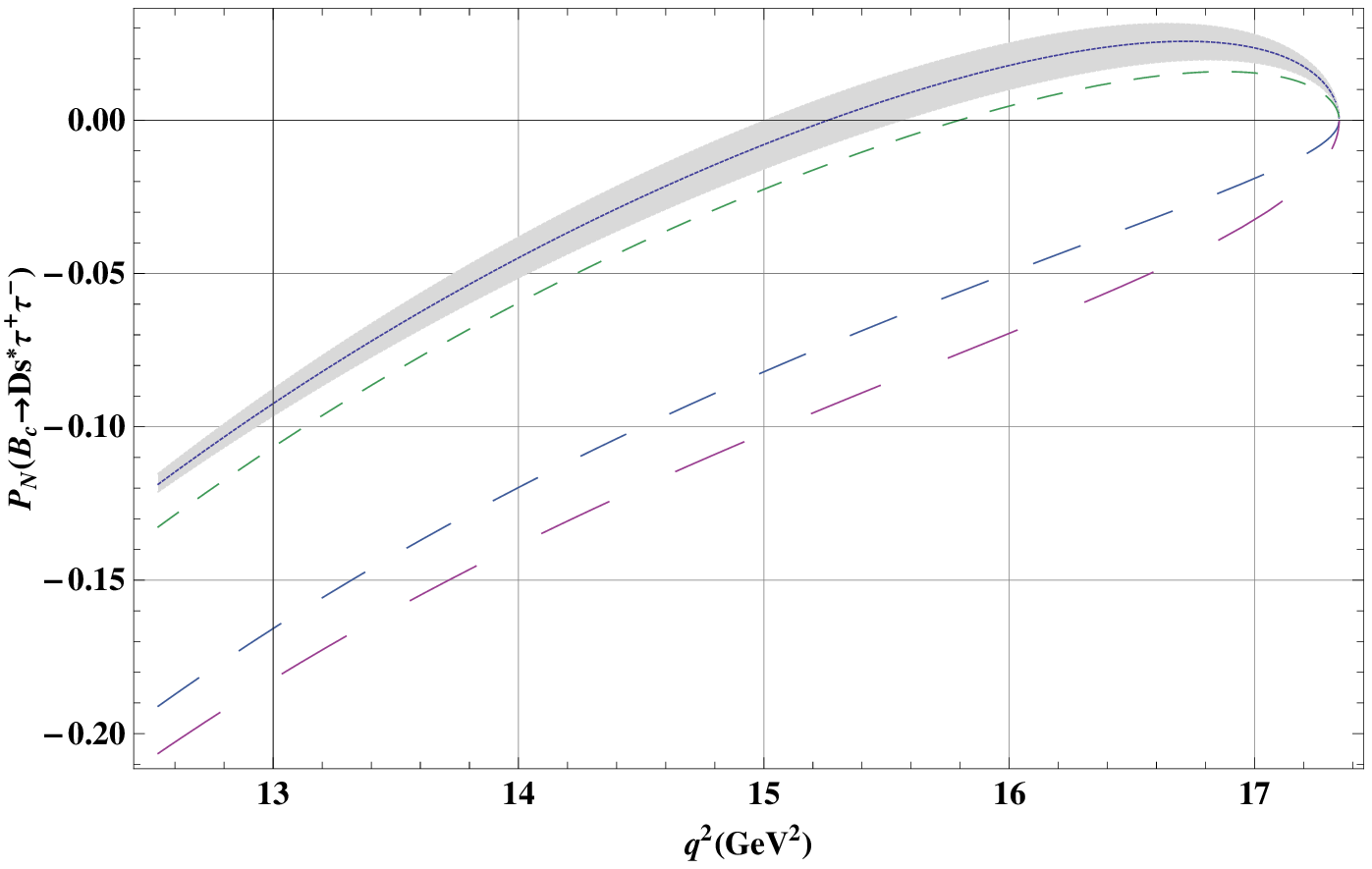}
\put (-350,240){(a)} \put (-100,240){(b)} \put (-350,0.2){(c)}
\put(-100,0.2){(d)} \vspace{-0.5cm} &  &
\end{tabular}%
\end{center}
\caption{The dependence of normal lepton polarization of $B_{c}\rightarrow D_{s}^{\ast
}\protect\tau ^{+}\protect\tau ^{-}$ on $q^{2}$ for different values of $%
m_{t^{\prime }}$ and $\left\vert V_{t^{\prime }b}^{\ast }V_{t^{\prime
}s}\right\vert $. The values of fourth generation parameters and the legends
are same as in Fig.1.  }
\label{NP tauon}
\end{figure}

The dependence of normal lepton polarization asymmetries for
$B_c\rightarrow D_{s}^{\ast }\ell^{+}\ell^{-}$ on the momentum transfer
square are presented in Figs. \ref{NP-muon} and \ref{NP tauon}. In terms of Eq.
(\ref{normal-polarization}), it can be seen that it is proportional to the mass of the
final state lepton and for $\mu $ its values are expected to be
small and Fig. \ref{NP-muon} displays it in the SM as well as in the SM4 for
the different values of fourth generation parameters. In SM4, there is a
slight shift from the SM value which, however, due to its small value
it is hard to measure experimentally. Now, for the $\tau ^{+}\tau ^{-}$ channel, Eq. (\ref{normal-polarization})
we will have a large value of normal lepton polarization compared to
the $\mu ^{+}\mu ^{-}$ case in the SM. Fig. \ref{NP tauon} shows that there is a
significant decrease in the value of $P_{N}$ in SM4 compared to the SM
and its experimental measurement will give us some clear hints of the fourth generation of quarks.

\begin{figure}[tbp]
\begin{center}
\begin{tabular}{ccc}
\vspace{-0.3cm} \includegraphics[scale=0.6]{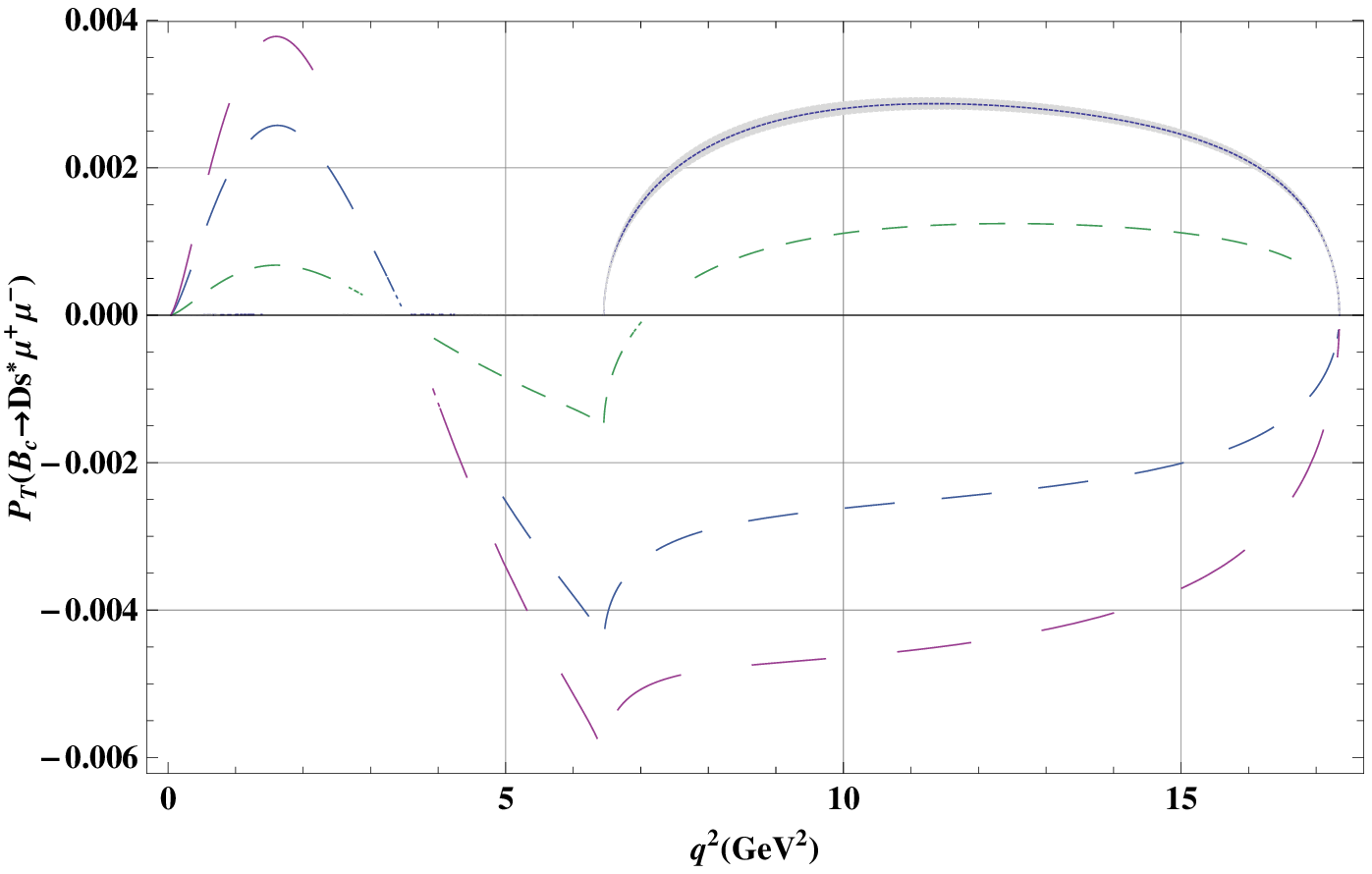} %
\includegraphics[scale=0.6]{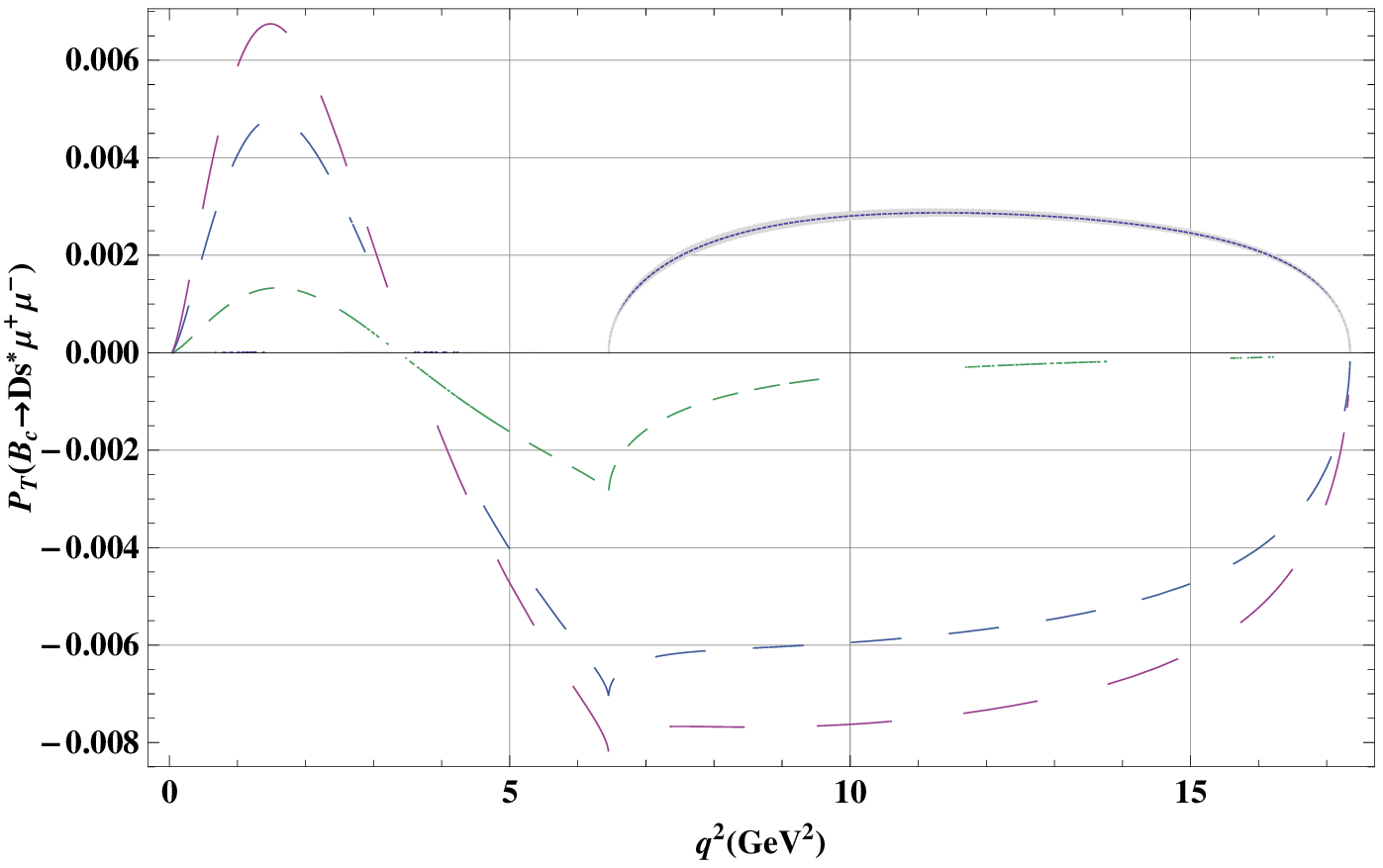} &  &  \\
\includegraphics[scale=0.6]{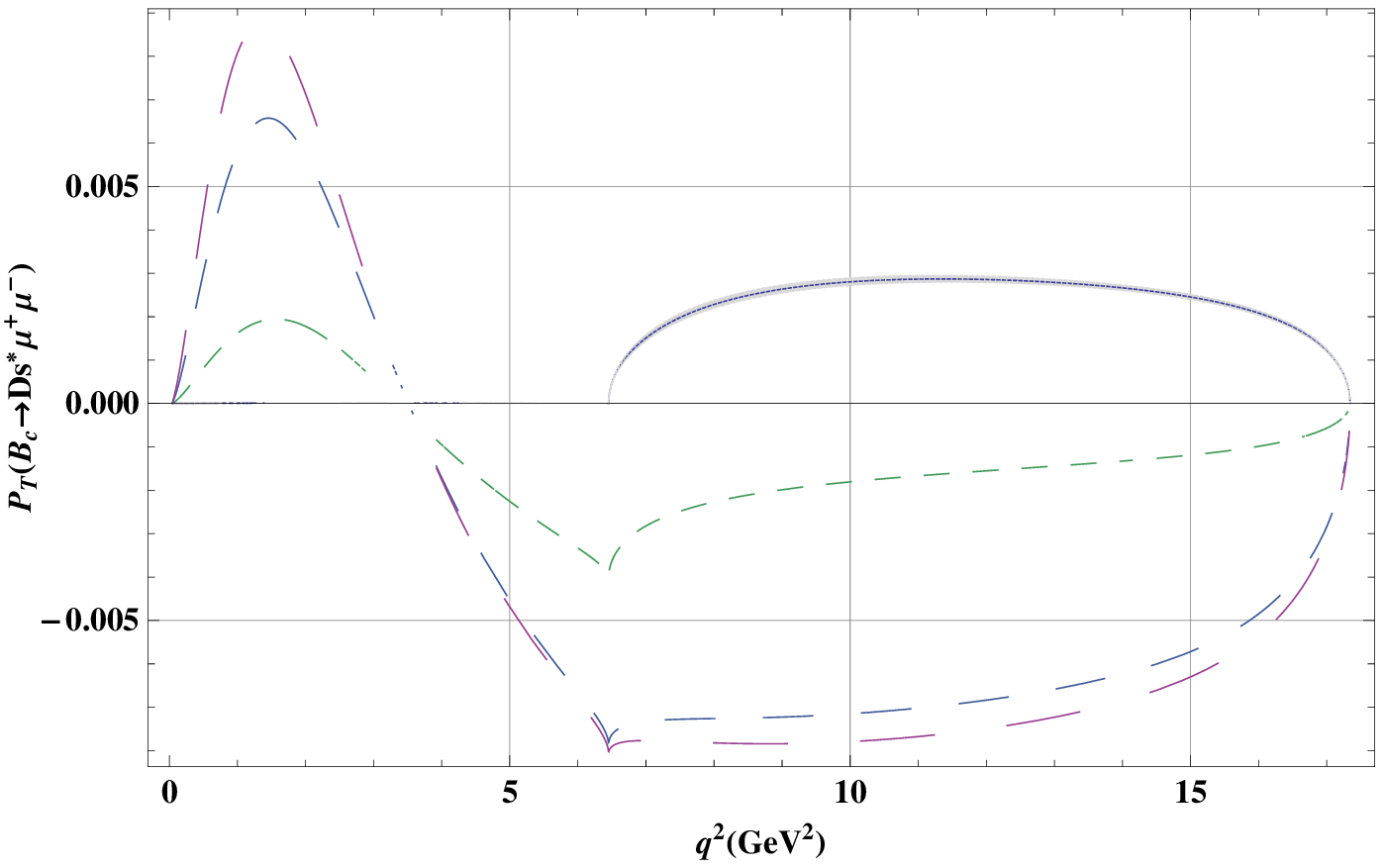} \includegraphics[scale=0.6]{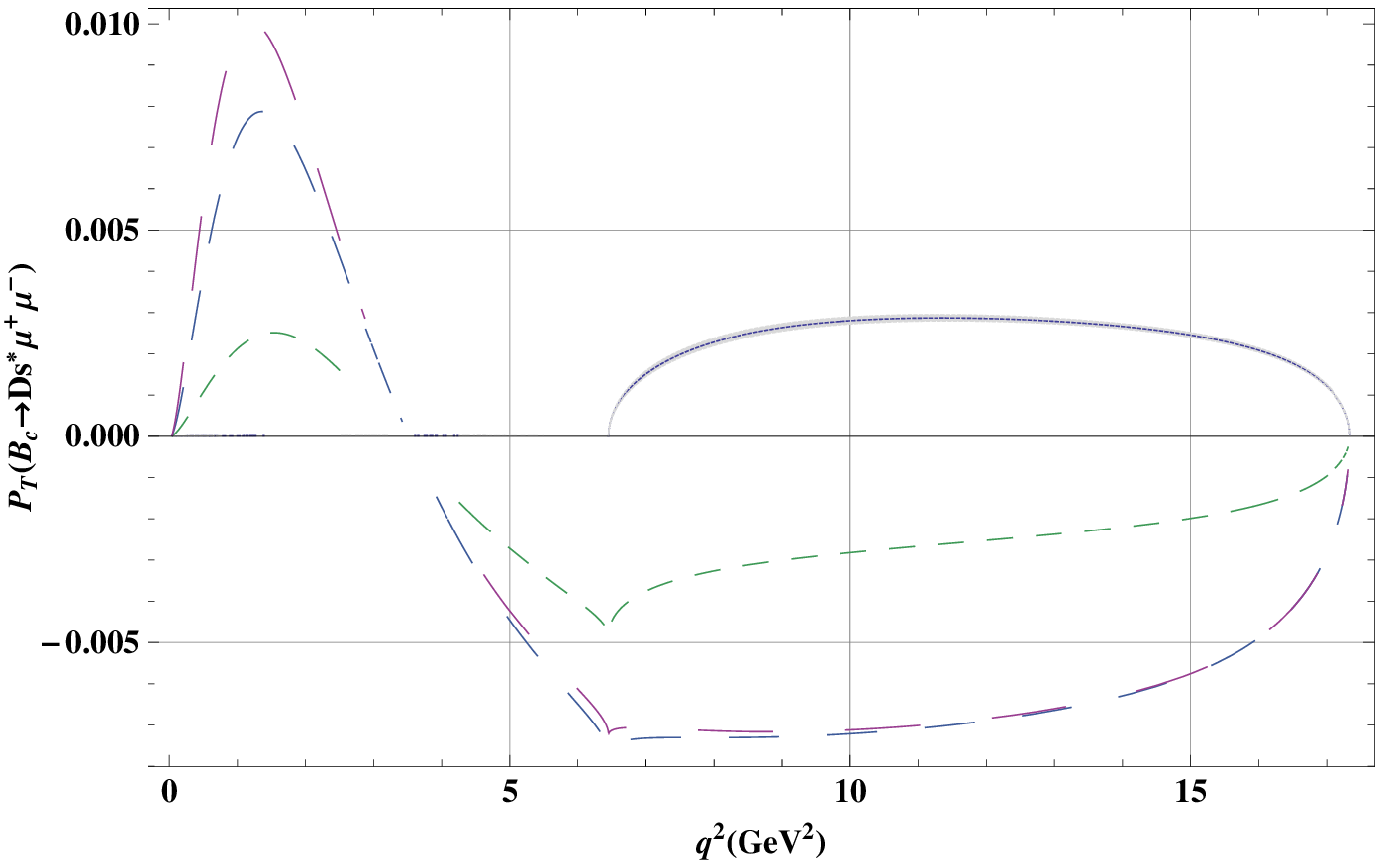}
\put (-350,240){(a)} \put (-100,240){(b)} \put (-350,0.2){(c)}
\put(-100,0.2){(d)} \vspace{-0.5cm} &  &
\end{tabular}%
\end{center}
\caption{The dependence of transverse lepton polarization of $B_{c}\rightarrow D_{s}^{\ast
}\protect\mu ^{+}\protect\mu ^{-}$ on $q^{2}$ for different values of $%
m_{t^{\prime }}$ and $\left\vert V_{t^{\prime }b}^{\ast }V_{t^{\prime
}s}\right\vert $. The values of fourth generation parameters and the legends
are same as in Fig.1. }
\label{TP-muon}
\end{figure}

\begin{figure}[tbp]
\begin{center}
\begin{tabular}{ccc}
\vspace{-0.3cm} \includegraphics[scale=0.6]{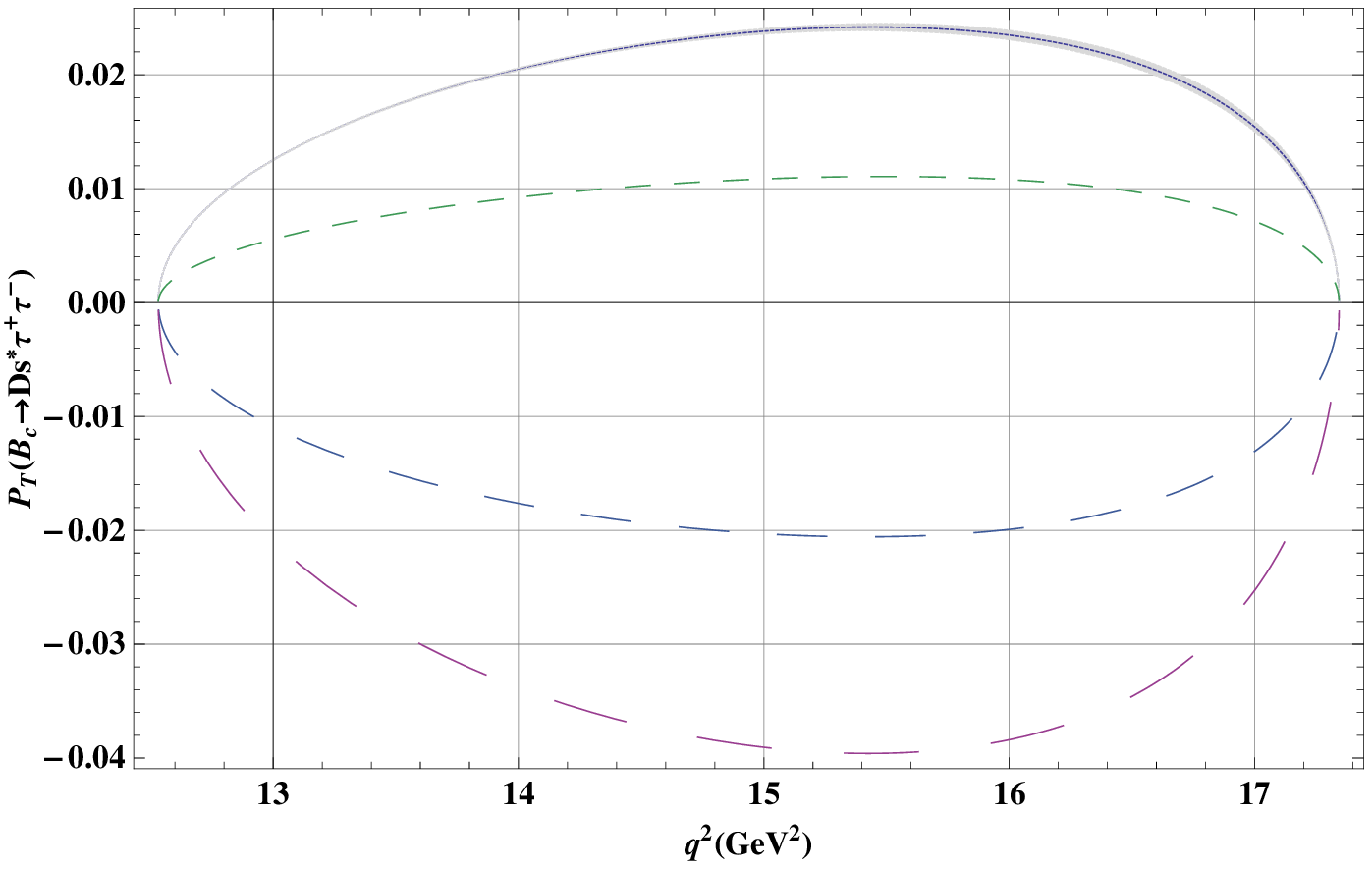} %
\includegraphics[scale=0.6]{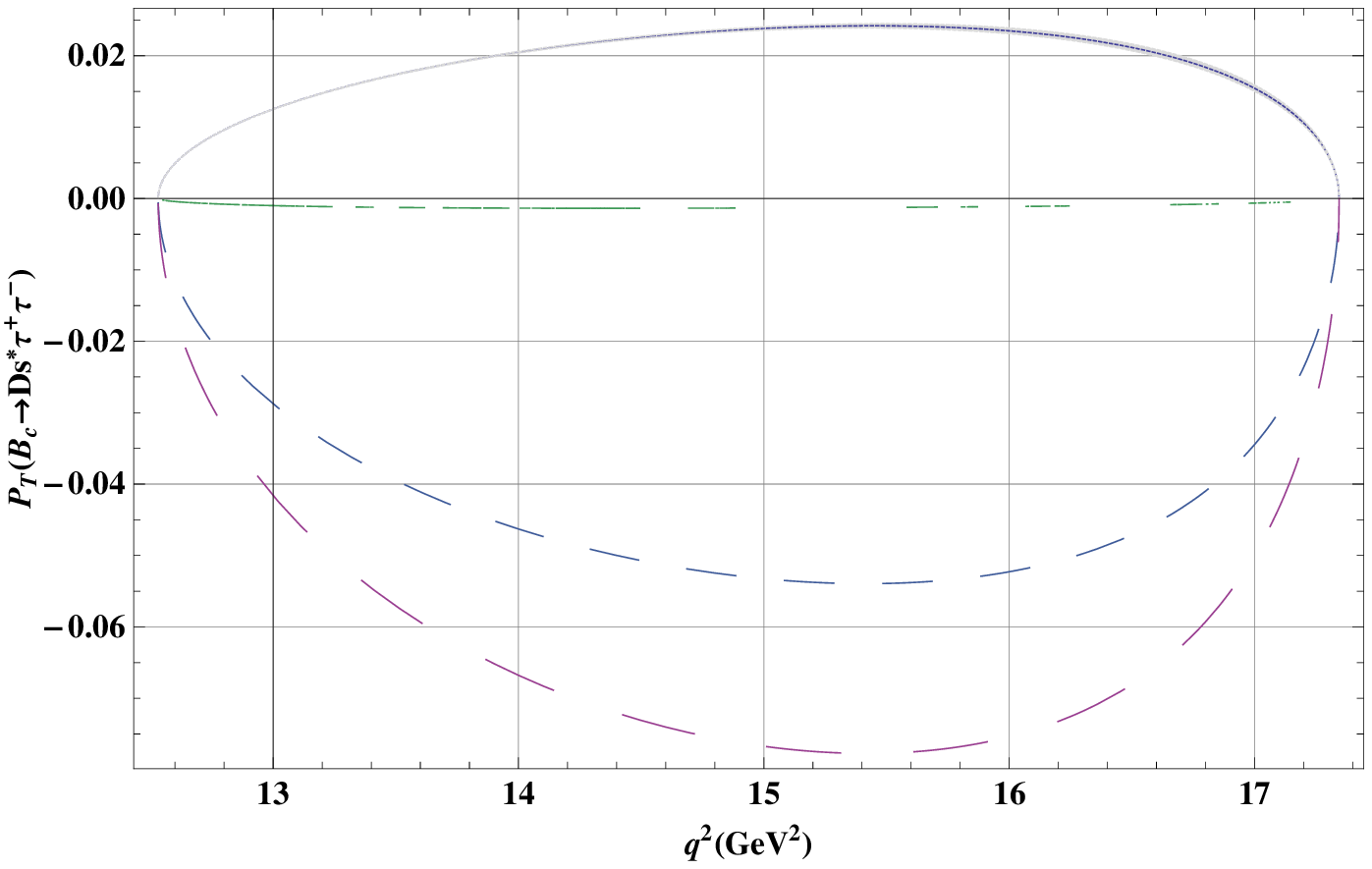} &  &  \\
\includegraphics[scale=0.6]{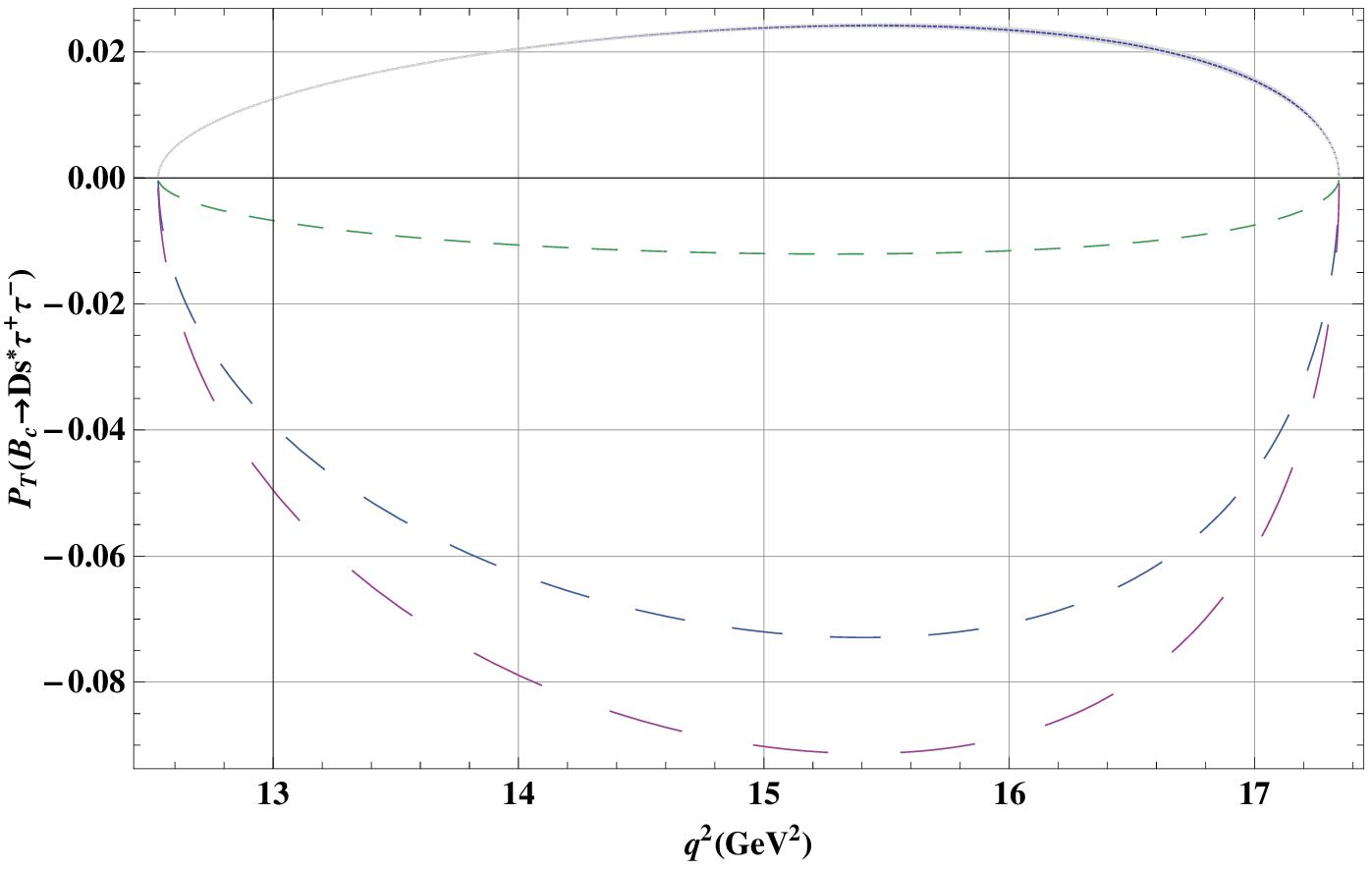} \includegraphics[scale=0.6]{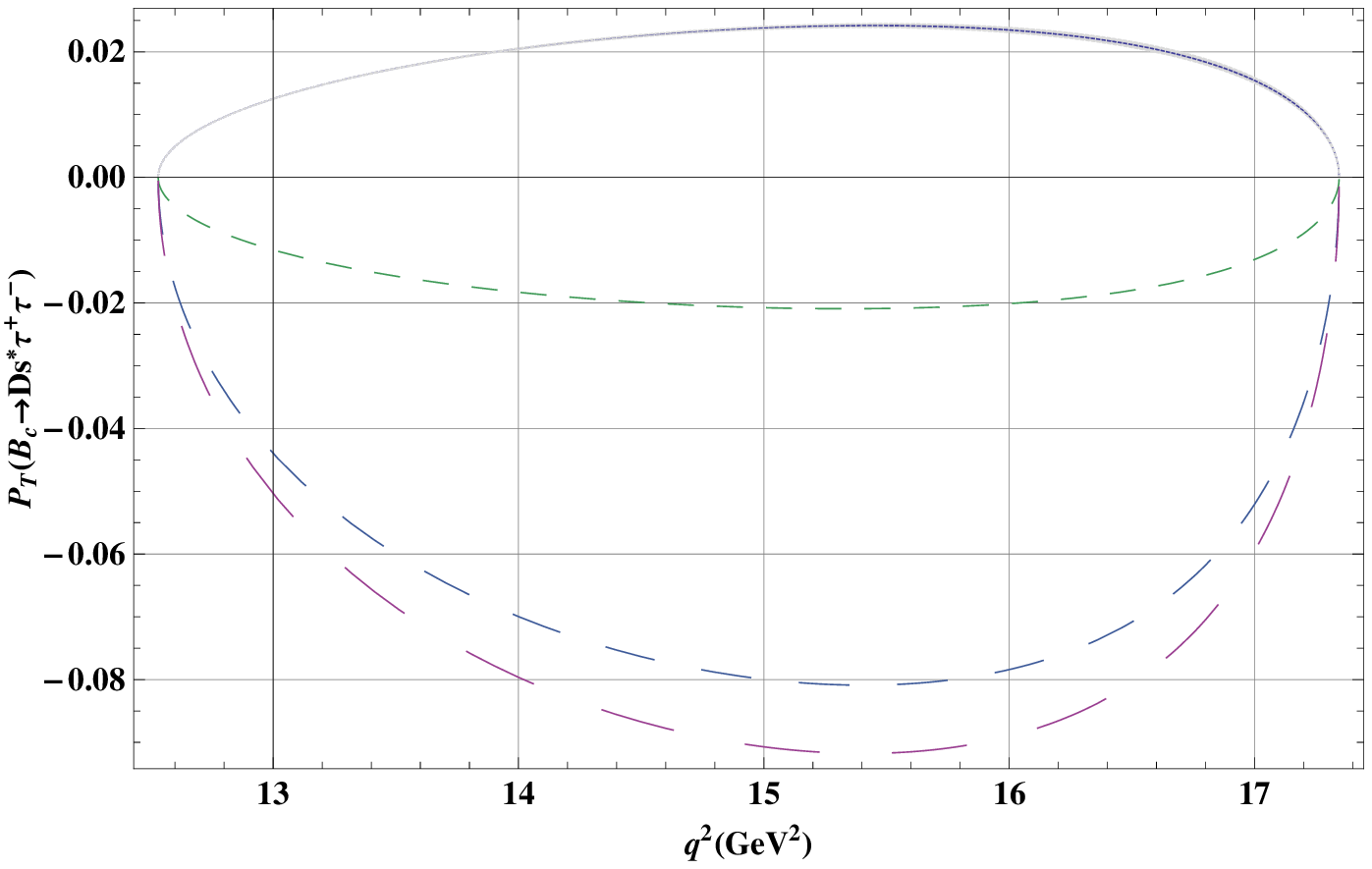}
\put (-350,240){(a)} \put (-100,240){(b)} \put (-350,0.2){(c)}
\put(-100,0.2){(d)} \vspace{-0.5cm} &  &
\end{tabular}%
\end{center}
\caption{The dependence of transverse lepton polarization of $B_{c}\rightarrow D_{s}^{\ast
}\protect\tau ^{+}\protect\tau ^{-}$ on $q^{2}$ for different values of $%
m_{t^{\prime }}$ and $\left\vert V_{t^{\prime }b}^{\ast }V_{t^{\prime
}s}\right\vert $. The values of fourth generation parameters and the legends
are same as in Fig.1. }
\label{TP tauon}
\end{figure}

Figs. \ref{TP-muon} and \ref{TP tauon}  show the value of transverse lepton polarization both in the
SM as well as in SM4 for $B_c\rightarrow D_{s}^{\ast } \ell^{+}\ell^{-}$ decay. One
can see that it is zero in the SM but non zero in sequential fourth
generation SM (SM4). This non zero value comes from the interference of the
Wilson coefficient for SM4 which are complex in SM4, see Eqs. (\ref{wilson-tot}, \ref%
{Transverse-polarization}). The transverse lepton polarization is proportional to the
lepton mass and as the imaginary part of the Wilson coefficients therefore, it is expected
to be small which can also be seen from Figs. \ref{TP-muon} and \ref{TP tauon}.

\section{Summary and Conclusion:}

We have carried out the study of invariant mass spectrum, forward-backward asymmetries,
polarization asymmetries of final state $D_{s}^{\ast}$ meson and lepton in
$B_{c}\rightarrow D_{s}^{\ast} \ell ^{+}\ell^{-}$ ($\ell=\mu ,\tau $) decays in
the Standard Model with extra sequential generation of quarks (SM4). Particularly, we analyze the
effects of fourth generation up-type quark mass $m_{t}^{\prime}$ and corresponding CKM matrix element
$\left\vert V_{t^{\prime }b}^{\ast }V_{t^{\prime
}s}\right\vert $ to this process and our main outcomes can be summarized as
follows:
\begin{itemize}
\item We have found that the branching ratios deviate sizably from that of the SM
in almost all momentum transfer region.
The study shows that the $\mathcal{BR}$ is an
increasing function of the fourth generation
parameters $m_{t^{\prime}}$ and $V_{t^{\prime} b}V_{t^{\prime} s}$. At maximum values of these
parameters, i.e. $|V_{t^{\prime} b}V_{t^{\prime} s}|=0.012$ and
$m_{t^{\prime}}=600$ GeV, the values of $\mathcal{BR}$ increase
approximately by 3 times of their SM values when the
final leptons are muons or tauons. Hence the accurate measurement of the
$\mathcal{BR}$ for these decays is very important tool to explore
physics beyond the SM.

\item The value of the forward-backward asymmetry decreases significantly
from that of the SM value in SM4 when the mass of the fourth
generation quark varies from $300$ GeV to $600$ GeV. The value of
the zero position of forward-backward asymmetry shifted towards the
left for all values of $\left\vert V_{t^{\prime }b}^{\ast
}V_{t^{\prime }s}\right\vert $ in $B_{c }\rightarrow D_{s}^{\ast
}\mu ^{+}\mu ^{-}$ decay. This shifting is significant for
large values of the fourth generation CKM matrix elements
$\left\vert V_{t^{\prime }b}^{\ast }V_{t^{\prime }s}\right\vert $
and fourth generation top quark mass $m_{t^\prime}$. As it is almost
free from the hadronic uncertainties therefore this shifting will help us to find
clues of the SM4.

\item The polarization effects of final state $D_{s}^{\ast}$ meson and lepton are
calculated in the sequential fourth generation SM4. It is found that the SM4 effects are very
promising, which could be measured at present and future experiments like LHCb
where large numbers of $\ b\bar{b}$ pairs are expected to be produced.
\end{itemize}

In short, the precision measurements of these observables at Tevatron and LHC will help us to
find the indications of new physics encoded in the fourth generation
parameters such as $V_{t^{\prime} b}V_{t^{\prime} s}$ and
$m_{t^{\prime}}$.
\section*{Acknowledgements}

Helpful discussions with Prof. Riazuddin and Prof. Fayyazuddin are
greatly acknowledged. M. J. A acknowledge the grant provided by
Quaid-i-Azam University from University Research Funds.

\end{document}